\documentclass[a4paper,twocolumn,11pt,accepted=2022-12-11]{quantumarticle}
\pdfoutput=1
\usepackage[utf8]{inputenc}
\usepackage{braket}
\usepackage{amsmath,amssymb,graphicx}
\usepackage{caption}
\usepackage{subcaption}
\usepackage{qcircuit}
\usepackage{xcolor}
\usepackage{verbatim}
\usepackage[normalem]{ulem}
\usepackage{graphicx}
\usepackage[rightcaption]{sidecap}

\newcommand{\norm}[1]{\left\lVert#1\right\rVert}
\newcommand{\R}{\mathbb{R}}

\def\ket#1{\mathinner{|{#1}\rangle}}
\newcommand{\revised}[1]{\textcolor{black}{#1}}

\usepackage{float}
\usepackage{multirow}

\begin{document}

\title{Quantum Methods for Neural Networks and Application to Medical Image Classification}

\author{Jonas Landman}
\affiliation{QC Ware, Palo Alto, USA and Paris, France}
\affiliation{IRIF, CNRS - University of Paris, France}
\email{jonas.landman@qcware.com}
%\orcid{0000-0000-0000-0000}

\author{Natansh Mathur}
\affiliation{QC Ware, Palo Alto, USA and Paris, France}
\affiliation{Indian Institute of Technology Roorkee, India}
%\orcid{0000-0000-0000-0000}

\author{Yun Yvonna Li}
\affiliation{F. Hoffmann La Roche AG}
%\orcid{0000-0000-0000-0000}

\author{Martin Strahm}
\affiliation{F. Hoffmann La Roche AG}
%\orcid{0000-0000-0000-0000}

\author{Skander Kazdaghli}
\affiliation{QC Ware, Palo Alto, USA and Paris, France}
%\orcid{0000-0000-0000-0000}

\author{Anupam Prakash}
\affiliation{QC Ware, Palo Alto, USA and Paris, France}
%\orcid{0000-0000-0000-0000}

\author{Iordanis Kerenidis}
\affiliation{QC Ware, Palo Alto, USA and Paris, France}
\affiliation{IRIF, CNRS - University of Paris, France}
%\orcid{0000-0000-0000-0000}
\maketitle

\begin{abstract}
  Quantum machine learning techniques have been proposed as a way to potentially enhance performance in machine learning applications. 
  In this paper, we introduce two new quantum methods for neural networks. 
  \revised{The first method is quantum-assisted neural networks, where a quantum computer is used to perform inner product estimation for inference and training of classical neural networks.
  The second is a quantum orthogonal neural network, which is based on a quantum pyramidal circuit as the building block for implementing orthogonal matrix multiplication. We provide an efficient way for training such orthogonal neural networks; novel algorithms are detailed for both classical and quantum hardware, where both are proven to scale asymptotically better than previously known training algorithms. }
 Extensive experiments applied to medical image classification tasks using current state of the art quantum hardware are presented, where different quantum methods are compared with classical ones, on both real quantum hardware and simulators. Our results show that the proposed quantum networks generate similar level of accuracy compared with classical neural networks while demonstrating competitive scalability, supporting the promise that quantum methods can be useful in solving visual tasks, given the advent of better quantum hardware.
\end{abstract}

\section{Introduction}

The possibility to perform machine learning tasks using quantum computers has been an area of extensive research in the past few years, and a number of different avenues of research have emerged. At the high level, machine learning is made of three key ingredients: data domain, hypothesis family, and the learning algorithm. In this context, \emph{quantum machine learning} can have different meanings depending on which parts remain classical or become quantum. A short overview is provided in the following, where the different variants of quantum machine learning are outlined.

\subsubsection*{Related Work}

At the advent of quantum computing, research has focused on long-term quantum machine learning applications with a traditional computer science and linear algebraic approach, assuming the existence of a full-scale fault-tolerant quantum computer. The data sets and the hypothesis family are classical, and a quantum computer is used to assist in the learning phase, aiming at speeding up the process.
Indeed, it has been shown that quantum algorithms using ideas from the HHL algorithm \cite{HHL} and Phase Estimation could speed up core applications of classical machine learning, for example Principal Component Analysis, Support Vector Machines, Recommendation systems, Topological Data Analysis, or clustering \cite{lloyd2013quantum, lloyd2014quantum,kerenidis2016quantum,kerenidis2019qmeans, lloyd2016quantum}. These quantum algorithms perform to a large extent the same linear algebraic operations as the corresponding classical algorithms, while providing a provable speedup on the computation time in certain cases. The speedup comes from the ability of a quantum computer to efficiently perform certain linear algebraic tasks, for example inversion of a matrix. The idea of using a quantum computer to speedup computation via HHL-based techniques was also extended to neural networks, for example in  \cite{QNN2018, QCNN2019, QNN2020}, where again the quantum computer is used in order to perform efficiently matrix-matrix multiplications for inference and learning. 
%In this framework, one usually loads classical data into quantum states using an amplitude encoding, an encoding scheme where the coordinates of the input data vectors are encoded as amplitudes in a quantum superposition, using what is called a QRAM architecture. In this context, \emph{quantum machine learning algorithms} or \emph{quantum neural networks} are quantum algorithms that follow closely the computation of the corresponding classical algorithms with proven running times and performance. 

With the advent of the first small-scale noisy quantum computers, the study of quantum machine learning and quantum neural network methods has again received a lot of attention. While current quantum hardware is far from being powerful enough to compete with classical machine learning algorithms, the first proof of concept demonstrations of interesting quantum machine learning algorithms have started to appear \cite{QNN2018,CNN2019,Image2020,Dressed2020, Semisupervised2020,Polyadic2020,Supervised2018,Hierarchical2018, Kiani20}.
The approach taken by these publications is suited for near term and noisy quantum devices, and is based on Variational Quantum Circuits (VQCs) \cite{cerezo2020variational,bharti2021noisy}, which are shallow quantum circuits with parametrized gates. Such parametrized quantum circuits can be trained so that the final measurement on such quantum circuits would give the answer to a machine learning task. While the data can be either classical or quantum \cite{noirhomme2011far}, the hypothesis family and the learning algorithms are fundamentally different from the classical ones. The circuits usually consist of an encoding component with gate parameters dependent on the input data vectors, and an ansatz component where gate parameters, usually angles in rotation gates, are trained using a learning algorithm \cite{perez2020data}. The angles are updated using classical gradient descent method until convergence, which requires computation of these gradients using the \emph{parameter shift rule} \cite{mitarai2018quantum, schuld2019evaluating}. 
These quantum learning methods using parametrized quantum circuits with trainable parameters and gradient computations have originally been coined \emph{quantum neural networks}, while more recently the terms \emph{quantum models} or \emph{quantum kernels} have been proposed as more appropriate \cite{schuld2021quantum, schuld2021effect}. Indeed, while these VQCs share similarities with classical neural networks, one cannot easily find a strict correspondence between a classical neural network and a VQC. Take for instance, Quantum Convolutional Neural Networks defined in \cite{cong2019quantum}, which outline some common high-level properties shared between the proposed VQC and a classical Convolutional Neural Network, namely that they are both translation invariant and share some notion of pooling, etc. However, the quantum computation found in \cite{cong2019quantum} does not strictly correspond to a convolution layer in classical CNN. 

To be more precise, consider a classical neural network with a single fully connected layer with $n$-dimensional inputs and $n$-dimensional outputs applied to an $n$-dimensional data point, the mathematical operation in consideration is a multiplication of an $n \times n$ matrix (the weight matrix) with an $n$-dimensional vector (the data), with a non-linearity applied to the output of this multiplication. In the case of a VQC using $n$ qubits as a quantum neural network, it is customary to encode a classical $n$-dimensional data point onto the state of the $n$ qubits, which is mathematically a $2^n$-dimensional unit vector and that there are different ways of performing this step. Then the ansatz of the quantum circuit is applied, which corresponds to a multiplication of the $2^n$-dimensional vector with a $2^n \times 2^n$ unitary matrix. Operations such as measuring or tracing out can be used as a form of non-linearity to some of the output qubits. 

While these methods based on VQCs are quite promising and try to take advantage of this exponentially larger space for learning, they also show the need both for more solid theoretical evidence that such quantum models can be easily trained to provide high accuracy, and for larger quantum hardware to test and fine tune these methods on real-world data. In addition,  one needs to consider phenomena such as barren plateaus \revised{\cite{mcclean2018barren, marrero2020entanglement}} 
that can occur while performing optimization in this exponentially large space while designing the architectures,  cost-functions, and initialization of the parameters \cite{QCNNplateaus2020, Train2020}. In short, further work is needed to understand the power and limitations of general VQC for machine learning applications.

\subsubsection*{Our Contribution}

In our work, we take an approach to quantum machine learning that bring the two approaches described above together, namely the first based on the ability of a quantum computer to efficiently perform linear algebraic tasks, and the second based on VQC. The goal is to keep the theoretical guarantees of the first approach, while making the algorithm available to noisy intermediate-scale quantum computers. In order to achieve this goal, we designed quantum circuits that can be implemented on noisy quantum computers to perform linear algebraic operations such as orthogonal matrix multiplication, inner product estimation, and data loading. These are parametrized quantum circuits whose angles can be used to load classical vectors or matrices into the quantum circuit. These quantum circuits are used as layers of the neural network, and their parameters are trained using a classical gradient descent algorithm. These quantum circuits used as individual layers of a neural network can be thought of as VQCs, since each consist of parametrized gates with trainable angles. Nevertheless, the proposed quantum circuits do possess some very specific properties stemming from the specific data encoding strategy and the structure of the circuits themselves, which enable us both to find efficient ways to compute the gradients during learning, and to relate the quantum computation to that of a corresponding classical neural network layer, thus making the comparison to classical neural networks more meaningful. These properties are discussed in detailed in the methods and the results sections below.

We start by presenting two quantum methods for neural networks.
The first one (Section \ref{sec:quantum_assisted_nn}) utilizes the quantum computer to assist the learning process of any classical neural network with fully connected linear layers. More precisely, we use a quantum algorithm for the estimation of inner products to assist in the training and inference of classical neural networks by adapting the work in \cite{QNN2020, QCNN2019} to be amenable to current quantum hardware. 
Since we only use the quantum computer to compute the inner products in the neural network, the data and hypothesis family are strictly classical. As for the learning algorithm, including backpropagation and gradient descent, we use a quantum computer to perform part of the computations.
A similar approach has been used before for similarity-based learning in \cite{NearestCentroid2021}, where a quantum Nearest Centroid classifier was run on an 11-qubit trapped-ion machine of IonQ. Here we use it for the first time with neural networks.

The second method (Section \ref{sec:quantum_orthoNN}) uses a parametrized quantum circuit as a single quantum neural network layer, which again has an encoding part and an ansatz part. As we will see in more detail in the following sections, the encoding part constructs a unary amplitude encoding of the classical data point and the ansatz consists of a pyramid-shape circuit with a single type of parametrized two-qubit gates that perform rotation in the two-dimensional subspace. These characteristics result in the property that instead of utilizing the full operation of a quantum neural network layer which corresponds to a multiplication of a $2^n \times 2^n$ unitary matrix with a $2^n$-dimensional vector as is the case for a general VQC, the quantum circuit focuses on a more restricted operation which can be interpreted as an $n \times n$-dimensional orthogonal matrix-vector multiplication where the matrix is in fact the submatrix of the unitary restricted to the unary basis. This directly corresponds to a classical layer with $n$-dimensional input and an $n \times n$ weight matrix. However, even though the dimension of the matrices are the same for both classical and quantum cases, classically training happens on the elements of the weight matrix, in the quantum case the training happens directly on the angles of the quantum gates, thus providing different models.
Therefore, data remains classical, the hypothesis family is quantum while staying equivalent to a classical one with the same dimension, and the learning algorithm is quantum as well.

The quantum layer found in our second method has the added property that the corresponding weight matrix remains orthogonal. In classical deep learning research, orthogonal neural networks (OrthoNNs) have been proposed \cite{jia2019orthogonal, wang2020orthogonal, bansal2018can} as a new type of neural networks the weight matrix for every layer should remain orthogonal. In the evolution of neural network architectures, adding constraints to the weight matrices has often been an effective path. The orthogonality property is proven to enable higher accuracy and avoid vanishing or exploding gradients for deep architectures. Several classical gradient descent methods have been proposed to preserve orthogonality while updating the weight matrices using stochastic gradient descent, but these exact techniques suffer from longer running time (cubic in the input size) while others only can approximate orthogonality. Our quantum approach preserves perfect orthogonality of these weight matrices through each step of back-propagation of the quantum circuit, while having the same running time as the ordinary classical gradient descent methods without the orthogonality property, which is quadratic in the input size. For both methods, we analyze the theoretical quantum advantages compared to classical methods in Section \ref{methods}.

\subsubsection*{Experimental results}

 In Section \ref{sec:Applications} we aim to understand how our proposed quantum methods perform in practice by focusing on medical image classification, which is an important application of machine learning in the healthcare domain and beyond. Despite the success of artificial neural networks to classify, segment, and perform other image related tasks, the complexity of such models and the cost of training continues to increase. One of the latest visual transformer architectures trained on imageNet boasts of more than 2 billion parameters, trained using more than 10 000 TPU core-days \cite{Zhai}. This begs the question whether alternative technologies can provide better solutions in terms of scalability. 

We perform extensive simulations as well as a hardware demonstration, which remains a simple proof of concept. Hardware experiments both for training a quantum neural network and applying forward inferences have been performed on IBM superconducting quantum hardware. We also performed the same training and inference experiments using classical methods as well as simulators of the quantum neural networks for benchmarking the quantum methods.

Our experimental results show that quantum neural networks provide comparable accuracy to equivalent classical neural networks for medical image classification.
From both simulation and hardware experiments, we see that for a number of classification tasks the quantum neural networks can be indeed trained to the same level of accuracy as their classical counterparts, while for more difficult tasks, accuracy levels start to drop due to the quantum hardware limitations. From the simulations of the quantum circuits, we also observe agreement between the theoretical scalability of the quantum algorithm and experimental results, giving us more confidence in the utility of these algorithms when bigger hardware becomes available.

\section{Quantum methods for neural networks}
\label{methods}

In section \ref{sec:quantum_data_loaders}, we introduce the notion of \emph{Data Loaders} \cite{NearestCentroid2021}, an optimized quantum circuit at the core of our linear algebra approach. Then, in Section \ref{sec:quantum_assisted_nn}, quantum inner product estimation is derived from the data loaders, and used to assist training of classical neural networks, which we call quantum-assisted neural networks. Finally, in Section \ref{sec:quantum_orthoNN}, we present a novel quantum neural network that we call Quantum Orthogonal Neural Network (QOrthoNN). We analyze in depth how they work and how to efficiently train them provably faster than previously known methods. Further details are provided in the Appendix.

\subsection{Quantum data-loaders}\label{sec:quantum_data_loaders}

As introduced in \cite{NearestCentroid2021}, we choose \emph{unary amplitude encoding} to encode classical vectors as quantum states, in order to perform fast linear algebra operations with quantum circuits. Given a vector $x \in \R^d$ we use exactly $d$ qubits, exactly one qubit per feature. We outline three different ways of performing unary amplitude encoding of the classical data points below. 

We use a two-qubit parametrized gate, called Reconfigurable Beam Splitter gate ($RBS$), also known as partial-SWAP or fSIM gate, which is defined as 
\begin{equation} \label{RBS}
RBS(\theta) = \left( \begin{array}{cccc}
1 & 0 & 0 & 0 \\
0 & \cos \theta & \sin \theta & 0 \\
0 & -\sin\theta & \cos\theta & 0 \\
0 & 0 & 0 & 1  \end{array} \right)
\end{equation} 

\begin{figure}[h]
    \centering
    \includegraphics[width=100px]{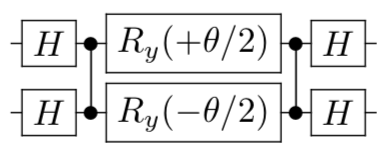}
    \caption{A possible decomposition of the $RBS(\theta)$ gate.}
    \label{fig:RBS_implementation}
\end{figure}

If not native to the quantum hardware, the $RBS$ gate can be easily implemented on hardware through the decomposition in Fig. \ref{fig:RBS_implementation}, where $H$ is the Hadamard gate, $R_{y}(\theta)$ is a single qubit Pauli rotation with angle $\theta$ around the $y$-axis, and the two two-qubit gates represent the $CZ$ gate that flips the sign when both qubits are in state 1. This is a more efficient decomposition than the one in \cite{NearestCentroid2021}, since it only uses two instead of three two-qubit gates. Note that this gate performs a Givens rotation, and a more detailed discussion can be found in \cite{KP22}.

The first step of the data loading, given access to a classical data point $x = (x_1,x_2,\ldots, x_d) \in \R^d$, is to pre-process the classical data efficiently, i.e. spending only $\widetilde{O}(d)$ total time (where the logarithmic factors are hidden), in order to create a set of parameters $\theta = (\theta_1,\theta_2,\ldots, \theta_{d-1}) \in \R^{d-1}$, that will be the parameters of the $(d-1)$ two-qubit gates used in our quantum circuit, see Section \ref{sec:data_loading} or \cite{NearestCentroid2021} for details about how to find the parameters $\theta_k$. During pre-processing, we also keep track of the norms of the vectors; see Section \ref{sec:data_loading} for an example. Note that these angles parameters are different depending on which data loader circuit is used.

\begin{figure}[!h]
\includegraphics[width=0.5\textwidth]{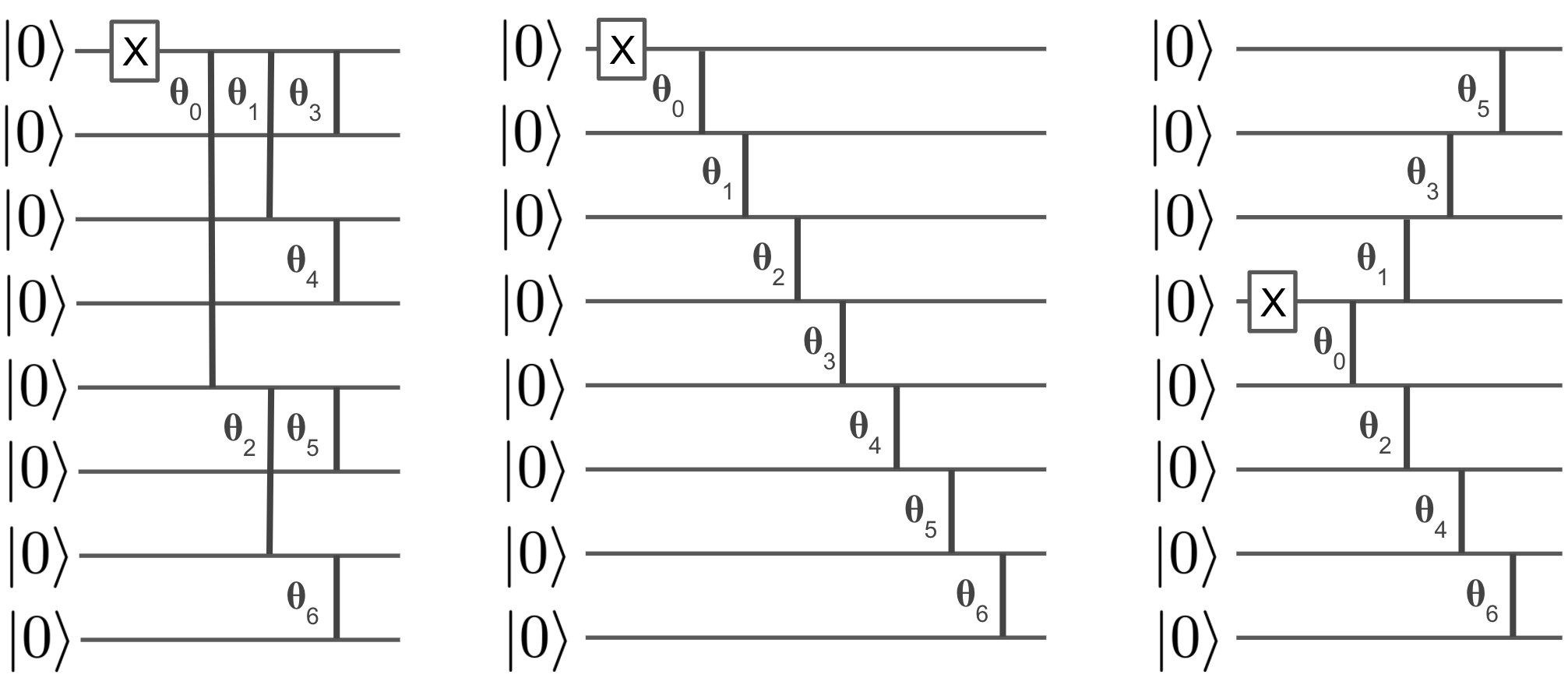}
\caption{The parallel, diagonal, and semidiagonal data loader circuit for an 8-dimensional data point. The $X$ corresponds to the single-qubit Pauli X gate, while vertical lines represent the two-qubit RBS gates.}
%while the $B-S$ symbols between two qubits correspond to the two-qubit RBS gate ($B$: first qubit, $S$: second qubit). \jonas{\emph{(reviewer: ``These ASCII diagram is harder to parse and not so nice to look at. Why change the figure style with respect to other Figs?")}}}
\label{loaders}
\end{figure}

After the data loader circuit, we obtain the state:
\begin{equation}\label{state}
\ket{x} = \frac{1}{\norm{x}}\sum_{i=1}^{d} x_i \ket{e_i} 
\end{equation}
where the states $\ket{e_i}$ are unary representations of the numbers $1$ to $d$ with \revised{$e_i = 0^{i-1}10^{d-i}$}. We recall that this superposition uses $d$ qubits. Note the presence of a normalization factor to respect quantum amplitudes unit norm. Three different types of data loader circuits appear in Fig.\ref{loaders}.

The shallowest data loader \cite{NearestCentroid2021} is a parallel version which loads $d$-dimensional data points using $d$ qubits, $d-1$ $RBS$ gates and circuits of depth only $\log d$ (see first circuit in Fig.\ref{loaders}). 

% \begin{SCfigure}[0.9][h]
% \label{parallel}
% \caption{The parallel data loader circuit for an 8-dimensional data point. 
% %The angles of the $RBS(\theta)$ gates starting from left to right and top to bottom correspond to $ (\theta_1,\theta_2,\ldots, \theta_{7})$.
% }
% \includegraphics[width=0.2\textwidth]{parallel}
% \end{SCfigure}

While this loader has the smallest depth, it also requires connectivity beyond the available connectivity on the IBM hardware we used for the numerical experiments in Section \ref{sec:Applications}. For experiments on quantum computers with a small number of qubits, as the ones available nowadays, the difference in the depth between the different data loader circuits is not significant. However, as the number of qubits increases, being able to use the parallel data loader can be very beneficial: for example, for a 1024-dimensional data point, the circuit depth can be reduced from 1024 to 10. 

The two data loaders we use in our demonstration have worse asymptotic depth, but respect the nearest neighbors connectivity of the IBM hardware.
The first is a simple diagonal unary loader that still uses $d$ qubits and $d-1$ $RBS$ gates between neighboring qubits, but it has circuit depth of $d-1$ (see second circuit in Fig.\ref{loaders}). 
% \begin{SCfigure}[0.9][h]
% \label{diagonal}
% \caption{The diagonal data loader circuit for an 8-dimensional data point.}
% \includegraphics[width=0.25\textwidth]{diagonal}
% \end{SCfigure}
The second one is another unary loader whose depth now decreases to $d/2$, and that we refer to as semi-diagonal (see third circuit in Fig.\ref{loaders}). 

We provide for completeness a way for finding these parameters for the diagonal loader in Section \ref{sec:data_loading}. 
Such unary amplitude encoding loaders have been used before in different scenarios \cite{NearestCentroid2021, unary2019}.   

\subsection{Quantum-assisted Neural Networks}\label{sec:quantum_assisted_nn}

Our first method uses a quantum computer to assist both training and inference of a classical neural network. \revised{It uses two quantum data loader circuits to compute the inner product between two vectors of dimension $d$ \cite{NearestCentroid2021}. The depth of the quantum circuit can be $O(\log(d))$, when using the parallel loaders, or $O(d)$, when using the diagonal or semi-diagonal data loader.}

In \cite{QNN2020, QCNN2019}, it was shown how quantum linear algebra methods on fault-tolerant quantum computers can potentially offer speedups in the training of classical neural networks, for both fully connected and convolution networks. These quantum methods are still outside the capabilities of current quantum hardware, so we designed and implemented in this work Noisy Intermediate Scale Quantum (NISQ) friendly subroutines for performing matrix-matrix multiplications through simpler quantum circuits that estimate inner products between two vectors. These circuits have the same behavior as the more complex quantum linear algebraic procedures, where the quantities that need to be computed can be estimated through sampling a quantum state that encodes these quantities in its amplitudes. This allows us to study how this estimation, instead of an exact computation as is the case in practice classically, affects the performance of the quantum deep learning methods; it also allows us to find ways to mitigate the errors of the quantum hardware.

It is important to note here that simply sampling a quantum state is enough to estimate the squares of the amplitudes of the desired state, which can be enough if we already know that the sign of the quantity we are interested. For estimating the inner product of two vectors, which can be either positive or negative, one needs to find efficient ways to estimate the sign as well.

\begin{figure}[!h]
\includegraphics[width=0.46\textwidth]{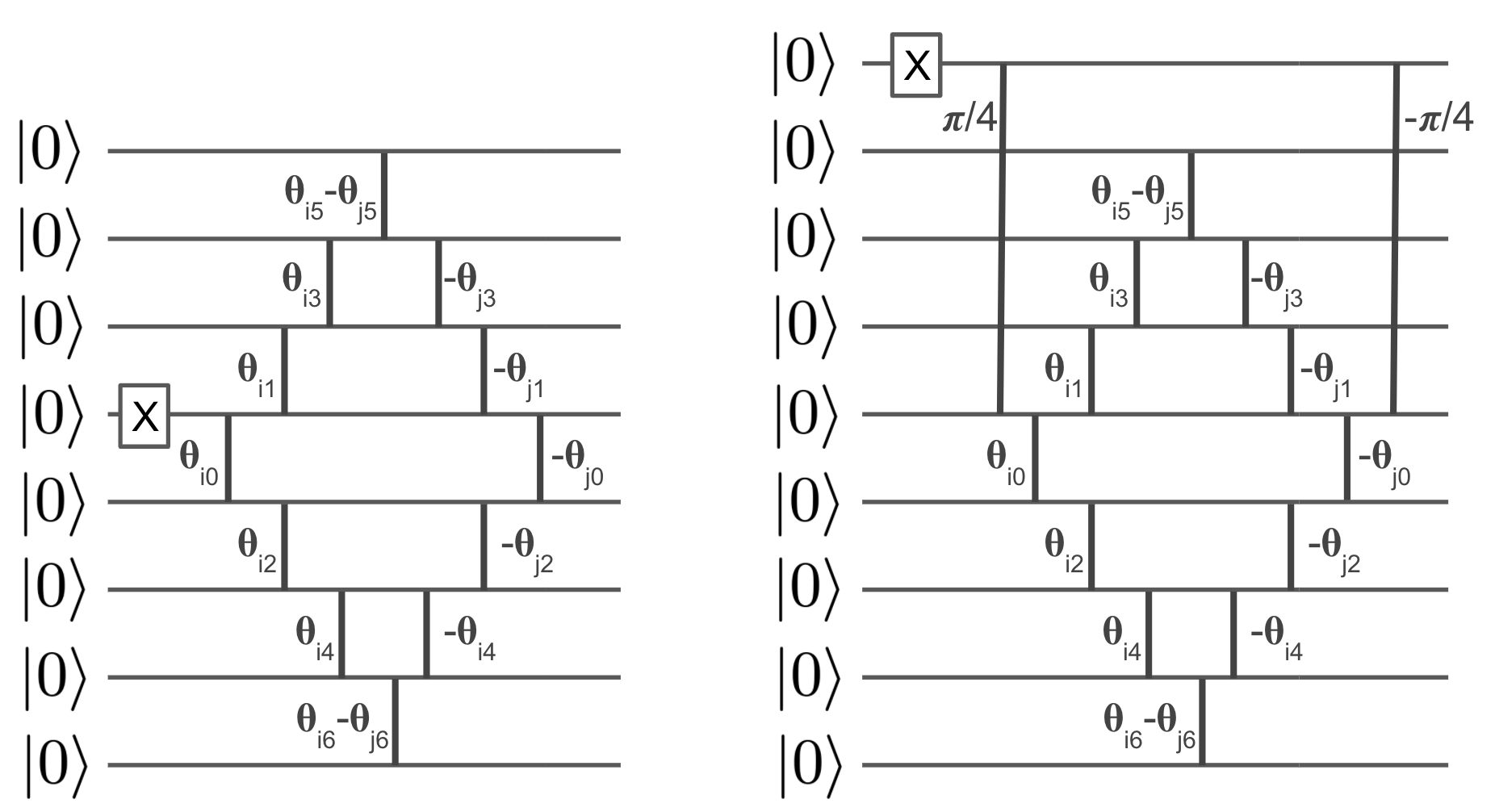}
\caption{The square inner product and the inner product estimation circuits with semi-diagonal data loaders for 8-dimensional data points $x$ and $w$. The $RBS$ gates on the left part of the circuits have parameters that correspond to the first vector $x$, while the $RBS$ gates on the right part of the circuits have parameters that correspond to the second vector $w$.}
\label{fig:IP_circ}
\end{figure}

We start the process with a simpler circuit that can be used for computing the square of the inner product between two normalized vectors $x$ and $w$, as shown in the left circuit in Fig.\ref{fig:IP_circ}, using the semi-diagonal data loader. The first part of the circuit is the data loader for some vector $x$, which in our case will be the normalized data point, and the second part is the adjoint data loader circuit for some vector $w$, which in our case will be a row of the weight matrix. The parameters of the gates in the first part of the circuit are fixed and correspond to the input data points, while the parameters of the gates of the second part are updated as we update the weight matrix through gradient descent computation. Note that in \cite{NearestCentroid2021} the parallel loader was used, since the experiments were performed in a fully connected trapped ion quantum computer. In this work, we use the semi-diagonal loaders to accommodate the limited connectivity of the superconducting hardware.

The final state of this circuit has the form
\begin{equation}
\revised{w\cdot x \ket{e_1} + \sqrt{1-(w\cdot x)^2}\ket{0G}}
\end{equation}
where $\ket{0G}$ is a state orthogonal to $e_1$, in particular whose first qubit is in state $\ket{0}$. This implies that when we measure the first qubit of this quantum state, the probability we obtain the outcome $1$ is exactly $(w\cdot x)^2$, and the probability of obtaining the outcome 0 is $1-(w\cdot x)^2$. Thus, if we run the same quantum circuit $N_{shot}$ times and count the number of times the measurement of the first qubit gave outcome 1, which we denote by $N_1$, then we can estimate the square inner product between the two vectors as $N_1/N_{shot}$. A simple Chernoff bound can be used to bound the tails of the binomial distribution and shows that if we want the estimate to be within $\epsilon$ of the correct answer with high probability, then we need to take $N_{shot}=O(1/\epsilon^2)$.\\

The above circuit is sufficient in case we know that the inner products are positive, which is for example the case of inner products between data points where each data point has only positive components \cite{NearestCentroid2021}. In our case, the inner products can also be negative, since they will be between data points and weight vectors which can be trained to have negative components; hence we need to extend the above circuit. While there are different possible ways to do so, we decided on the way shown in the second circuit in Fig.\ref{fig:IP_circ}, which adds one extra qubit in the quantum circuit and only a constant number of gates, independent of the dimension of the data points. This circuit only perform measurements in the computational basis, which is simple to implement.
%also does not need multiple measurements that form a complete operator basis on the Hilbert space of the system. \yvonna{not sure if I understand this last sentence about multiple measurement to form a complete operator basis, do we need to explain a bit?}

In short, in order to compute the inner product of two vectors $x$ and $w$, the extra qubit, the first from the top in the second circuit in Fig. \ref{fig:IP_circ}, is initialized with an $X$ gate to $\ket{1}$ and used as a control qubit through a $RBS$ gate between the first two qubits from the top with $\theta=\pi/4$. This way when the first qubit is 1 nothing happens, and when it is 0, the quantum circuit described above consisting of a quantum data loader for $x$ and the adjoint quantum data loader for $w$ is performed. By performing a final $RBS$ gate with $\theta=\pi/4$ we end up with a final state of the form
\begin{equation}
\left( \frac{1}{2} - \frac{1}{2}w\cdot x \right) \ket{e_1} + \ket{0G}
\end{equation}
where the state $\ket{0G}$ is an unnormalized state orthogonal to $e_1$, whose first qubit is in state $\ket{0}$. Thus, as described above, measuring the first qubit enough times and counting the number of times the outcome is 1, we can estimate the quantity $\left( \frac{1}{2} - \frac{1}{2}w\cdot x \right)^2$ from which, using the fact that $\left( \frac{1}{2} - \frac{1}{2}w\cdot x \right) \geq 0$, we can provide an estimate of $w\cdot x$ directly.
Note also that the connectivity of the new circuit necessitates one qubit with three neighbors while all other qubits can be on a line, which is available on the latest IBM machines. 

The proposed quantum circuits can be used to assist in the training and inference of classical neural networks when there is need for multiplication between data points and weights, and  replace the classical operation with the quantum circuit. The advantage these circuits offer is two-fold. First, with the advent of faster quantum computers that have the possibility to apply gates in parallel (see for example \cite{parallelGates}), one could take advantage of the fact that these quantum circuits require only logarithmic depth (using the parallel loaders) in order to provide a sample from the random variable we are estimating, and thus for larger feature spaces one could in principle expect a speedup in the estimation compared to a classical inner product computation on a single CPU. As state-of-the-art classical computers offer extremely efficient ways for performing a large number of such inner product calculations, using specialized hardware such as GPUs and TPUs, it will be quite unlikely in the near future to have a single quantum processing unit compete in speed on this particular task against such specialized hardware.
On the other hand, larger quantum speedups may become available when one can use the more advanced quantum linear algebra procedures as described in \cite{QNN2020, QCNN2019, ZNL21} on larger-scale quantum computers. Our circuits are more NISQ-friendly as we will see from the hardware demonstration later on. 
A second and more interesting property may come from the fact that one can perform training in an optimization landscape of the parameters of the angles of the quantum gates and not directly the elements of the weight matrices, which can provide different and potentially better models. We will provide more details on this when describing our second method. The theoretical analysis of the differences between two of these optimization landscapes and the corresponding trained models is left for future work.

\subsection{Quantum Orthogonal Neural Networks}\label{sec:quantum_orthoNN}

%In this Section, we present  modular quantum algorithm to compute an orthogonal neural network. It is equivalent to the classical algorithm in the prediction phase, but differs in the training phase. Some details can be found in the Appendix. 

In this section, we present our second quantum method, which we call Quantum Orthogonal Neural Networks (QOrthoNN). We start with a brief overview of classical orthogonal neural networks. Next, parametrized quantum circuits used for the encoding and the ansatz part of the quantum layers are defined, and the forward and back-propagation algorithms are analyzed. Lastly, we provide some general remarks and properties of our construction.

\subsubsection{Classical Orthogonal Neural Networks}

The main idea behind classical Orthogonal Neural Networks (OrthoNN) is the addition of an orthogonality constraint to the weight matrices corresponding to the layers of a neural network, which has theoretical and practical benefits in the generalization error \cite{jia2019orthogonal}. Orthogonality lowers redundancy in trained weights and preserves the magnitude of the weight matrix's eigenvalues to avoid vanishing gradients. In terms of complexity, for a single layer, the forward pass of an OrthoNN is simply a matrix multiplication, hence has a running time of $O(n^2)$ if the orthogonal matrix has a size of $n \times n$. It is also interesting to note that OrthoNNs have been generalized to Convolutional Neural Networks \cite{wang2020orthogonal}. 

The main difficulty of OrthoNNs is to preserve the orthogonality of the matrices while updating them during gradient descent. Several algorithms have been proposed to this end \cite{wang2020orthogonal, bansal2018can, lezcano2019cheap}, but they all point to the fact that perfect orthogonality is computationally hard to conserve, i.e. cubic with respect to the input size. Therefore, previous works aimed for approximations, and the matrices are often pushed toward orthogonality using regularization techniques during weights update. For completeness, we provide details on two training methods from the literature in Appendix \ref{sec:ortho_nn_backprop_details}: the first is called \emph{Singular Value Bounding} (SVB) and uses a Singular Value Decomposition (SVD) at every step to approximately orthogonalize the weight matrix, and the second ensures perfect orthogonality by performing the gradient descent in the manifold of orthogonal matrices, called the Stiefel Manifold. Both algorithms run in time $O(n^3)$ for a weight matrix of size $n \times n$. %\yvonna{is it not enough to cite the reference to these two methods. Do we need to include them in the appendix?}

\subsubsection{The pyramidal quantum circuit}\label{sec:ortho_pyramid_details}

\begin{figure}[h]
\centering
\begin{subfigure}{.30\textwidth}
  \centering
  \includegraphics[width=\linewidth]{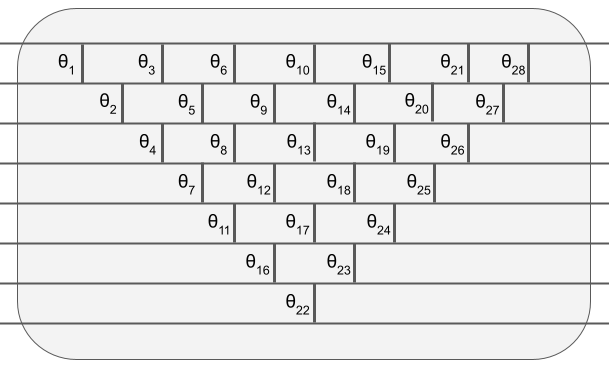}
  \caption{}
  \label{fig:QONNcircuit}
\end{subfigure}%
\begin{subfigure}{.20\textwidth}
  \centering
  \includegraphics[width=\linewidth]{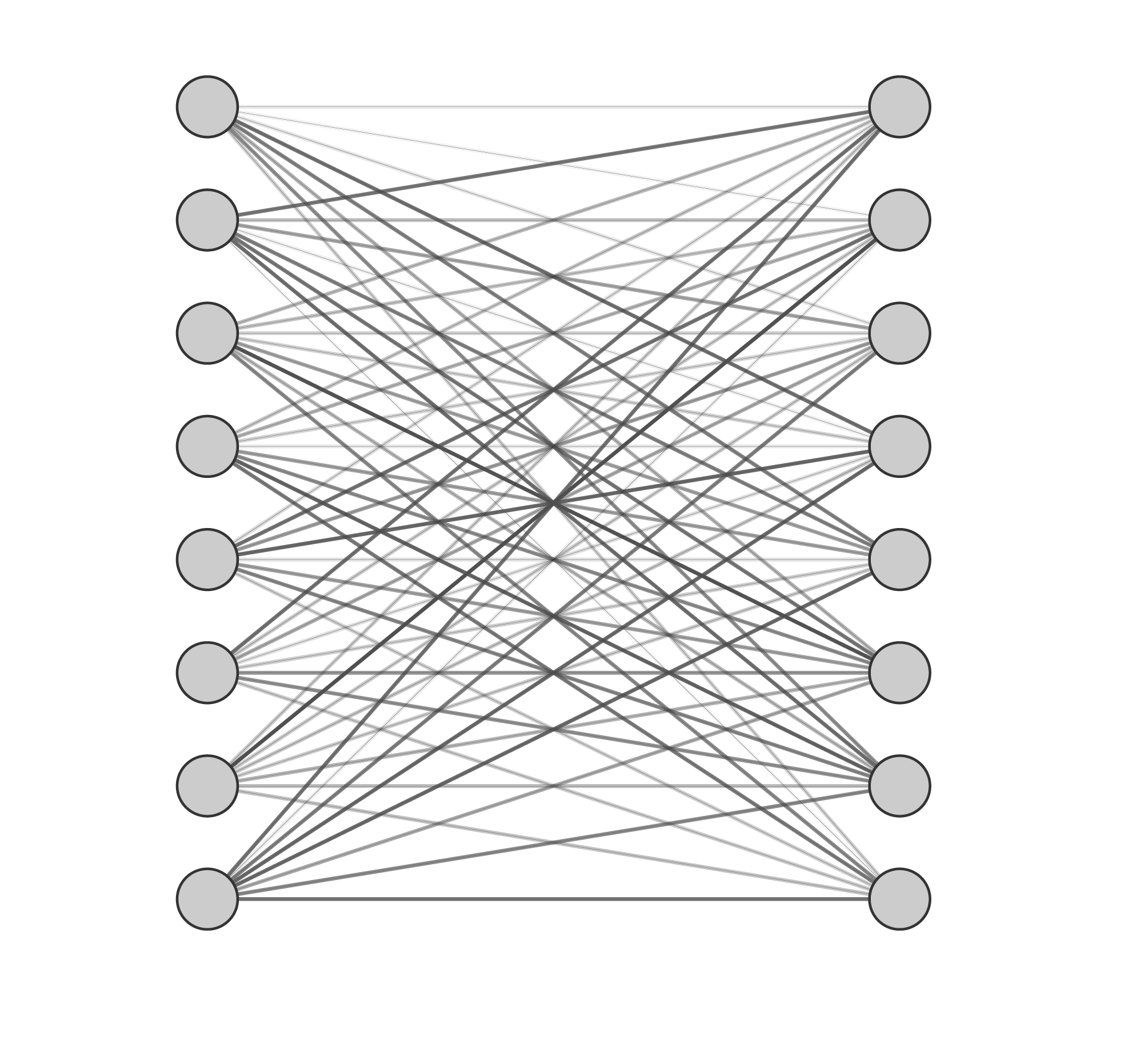}
  \caption{}
  \label{fig:8-8-nn}
\end{subfigure}
\caption{ (a) Parametrized quantum circuit for an 8$\times$8 quantum orthogonal layer (ansatz part). Each vertical line corresponds to an $RBS$ gate with its angle parameter. And (b), the equivalent classical orthogonal neural network 8$\times$8 layer.}
\label{fig:QONNcircuit_comparison}
\end{figure}

In this section, a parametrized quantum circuit that is used as the ansatz of a quantum orthogonal layer of a neural network is presented. The circuit is a pyramidal structure made of $RBS$ gates, each with an independent angle, as represented in Fig.\ref{fig:QONNcircuit}. 
%In Sections \ref{sec:data_loading} and \ref{sec:QONN_forward}, more details will be provided concerning respectively the input loading, and the equivalence with a neural network's orthogonal layer. 

\begin{figure}[h]
\centering
\begin{subfigure}{.3\textwidth}
  \centering
  \includegraphics[width=\linewidth]{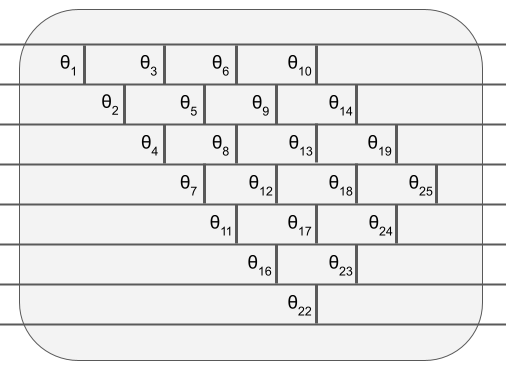}
  \caption{}
  \label{fig:QONNcircuit_rectangular}
\end{subfigure}%
\begin{subfigure}{.2\textwidth}
  \centering
  \includegraphics[width=\linewidth]{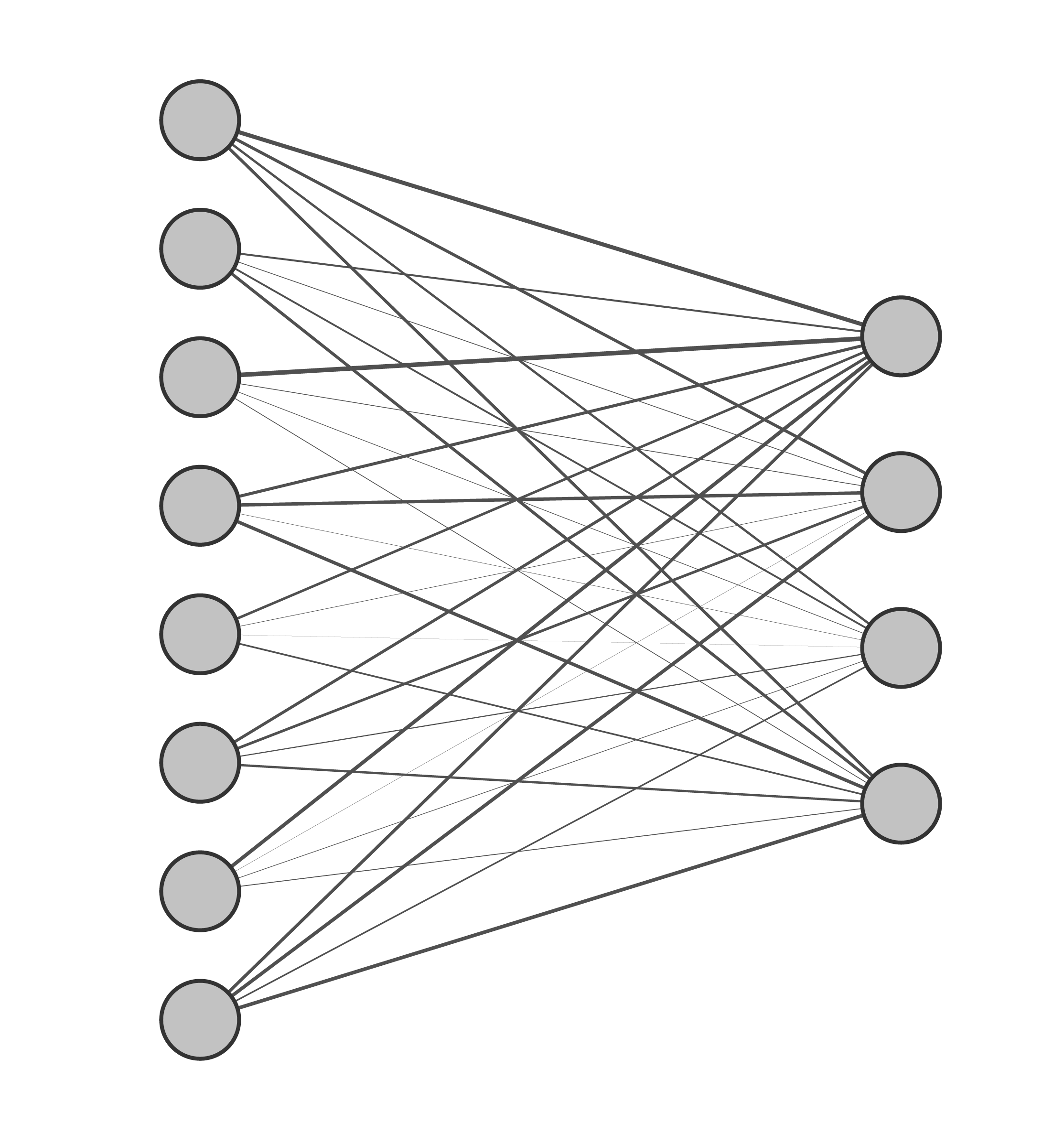}
  \caption{}
  \label{fig:8-8-nn_rectangular}
\end{subfigure}
\caption{ (a) Quantum circuit for a \emph{rectangular} 8$\times$4 fully connected orthogonal layer, where only the last four qubits from the output are used, and  (b) the equivalent 8$\times$4 classical orthogonal neural network. They both have 22 free parameters.}
\label{fig:rectangular}
\end{figure}

 We refer to the circuit as \emph{square} when the number of input and output qubits are equal, and as \emph{rectangular} otherwise (see Fig.\ref{fig:QONNcircuit_rectangular}) where a number of qubits in the output can be different from the input. For a layer with more outputs than inputs, we can use the adjoint of the quantum pyramid circuit drawn in Fig.\ref{fig:QONNcircuit_rectangular}.

The important property to note is that the number of parameters of the quantum pyramidal circuit corresponding to a neural network layer of size $n \times d$ is $(2n-1-d)d/2$, which is exactly the same as the number of degrees of freedom of an orthogonal matrix of dimension $n \times d$ where we have $d$ constraints for the unit norms of each column and $d(d-1)/2$ constraints for the orthogonality of all columns between them, hence $nd-d(d-1)/2-d = (2n-1-d)d/2$ degrees of freedom. 
For simplicity, we pursue our analysis using only the \emph{square case}, but everything can be easily extended to the rectangular case. And in case of square weight matrix, the full pyramidal structure of the quantum circuit described above imposes the number of free parameters to be $N=n(n-1)/2$, the exact number of free parameters to specify a $n\times n$ orthogonal matrix. \\

We now state a few important remarks concerning these circuits.

\begin{itemize}

\item 
From the definition of the RBS gates, we can see that when the pyramid circuit is applied to some unary computational basis state (e.g. a state of the form $\ket{00010}$), then the output quantum state can only be a superposition of unary basis states, in other words the pyramid circuit keeps the number of 1s and 0s unchanged. Thus, if we apply the pyramid circuit to a unary amplitude encoded state, then the output is a quantum state in the $n$-dimensional space spanned by the unary computational basis states. In other words, this operation can be seen as a multiplication of an $n \times n$ orthogonal matrix with an $n$-dimensional vector. This also allows for simulation of the application of the pyramid circuit on a unary amplitude encoded state classically in time $O(n^2)$ (see  Appendix \ref{Cpyr}).

\item
More generally, if the pyramid circuit is applied to some general computational basis state, for example a state of the form $\ket{01011}$, then the output quantum state can only be a superposition of basis states with the same hamming weight as the input state. When the input states are arbitrary superpositions of $n$-qubit states with fixed hamming weight $k$ then the operation of the pyramid circuit can be seen as a multiplication of an $n^k \times n^k$ orthogonal matrix, which in fact this is the $k$-th order compound matrix (see \cite{KP22} for more details), with an $n^k$-dimensional vector, hence the time to classically simulate the operation grows as $O(n^{2k})$. The quantum circuit remains the same; see \cite{KP22} for a more detailed description. The above results follow as well from the fact that the RBS gate is an example of a match gate, and it is known that the application of a quantum circuit with only match gates on a computational basis state can be simulated classically with a polynomial overhead.

\item 
Any orthogonal matrix can be mapped onto a pyramidal circuit. To be more precise, as the determinant of one RBS matrix is $\cos^2{\theta} + \sin^2{\theta}=1$, the matrices that are corresponding to circuits made of RBS gates are in the special orthogonal group $SO(n)$ and not the general orthogonal group, for which the determinants can be $\pm 1$ as well. The general orthogonal group can also be obtained by considering circuits that apply a $Z$-gate at the end of the pyramid circuit on the last qubit. This extra operation restricted to the unary basis is the identity matrix with the very last element equal to -1. The determinant of this matrix is $-1$ and hence the determinant of the entire circuit becomes now $-1$. Thus given an orthogonal matrix with determinant $-1$, we can easily factorize it into a product of two matrices, the first one with determinant $1$ that can be mapped onto a pyramid, and the second one corresponding to the $Z$ gate operation with determinant $-1$.

\item
The pyramidal circuit has linear depth (for $n$ qubits, the depth is $2n-3$) and is convenient for near term quantum hardware platforms with restricted connectivity. In fact, this distribution of the $RBS$ gates requires only nearest neighbor connectivity between qubits and the RBS gates are native or almost native on most quantum hardware platforms. 
    
\end{itemize}

\subsubsection{Loading the Data}\label{sec:data_loading}

Before applying the quantum pyramidal circuit, we need to upload the classical data into the quantum circuit via encoding part of the quantum orthogonal layer. Following Section \ref{sec:quantum_data_loaders}, we use one qubit per feature of the input data. 
Again, we use unary amplitude encoding of the input data (see Figure \ref{loaders}), where an input sample $x=(x_0,\cdots,x_{n-1}) \in \R^n$ such that $\norm{x}_2=1$, is loaded into the circuit as
$\ket{x} = \sum^{n-1}_{i=0} x_i\ket{e_i}$.
%Recent work \cite{dataloader} proposed a logarithmic depth data loader circuit for loading such states. 
To fit to the particular structure of our quantum pyramidal circuit, we use the diagonal data loader circuit from Section \ref{sec:quantum_data_loaders}, which requires a linear depth cascade of $n$-1 $RBS$ gates and only adds 2 extra time steps to our circuit.

\begin{figure}[!h]
    \centering
    \includegraphics[width=0.46\textwidth]{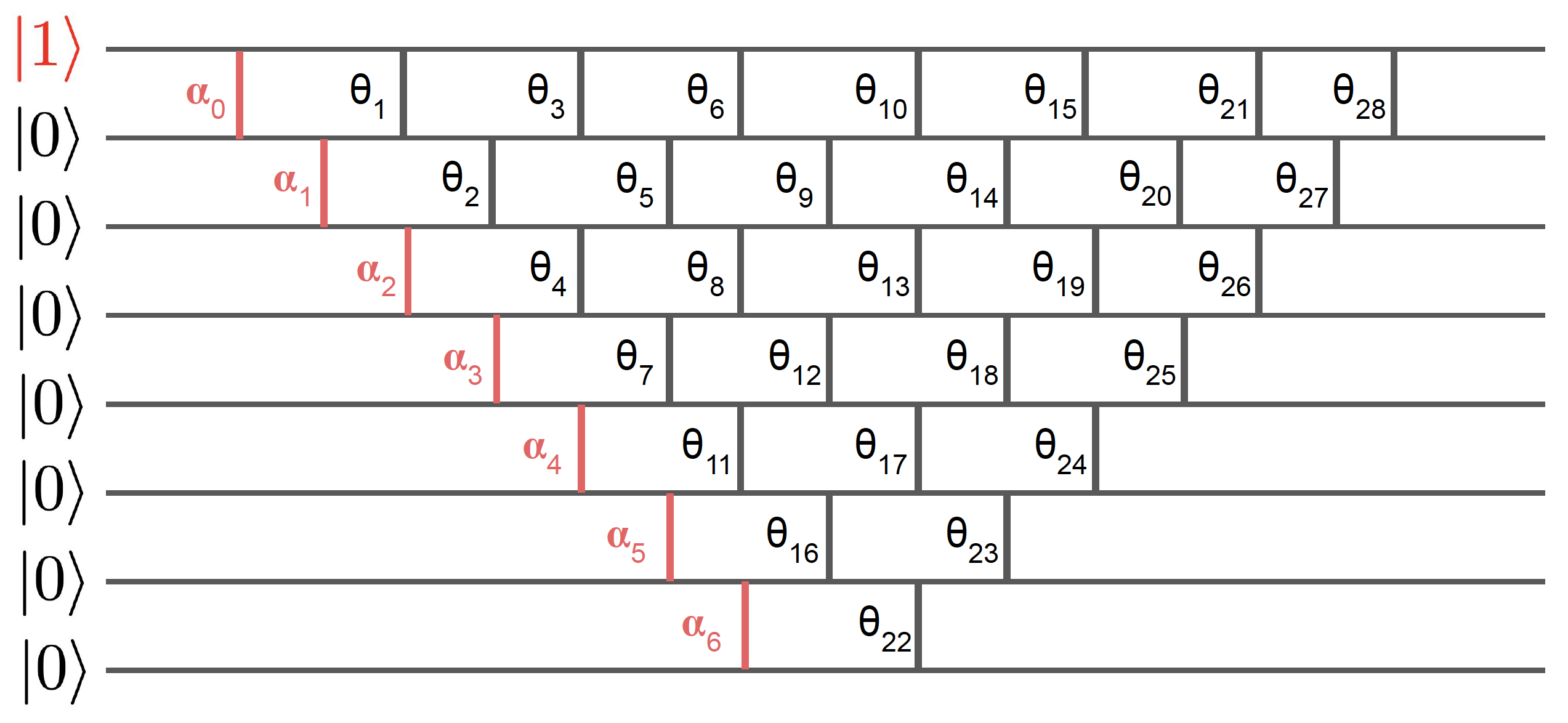}
    \caption{The 8 dimensional linear data loader circuit (in red) is efficiently embedded before the pyramidal circuit to form the quantum orthogonal layer. The input state is the unary state loaded through the diagonal loader (in red). The angles parameters $\alpha_0,\cdots,\alpha_{n-2}$ are classically pre-computed from the input vector.}
    \label{fig:data_loader}
\end{figure}

The diagonal data loader prepended to the pyramidal circuit initializes the input qubits in the $\ket{0}$ state, then flips the first qubit using an $X$ gate, in order to obtain the unary state $\ket{10\cdots 0}$ as shown in Fig.\ref{loaders}. Then a cascade of $RBS$ gates are used to create the state $\ket{x}$ using a set of $n-1$ angles $\alpha_0,\cdots,\alpha_{n-2}$. We choose the angles so that after the first $RBS$ gate of the loader the qubits would be in the state 
$
x_0\ket{100\cdots} 
+ \sin(\alpha_0)\ket{010\cdots}
$
and after the second one in the state 
$
x_0\ket{100\cdots} 
+ x_1\ket{010\cdots}
+ \sin(\alpha_0)\sin(\alpha_1)\ket{001\cdots}
$
and so on, until obtaining $\ket{x}$ as in Eq.(\ref{state}). To this end, we perform classical preprocessing to compute recursively the $n$-1 loading angles, in time $O(n)$. We choose $\alpha_0 = \arccos(x_0)$, $\alpha_1 = \arccos(x_1\sin^{-1}(\alpha_0))$, $\alpha_2 = \arccos(x_2\sin^{-1}(\alpha_0)\sin^{-1}(\alpha_1))$ and so on.

The ability of loading data in such a way relies on the assumption that each input vector is normalized, i.e. $\norm{x}_2=1$. This normalization constraint could seem arbitrary and impact the ability to learn from the data. In fact, in the case of an orthogonal neural network, this normalization shouldn't degrade the training because  weight matrices are in fact orthogonal and even orthonormal due to properties of quantum unitaries. This means that the operations are norm-preserving. Hence, changing the norm of the input vector, by diving each component by $\norm{x}_2$, in both classical and quantum settings should not be a problem. 
%\yvonna{repetitive? Note that for the classical OrthoNNs, depending on the strategy used, the orthonormality is sometimes approximated.} 
%\yvonna{The normalization would impose that each input has the same norm, or the same "luminosity" in the context of images, which can be helpful or harmful depending on the case. This sentence could be rephrased/removed? Most image processing pipeline include image normalization as a preprocessing step, and the relative luminosity of different images can still be preserved if normalization is applied across images.}

Note that the way we load the data using a unary amplitude encoding is fundamentally different from using independent single qubit rotations as is common in the literature \cite{schuld2021effect}. One property particular to this encoding strategy is the preservation of \revised{the $\ell_2$} distance between the normalized data points. For example, two images that are almost the same apart from a single pixel, will be mapped with the unary amplitude encoding to quantum states which are close to each other, which is not true with the independent single-qubit rotations. 
\revised{Namely, we have: 
\begin{equation}
\begin{split}
\norm{\ket{x}-\ket{x\prime}}^2_2 
= \norm{\sum_{i=1}^d (x_i-x_i^\prime)\ket{e_i}}_2^2\\
= \sum_{i=1}^d (x_i-x_i^\prime)^2
= \norm{x-x^\prime}^2_2    
\end{split}
\end{equation}}

\subsubsection{QOrthoNN: Forward Pass} \label{sec:QONN_forward}

In this section, we will detail the quantum orthogonal layer, namely the effect of the quantum pyramidal circuit on an input encoded through a unary amplitude encoding, as in Eq.(\ref{state}). %We will also see in the end how to simulate this quantum circuit classically with a polynomial overhead and thus be able to provide a fully classical scheme.

\begin{figure}[h]
    \centering
    \includegraphics[width=0.45\textwidth]{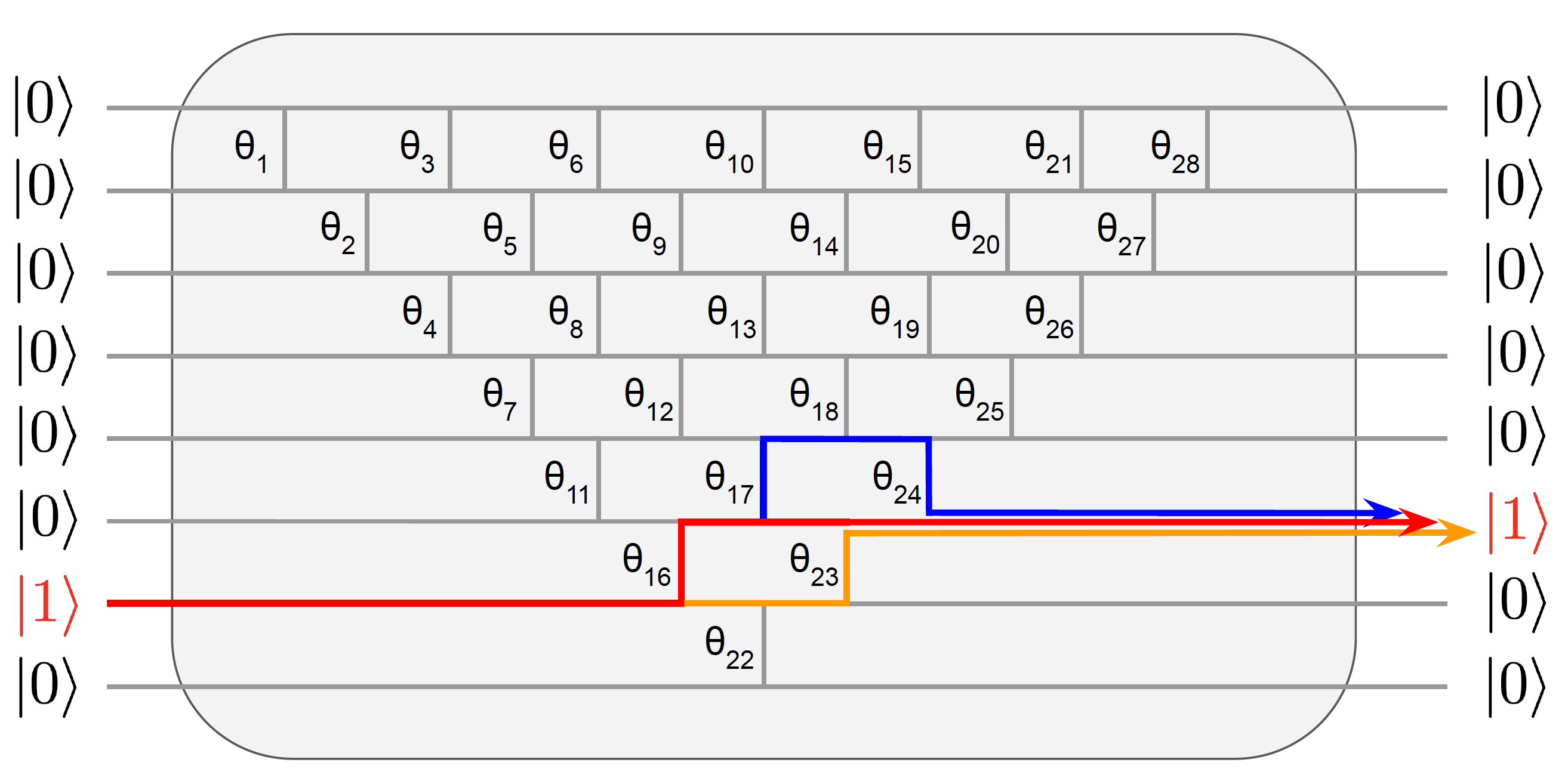}
    \caption{The three possibles paths from the $7^{th}$ unary state to the $6^{th}$ unary state, on an $8\times 8$ quantum pyramidal circuit.}
    \label{fig:QONNcircuit_path}
\end{figure}

Consider a unary computational basis input, where only the qubit $j$ is in state $\ket{1}$ (e.g. $\ket{00000010}$). By the end of the pyramid circuit, this unary input will be transformed into a superposition of unary states, each with an amplitude. If we consider again only one of these possible unary outputs, where only the qubit $i$ is in state $\ket{1}$, its amplitude can be interpreted as a conditional amplitude to transfer the $\ket{1}$ from qubit $j$ to qubit $i$. Intuitively, this value is the sum of the quantum amplitudes associated to each possible path that \emph{connects} the qubit $j$ to qubit $i$, as shown in Fig.\ref{fig:QONNcircuit_path}. 
Using this image of \emph{connectivity} between input and output qubits, we can construct a matrix $W \in \R^{n\times n}$, where each element $W_{ij}$ is the overall conditional amplitude to transfer the $\ket{1}$ from qubit $j$ to qubit $i$. 

Fig.\ref{fig:QONNcircuit_path} shows an example where exactly three paths can be taken to map the input qubit $j=6$ (the 7$^{th}$ unary state) to the qubit $i=5$ (the 6$^{th}$ unary state). Each path comes with a certain amplitude that can be positive or negative, which allows for interference. For instance, one of the paths (the red one in Fig.\ref{fig:QONNcircuit_path}) moves up at the first gate, and then stays put in the next three gates, with a resulting amplitude of $-\sin(\theta_{16}) \cos(\theta_{17}) \cos(\theta_{23}) \cos(\theta_{24})$. The sum of the amplitudes of all possible paths give us the element $W_{56}$ of the matrix $W$ (where, for simplicity, $s(\theta)$ and $c(\theta)$ respectively stand for $\sin(\theta)$ and $\cos(\theta)$):

\begin{multline}
W_{56} = 
- c(\theta_{16}) c(\theta_{22}) s(\theta_{23}) c(\theta_{24}) \\
-s(\theta_{16}) c(\theta_{17}) c(\theta_{23}) c(\theta_{24}) \\
+s(\theta_{16}) s(\theta_{17}) c(\theta_{18}) s(\theta_{24}) 
\end{multline}

In fact, the $n\times n$ matrix $W$ can be interpreted as the submatrix of the unitary matrix corresponding to the quantum circuit restricted to unary basis. In other words,
the full unitary $U_W$ in the Hilbert space of our $n$-qubit system is a $2^n\times 2^n$ matrix with the $n\times n$  matrix $W$ embedded in it as a submatrix on the unary basis. By loading the data as unary states and by using only $RBS$ gates that keep the number of 0s and 1s constant, we remain in the subspace spanned by the unary basis.

\begin{figure}[h]
    \centering
    \includegraphics[width=0.5\textwidth]{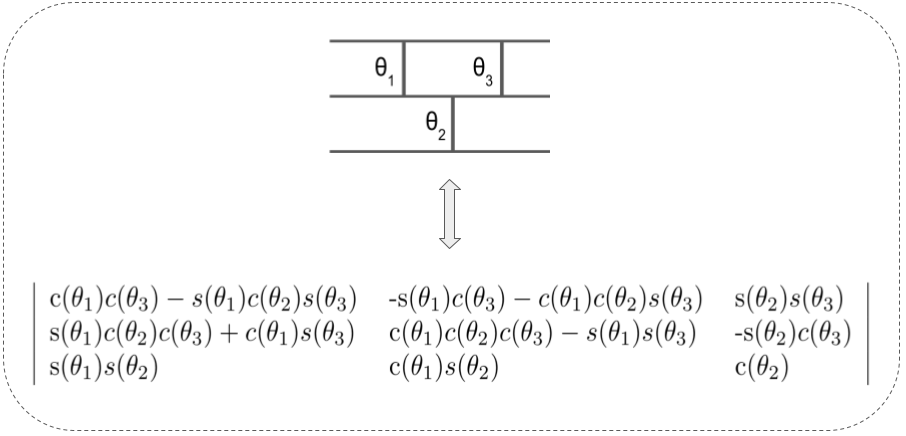}
    \caption{Example of a 3 qubits pyramidal circuit and the equivalent orthogonal matrix. $c(\theta)$ and $s(\theta)$ respectively stand for $\cos(\theta)$ and $\sin(\theta)$.}
    \label{fig:3x3_case}
\end{figure}

In Fig.\ref{fig:3x3_case}, a 3-qubit pyramidal circuit is described as a  $3\times 3$ matrix, that can be easily verified to be orthogonal.

In Fig.\ref{fig:QONNcircuit_path}, we consider the case of a single unary basis state for both the input and output. With actual data, as seen in Section \ref{sec:data_loading}, input and output states are in fact superpositions of unary states and the overall effect of the pyramidal circuit can be described by linearity of quantum mechanics.

\begin{figure}[h!]
    \begin{center}
    \centering
    \includegraphics[width=0.5\textwidth]{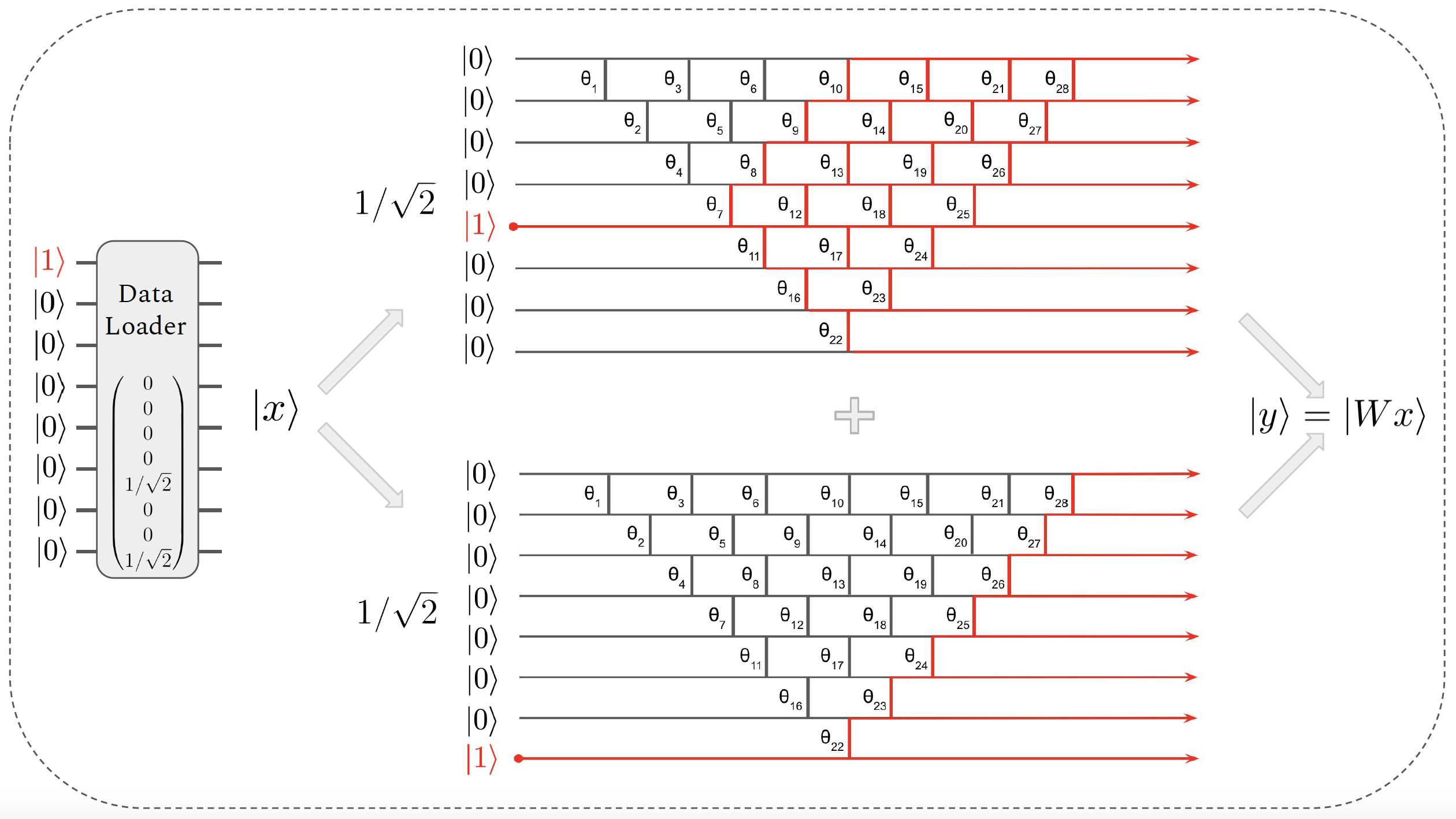}
    \caption{Schematic representation of a pyramidal circuit applied on a loaded vector $x$ with two non-zero values. The output is the unary encoding of $y=Wx$ where $W$ is the corresponding orthogonal matrix associated with the circuit.}
    \label{fig:full_schema}
    \end{center}
\end{figure}

For a general input vector $x \in \R^n$ encoded as a quantum state $\ket{x} = \sum^{n-1}_{i=0} x_i\ket{e_i}$ where $\ket{e_i}$ represents the i$^{th}$ unary state (see Section \ref{sec:quantum_data_loaders}), by definition of $W$, each unary basis $\ket{e_i}$ will undergo a proper evolution $\ket{e_i} \mapsto \sum_{j=0}^{n-1} W_{ij}\ket{e_j}$. This yields, by linearity, the following mapping 
\begin{equation}\label{eq:full_map}
\ket{x} \mapsto \sum_{i,j} W_{ij} x_i \ket{e_j}
\end{equation} 

As explained above, our quantum circuit is equivalently described by the sparse unitary $U_W\in\R^{2^n\times2^n}$ or by the matrix $W\in\R^{n\times n}$ on the unary basis subspace. This can be summarized with 
\begin{equation}\label{eq:equivalence}
    U_W\ket{x} = \ket{Wx}
\end{equation}
where $\ket{x}$ is a unary amplitude encoding.
We see from Eq.(\ref{eq:full_map}) and Eq.(\ref{eq:equivalence}) that the output is in fact $\ket{y}$, the unary encoding of the vector $y = Wx$, which is the output of a matrix multiplication between the $n\times n$ orthogonal matrix $W$ and the input $x\in\R^n$. As expected, each element of $y$ is given by $y_k = \sum_{i=0}^{n-1}W_{ik}x_i$. See Fig.\ref{fig:full_schema} for a diagram representation of this mapping.

Therefore, for any given neural network's orthogonal layer, there is a quantum pyramidal circuit that reproduces it. Conversely, any quantum pyramidal circuit applied on unary quantum states is implementing an orthogonal layer of some sort. 

Thus, the quantum pyramidal circuit can be used as a layer in a Quantum Orthogonal Neural Network. The running time of performing the forward pass is $O(n/\delta^2)$, where we consider the depth of the quantum circuit and also the fact that we need to repeat the quantum circuit $O(1/\delta^2)$ times in order to get the value of each output node with precision $\delta$. \revised{This scaling with respect to the $\delta$ is indeed a well known limitation that is present in many quantum algorithms where tomography has to be performed, but in the case of neural networks we have found that a small precision of even 0.1 can still converge (see Section \ref{sec:Applications}), since noise is already present in the data and the non-linearity eliminate the need for a very precise value of the inner product.}

Note that one can simulate classically this quantum circuit incurring a quadratic slowdown in time $O(n^2)$ (see Appendix \ref{Cpyr} for a mapping of the quantum pyramid circuit to a classical multi-layer neural network). Another reason to use the quantum circuit is that this layer could be combined with other quantum layers that increase entanglement and go beyond the unary subspace. Recent work has indeed shown how these pyramid quantum circuits are very useful beyond the unary basis for providing quantum machine learning applications with polynomial or potentially exponential speedups in determinant sampling and linear algebra \cite{KP22}.

As a side note, we also investigated if a circuit with only $\log(n)$ qubits could also implement an orthogonal matrix multiplication of size $n\times n$. Indeed, it would be a unitary matrix in $\R^{n\times n}$, but since the circuit should also have $n(n-1)/2$ free parameters to tune, or in other words parametrized gates, this would come at a cost of at least \revised{$\Omega(n^2/log(n))$} depth, quite unsuitable for NISQ devices.

In the Appendix \ref{sec:error} and \ref{sec:tomography} we respectively provide details regarding error mitigation and tomography, both methods being specific to our unary amplitude encoding convention.

\subsubsection{Multiple Quantum Layers}

\begin{figure}[h]
\centering
\begin{subfigure}[b]{0.5\textwidth}
   \includegraphics[width=\linewidth]{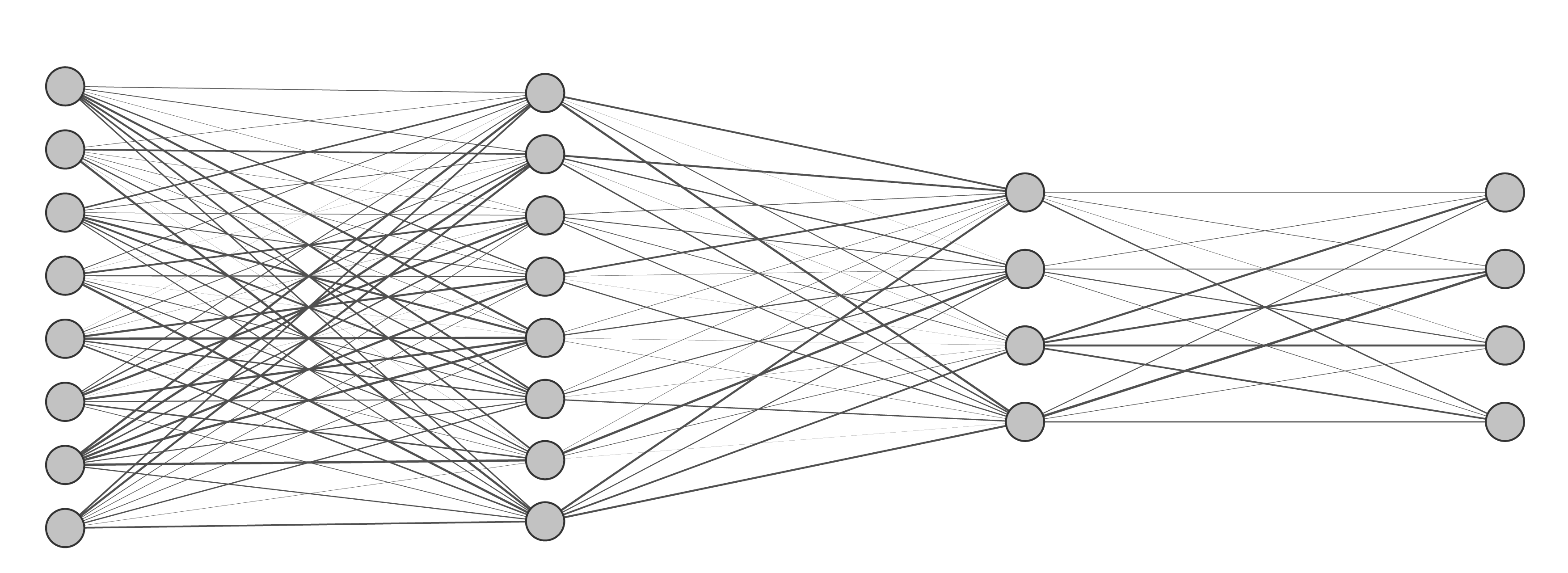}
   \caption{}
   \label{fig:8-8-5-5-nn} 
\end{subfigure}
\begin{subfigure}[b]{0.5\textwidth}
\includegraphics[width=\linewidth]{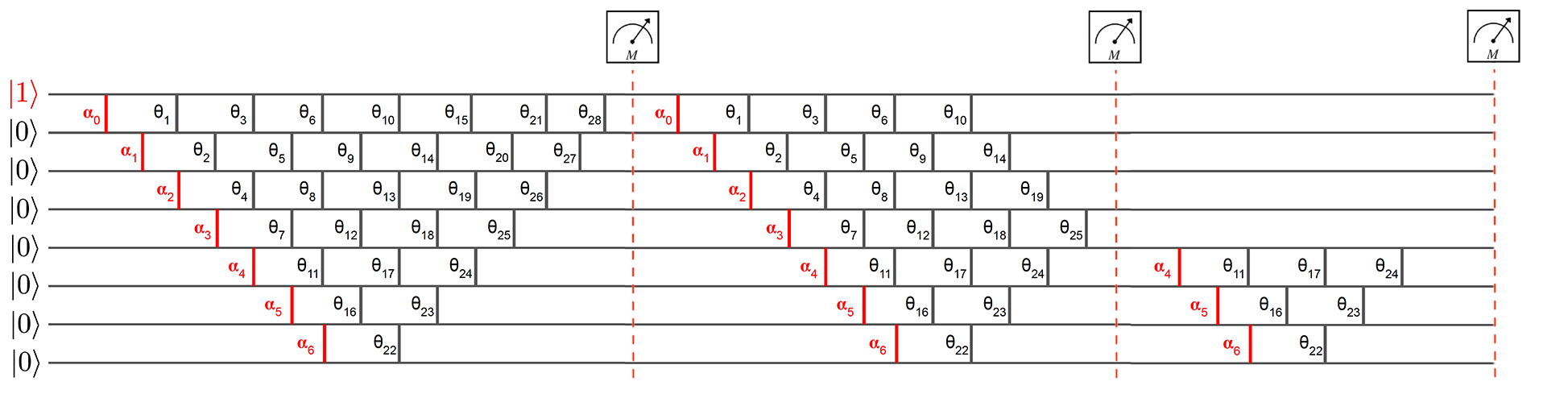}
   \caption{}
   \label{fig:8-8-5-5-Qnn}
\end{subfigure}
\caption{A full neural network with layers [8,8,4,4]. (a) Classical representation. (b) The equivalent quantum circuit is a concatenation of multiple pyramidal circuits. Between each layer, one performs a measurement and applies a non-linearity. Each layer starts with a new unary data loader.}
\end{figure}

In the previous sections, it has been shown how a single parametrized quantum circuit can perform one orthogonal layer. In classical deep learning, such layers are stacked to gain in expressivity and accuracy. Between each layer, a non-linear function is applied to the resulting vector. The presence of these non-linearities is key to the ability of the neural network to learn any function \cite{leshno1993multilayer}. 

One benefit of using our quantum pyramid circuit is the ability to easily concatenate them to mimic a multi layer neural network. After each layer, a tomography of the output state $\ket{z}$ is performed to retrieve each component, corresponding to its quantum amplitudes (see Section \ref{sec:tomography}). A non-linear function $\sigma$ is then applied classically to obtain $a = \sigma(z)$. The next layer starts with a new unary data loader (See Section \ref{sec:data_loading}). This hybrid scheme also allows keeping the depth of the quantum circuits acceptable for NISQ devices by applying the neural network layer by layer.

Note that the quantum neural networks we propose here are close in behavior to classical neural networks, which allows us to control and understand the quantum mapping and implement each layer and non-linearity in a modular way. These networks can be trained using similar gradient methods used by their classical counterparts, but they utilize a different optimization landscape that can provide different models (see Section \ref{sec:trainingquantum} for details). It will be interesting to compare our pyramidal circuit to a quantum variational circuit with $n$ qubits and $n(n-1)/2$ gates of any type, as we usually see in the literature. Using such variational circuits we would explore among all possible $2^n\times 2^n$ matrices instead of $n\times n$ classical orthogonal matrices, but so far there's no theoretical ground to explain why this should provide an advantage.

As an open outlook, one could imagine incorporating additional entangling gates after each pyramid layer (composed, for instance, of $CNOT$ or $CZ$). This would mark a step out of the unary basis and could effectively allow exploring more interactions in the Hilbert Space, with the caution that larger explored Hilbert spaces produce smaller gradients.

\subsubsection{QOrthoNN:  Backpropagation algorithm}\label{sec:trainingquantum}

In this section, we provide an efficient classical training algorithm for both classical and quantum orthogonal neural networks. This way, we do not need to use the parameter shift rule \cite{schuld2019evaluating, mitarai2018quantum}.

%\subsubsection{Classical Backpropagation Algorithm}\label{sec:trainingclassical}

\paragraph{Classical Backpropagation Algorithm\\}

Backpropagation for a fully connected neural network is a well known and efficient procedure to update the weight matrix at each layer \cite{hecht1992theory, rojas1996backpropagation}. At layer $\ell$, its weight matrices are noted by $W^{\ell}$ and its biases by $b^{\ell}$. Each layer is followed by a non-linear function $\sigma$, and can therefore be written as 
\begin{equation}
a^{\ell} = \sigma(W^{\ell}\cdot a^{\ell-1} + b^\ell) =  \sigma(z^{\ell})
\end{equation}
After the last layer, one can define a cost function $\mathcal{C}$ that compares the output to the ground truth. The goal is to calculate the gradient of $\mathcal{C}$ with respect to each weight and bias, namely 
$\frac{\partial \mathcal{C}}{\partial W^{\ell}}$ and $\frac{\partial \mathcal{C}}{\partial b^{\ell}}$. During backpropagation, we start by calculating these gradients for the last layer, then propagate back to the first layer. 

This method requires calculation of the \emph{error} vector at the layer $\ell$ defined by $\Delta^{\ell} = \frac{\partial \mathcal{C}}{\partial z^{\ell}}$. One can show the backward recursive relation $\Delta^{\ell} = (W^{\ell+1})^T\cdot \Delta^{\ell+1}\odot \sigma'(z^{\ell})$, where $\odot$ symbolizes the Hadamard product, or entry-wise multiplication. Note that the previous computation requires simply to apply the layer, \emph{i.e.} apply matrix multiplication, in reverse. We can then show that each element of the weight gradient matrix at layer $\ell$ is given by $\frac{\partial \mathcal{C}}{\partial W^{\ell}_{jk}} = \Delta^{\ell}_j\cdot a^{\ell-1}_1$. Similarly, the gradient with respect to the biases is defined as $\frac{\partial \mathcal{C}}{\partial b^{\ell}_{j}} = \Delta^\ell_j$. 

Once these gradients are computed, we update the parameters using the gradient descent rule with a given learning rate $\lambda$:
\begin{equation}\label{gradient_descent}
\centering
W^{\ell}_{jk} \gets W^{\ell}_{jk} -\lambda \frac{\partial \mathcal{C}}{\partial W^{\ell}_{jk}} \quad;\quad
b^{\ell}_{j} \gets b^{\ell}_{j} -\lambda \frac{\partial \mathcal{C}}{\partial b^{\ell}_{j}}
\end{equation}

%\subsubsection{OrthoNN training: Angle's Gradient Calculation and Orthogonal Matrix Update}\label{sec:trainingquantum}
\paragraph{QOrthoNN backpropagation: Angle's Gradient Calculation and Orthogonal Matrix Update\\}

Looking through the prism of the pyramidal quantum circuit, parameters to be updated are no longer individual elements of the weight matrices, but angles of the $RBS$ gates that parametrize to these matrices. Thus, we need to adapt the classical backpropagation method to the new setting based on the angles. Starting by introducing some notation for a single layer $\ell$, which is removed from future notation for simplicity, we further assume the number of output bits to be the same as input bits, while this method can easily be extended to the \emph{rectangular} case. 

\begin{figure}[t]
    \centering
    \includegraphics[width=0.48\textwidth]{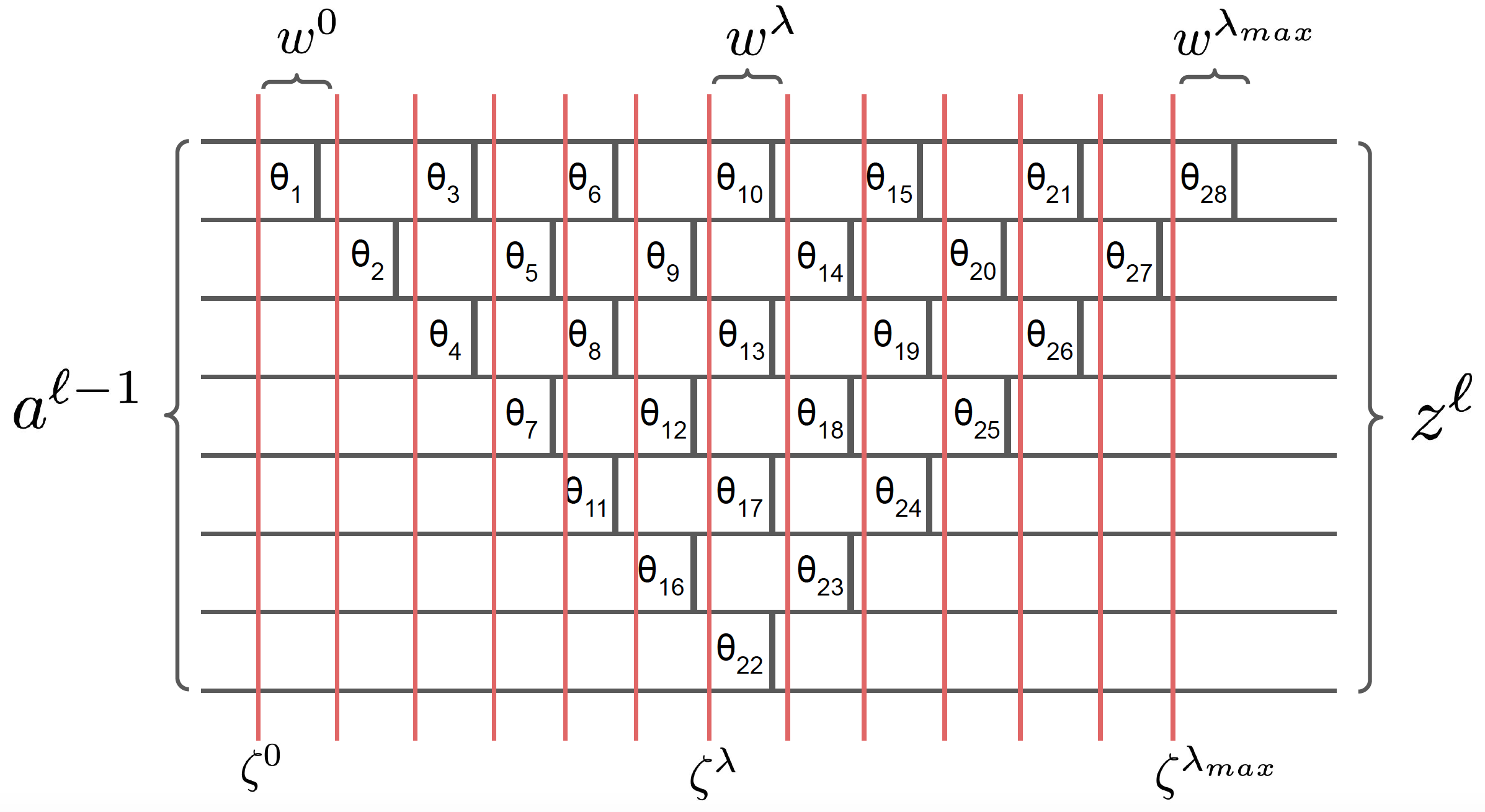}
    \caption{Quantum circuit for one neural network layer divided into \emph{timesteps} (red vertical lines) $\lambda \in [0,\cdots,\lambda_{max}]$. Each timestep corresponds to an \emph{inner layer} $\zeta^{\lambda}$ and an \emph{inner error} $\delta^{\lambda}$. The part of the circuit between two timesteps is a unitary matrix $w^{\lambda}$ in the unary basis.}
    \label{fig:QONNcircuit_timesteps}
\end{figure}

We first introduce the notion of \emph{timesteps} inside each layer, which correspond to the computational steps in the pyramidal structure of the circuit (see Fig.\ref{fig:QONNcircuit_timesteps}). 
For the square case with $n$ inputs, there will be $2n-3$ such \emph{timesteps}, each one indexed by an integer $\lambda \in [0,\cdots,\lambda_{max}]$. 
Applying a timestep consists in applying the matrix $w^{\lambda}$, made of all the RBS gates aligned vertically at this timestep, where $w^{\lambda}$ is the unitary in the unary basis (see Section \ref{sec:QONN_forward} for more details). After each timestep, the resulting quantum state is a vector in the unary basis named \emph{inner layer} and noted by $\zeta^\lambda$. This evolution can be written as $\zeta^{\lambda+1} = w^{\lambda}\cdot\zeta^{\lambda}$. We use this notation similar to the real layer $\ell$, with the weight matrix $W^{\ell}$ and the resulting vector $z^\ell$. %(see Section \ref{sec:trainingclassical}). 

In fact, we have the correspondences $\zeta^{0} = a^{\ell-1}$ for the first \emph{inner layer}, which is the input of the actual layer, and $z^\ell = w^{\lambda_{max}}\cdot\zeta^{\lambda_{max}}$ for the last one. We also have $W^\ell = w^{\lambda_{max}}\cdots w^1w^0$.

We use the same type of notation for the backpropagation errors. At each timestep $\lambda$ we define an \emph{inner error} $\delta^{\lambda} = \frac{\partial \mathcal{C}}{\partial \zeta^{\lambda}}$. This definition is similar to the layer error $\Delta^{\ell} = \frac{\partial \mathcal{C}}{\partial z^{\ell}}$. 
In fact, we also use the same backpropagation formulas, without non-linearities, to retrieve each \emph{inner error} vector $\delta^{\lambda} = (w^\lambda)^T\cdot \delta^{\lambda+1}$. In particular, for the last timestep which is the first timestep to be calculated in backpropagation, we have $\delta^{\lambda_{max}} = (w^{\lambda_{max}})^T\cdot \Delta^\ell$. Finally, we can retrieve the error at the previous layer $\ell-1$ using the correspondence $\Delta^{\ell-1} = \delta^{0} \odot \sigma'(z^\ell)$.

The reason for this breakdown into timesteps is the ability to efficiently obtain the gradient with respect to each angle. At timestep $\lambda$, for a single gate acting on qubits $i$ and $i+1$ with its angle noted by $\theta_i$   (note that the numbering is different from Fig.\ref{fig:QONNcircuit_timesteps}), the gradient $\frac{\partial \mathcal{C}}{\partial \theta_i}$ can be decomposed per component, indexed by the integer $k$, of the \emph{inner layer} and \emph{inner error} vectors 
$$\frac{\partial \mathcal{C}}{\partial \theta_i} = 
\sum_k
\frac{\partial \mathcal{C}}{\partial \zeta^{\lambda+1}_k}
\frac{\partial \zeta^{\lambda+1}_k}{\partial \theta_i}
= 
\sum_k
\delta^{\lambda+1}_k
\frac{\partial (w^\lambda_k\cdot\zeta^{\lambda})}{\partial \theta_i},$$ 
where $w^\lambda_k$ is the k$^{th}$ row of matrix $w^\lambda$.

Since each timestep $\lambda$ is only composed of parallel RBS gates, the matrix $w^\lambda$ consists of diagonally arranged $2\times2$ block submatrices given in Eq.(\ref{RBS}). Only one of these submatrices depends on the angle $\theta_i$ considered here, at the position $i$ and $i+1$ in the matrix. We can thus rewrite the above gradient as
$\frac{\partial \mathcal{C}}{\partial \theta_i} = 
\delta^{\lambda+1}_i
\frac{\partial}{\partial \theta_i}\left(w^\lambda_i\cdot\zeta^{\lambda}\right)
+
\delta^{\lambda+1}_{i+1}
\frac{\partial}{\partial \theta_i}\left(w^\lambda_{i+1}\cdot\zeta^{\lambda}\right)$, or:

\begin{multline}
\frac{\partial \mathcal{C}}{\partial \theta_i} = 
\delta^{\lambda+1}_i
\frac{\partial}{\partial \theta_i}\left(\cos(\theta_i)\zeta^{\lambda}_i+\sin(\theta_i)\zeta^{\lambda}_{i+1}\right)
\\+
\delta^{\lambda+1}_{i+1}
\frac{\partial}{\partial \theta_i}\left(-\sin(\theta_i)\zeta^{\lambda}_i+\cos(\theta_i)\zeta^{\lambda}_{i+1}\right)
\end{multline}

\begin{multline}\label{eq:gradient_formula_final}
\frac{\partial \mathcal{C}}{\partial \theta_i} = 
\delta^{\lambda+1}_i
(-\sin(\theta_i)\zeta^{\lambda}_i+\cos(\theta_i)\zeta^{\lambda}_{i+1})
\\+
\delta^{\lambda+1}_{i+1}
(-\cos(\theta_i)\zeta^{\lambda}_i-\sin(\theta_i)\zeta^{\lambda}_{i+1})
\end{multline}

To conclude the calculation of each angle gradient: during a forward pass, each of the $2n-3 = O(n)$ timesteps is applied sequentially, and resulting vectors, each \emph{inner layers} $\zeta^\lambda$, are stored in memory. During backpropagation, one obtains the \emph{inner errors} $\delta^\lambda$ by applying the timesteps in reverse. 
Finally, gradient descent on each angle $\theta_i$, while preserving the orthogonality of the overall equivalent weight matrix is obtained by

$$
\theta_i^{\ell} \gets \theta_i^{\ell} -\lambda \frac{\partial \mathcal{C}}{\partial \theta_i^{\ell}}.
$$

%\yvonna{we know mention this for the 3rd time: Consider removing this sentence: Since optimization is performed in the circuit angle landscape, and not on the weight matrix landscape, it can potentially produce different models, as we will see in the simulations later on. We leave open the theoretical study of the properties of both landscapes.}   

As one can see from the above description, this method is in fact a classical algorithm to obtain the angle's gradients, which allows us to train our OrthoNN and QOrthoNN efficiently classically while preserving the strict orthogonality. To obtain the gradient of each angle, one needs to store the $2n-3$ \emph{inner layers} $\zeta^{\lambda}$ during a forward pass. Next, given the error at the following layer, we perform a backward loop on each \emph{timestep} (see Fig.\ref{fig:quantum_vs_classical_representation}). At each \emph{timestep}, we obtain the gradient for each angle parameter by applying Eq.(\ref{eq:gradient_formula_final}). This requires $O(1)$ operations for each angle. Since there are at most $n/2$ angles per \emph{timesteps}, estimation of the gradients has a complexity of $O(n^2)$. After each \emph{timestep}, the next \emph{inner error} $\delta^{\lambda-1}$ is computed as well, using at most $4n/2$ operations. 

In the end, our classical algorithm allows us to compute the gradients of the $n(n-1)/2$ angles in total time $O(n^2)$, while respecting the strict orthogonality of the weight matrix. 
This is considerably faster than previous methods based on Singular Value Decomposition methods and provides a training method that is asymptotically as fast as for non-orthogonal neural networks, while preserving perfect orthogonality.

Our contributions are summarized in Table \ref{table:runningtimes}, where we have calculated both the time to perform one forward pass and one step of gradient descent for the case where a single neural network layer is considered with input and output of size $n$.

\begin{table*}[]
\begin{tabular}{|c|c|c|}
\hline
Algorithm                                                            & Forward Pass              & Weight Matrix Update      \\ \hline
Quantum Pyramidal Circuit (This work)                                & $2n/\delta^2 = O(n/\delta^2)$ & \multirow{2}{*}{$O(n^2)$} \\ \cline{1-2}
Classical Pyramidal Circuit (This work)                              & $2n(n-1) = O(n^2)$            &                           \\ \hline
Classical Approximated OrthoNN (SVB) \cite{jia2019orthogonal}        & $O(n^2)$                & $O(n^3)$                  \\ \hline
Classical Strict OrthoNN (Stiefel Manifold) \cite{jia2019orthogonal} & $O(n^2)$                & $O(n^3)$                  \\ \hline
Standard Neural Network (non orthogonal)                             & $O(n^2)$                & $O(n^2)$                  \\ \hline
\end{tabular}
\caption{Comparison of running times for classical and quantum Orthogonal Neural Networks. $n$ is the size of the input and output vectors, $\delta$ is the error parameter in the quantum implementation. See Appendix \ref{sec:ortho_nn_backprop_details} for details on related work.}
\label{table:runningtimes}
\end{table*}

\section{Application for Medical image classification on Quantum Hardware}\label{sec:Applications}

\subsection{Datasets and pre-processing}

In order to benchmark our quantum neural network techniques we focused on image classification and in particular we used datasets from MedMNIST, a collection of 10 pre-processed medical image open datasets \cite{medmnist}. The collection has been standardized for classification tasks on 10 different imaging modalities, each with medical images of $28 \times 28$ pixels. 

In this work, we looked specifically at two different datasets. The first is the Pneumonia-MNIST \cite{pneumonia}, a dataset of pediatric chest X-ray images. The task is binary classification between pneumonia-infected and healthy chest X-rays. The second is the Retina-MNIST, which is derived from DeepDRiD \cite{retina}, a dataset of retinal fundus images. The original task is ordinal regression on a 5-level grading of diabetic retinopathy severity, which has been adapted in this work to a binary classification task to distinguish between normal (class 0) and different levels of retinopathy (classes 1 to 4).  

In the Pneumonia-MNIST the training set consists of 4708 images (class 0: 1214, class 1: 3494) and the test set has 624 points (234 - 390). In the Retina-MNIST the training set has 1080 points (486 - 594) and the test set 400 (174 - 226). Note that in the Retina-MNIST we have considered the class 0 (normal) versus classes $\{1,2,3,4 \}$ together (different levels of retinopathy).

Though the image dimension of the MedMNIST (784 pixels) is small compared to the original medical images, one cannot load such data on the currently available quantum computers and thus a standard dimensionality reduction pre-processing has been done with Principal Component Analysis to reduce the images to 4 or 8 dimensions. Such pre-processing method indeed may reduce the possible accuracy of both classical and quantum methods but our goal is to benchmark such approaches on current quantum hardware and understand if and when quantum machine learning methods may become competitive.

\subsection{Hardware demonstration}

\subsubsection{Superconducting quantum computer}

\begin{figure}[!h]
    \centering
    \includegraphics[width=0.3\textwidth]{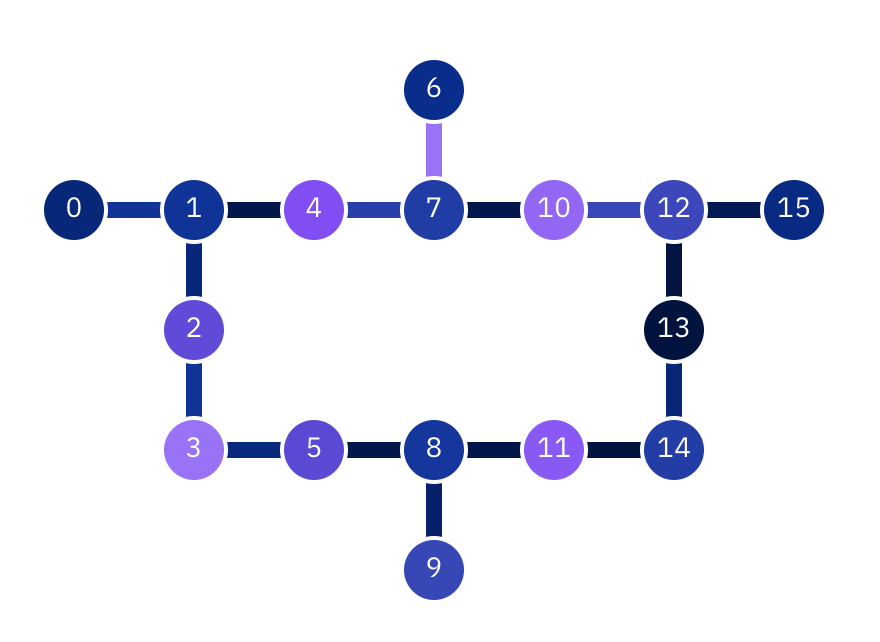}
    \caption{The 16-qubit ibmq\_guadalupe quantum computer.}
    \label{guadalupe}
\end{figure}

The hardware demonstration was performed on three different superconducting quantum computers provided by IBM, with the majority of the experiments performed on the 16-qubit {\em ibmq\_guadalupe} machine (see Fig.\ref{guadalupe}). 

The other two machines used were the 7-qubit {\em ibmq\_casablanca} and the 5-qubit {\em ibmq\_bogota}. All quantum hardware was accessed through the IBM Quantum Experience for Business cloud service.

In Appendix \ref{sec:additional_results} we give details regarding the quantum software and circuit compilation used for our experiments and simulations.

Note that the main sources of noise are the device noise and the finite sampling noise. In general, noise is not desirable during computations. In the case of a neural network, however, noise may not be as troublesome: noise can help escape local minima \cite{noiseNN2017}, or act as data augmentation to avoid overfitting. In classical deep learning, noise is sometimes artificially added for these purposes \cite{ying2019overview}. That being said, when the noise is too large, we also see a drop in the accuracy.

\subsubsection{Experimental results}

We tested our methods using a number of different datasets, classification tasks, and architectures that we detail below, and we performed both simulations and hardware experiments.

\begin{figure}[!h]
    \centering
    \includegraphics[width=0.85\linewidth]{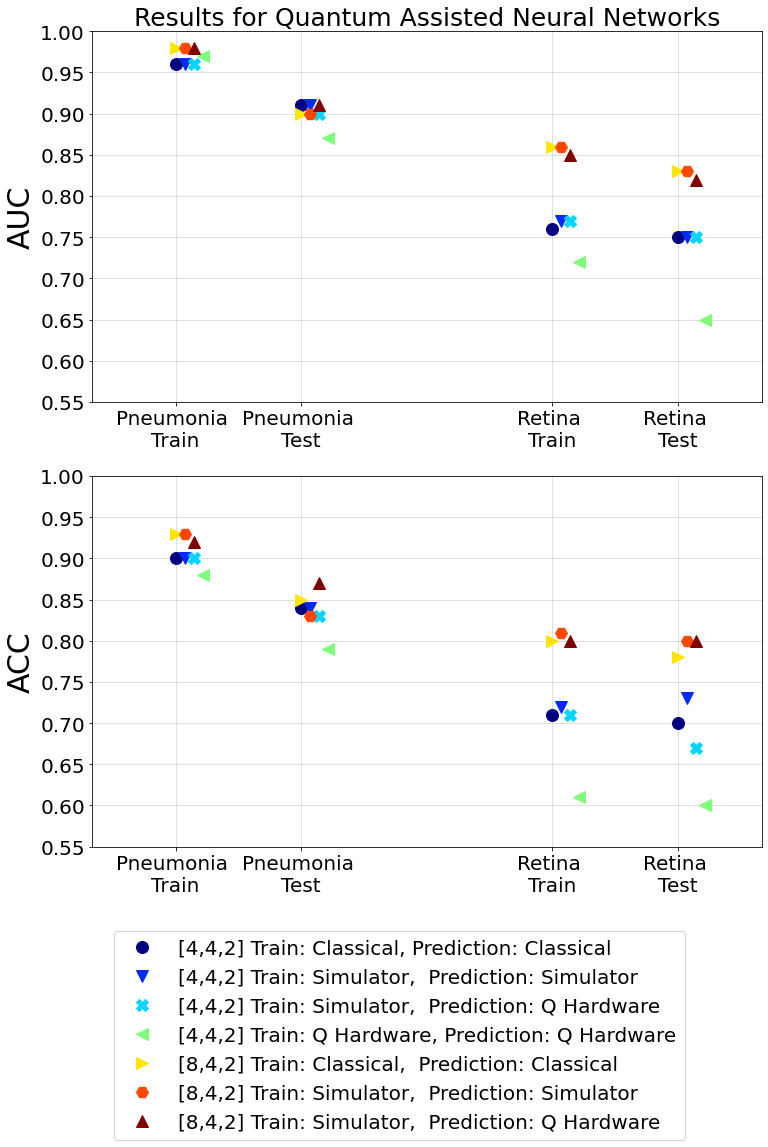}
    \caption{Results of experiments for the Quantum assisted Neural Network. Both [4,4,2] and [8,4,2] architectures were tested. For each, we compared different training and prediction methods, some are classical, some are simulated quantum circuits, others are real quantum circuits on hardware.}
    \label{fig:results1}
\end{figure}

\begin{figure}[h]
    \centering
    \includegraphics[width=0.85\linewidth]{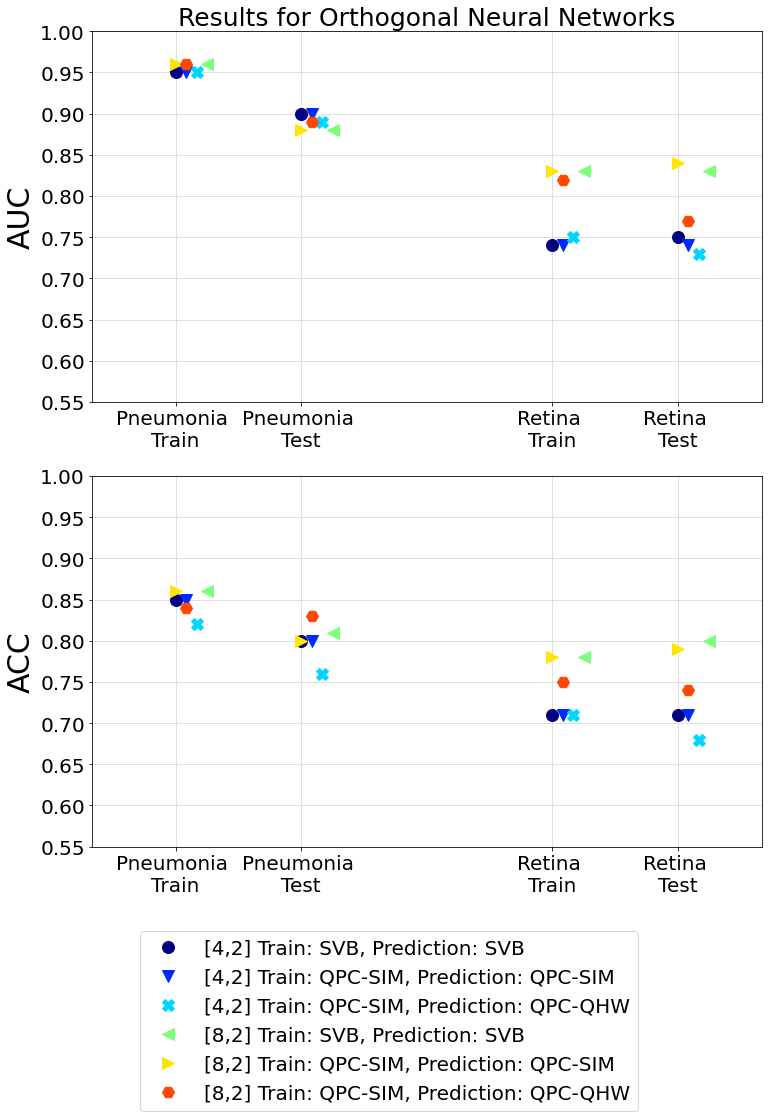}
    \caption{Results of experiments for the Orthogonal Neural Network. QPC stands for Quantum Pyramidal Circuit \revised{and QPC-SIM} is the classical algorithm simulating our quantum circuit. QHW is the quantum circuit on the real quantum hardware. SVB stands for the classical Singular Value Bounded algorithm.}
    \label{fig:results2}
\end{figure}

In Fig.\ref{fig:results1} and Fig.\ref{fig:results2} and Table \ref{state}, we show our results, where we provide the AUC (area under curve) and ACC (accuracy) for all types of neural network experiments, and for both the training and test sets from the Pneumonia and Retina datasets.

\begin{table*}[]
\resizebox{\textwidth}{!}{
\begin{tabular}{|cccc|cc|cc|cc|cc|}
\hline
\multicolumn{4}{|c|}{TYPES}              & \multicolumn{2}{c|}{PNEUMONIA AUC} & \multicolumn{2}{c|}{PNEUMONIA ACC} & \multicolumn{2}{c|}{RETINA AUC} & \multicolumn{2}{c|}{RETINA ACC} \\ \hline
method  & layers  & training & inference & TRAIN            & TEST            & TRAIN            & TEST            & TRAIN          & TEST           & TRAIN          & TEST           \\ \hline
qNN     & [4,4,2] & CLA      & CLA       & 0.96        & 0.91       & 0.90        & 0.84       & 0.76      & 0.75      & 0.71      & 0.70      \\
qNN     & [4,4,2] & SIM      & SIM       & 0.96        & 0.91       & 0.90        & 0.84       & 0.77      & 0.75      & 0.72      & 0.73      \\
qNN     & [4,4,2] & SIM      & QHW       & 0.96        & 0.90       & 0.90        & 0.83       & 0.77      & 0.75      & 0.71      & 0.67      \\
qNN     & [4,4,2] & QHW      & QHW       & 0.97             & 0.87            & 0.88             & 0.79            & 0.72           & 0.65           & 0.61           & 0.60           \\
qNN     & [8,4,2] & CLA      & CLA       & 0.98        & 0.90       & 0.93        & 0.85       & 0.86      & 0.83      & 0.80      & 0.78      \\
qNN     & [8,4,2] & SIM      & SIM       & 0.98        & 0.90       & 0.93        & 0.83       & 0.86      & 0.83      & 0.81      & 0.80      \\
qNN     & [8,4,2] & SIM      & QHW       & 0.98             & 0.91            & 0.92             & 0.87            & 0.85           & 0.82           & 0.80           & 0.80           \\ \hline
qOrthoNN & [4,2]   & SVB      & SVB       & 0.95        & 0.90       & 0.85       & 0.80       & 0.74      & 0.75      & 0.71      & 0.71      \\
qOrthoNN & [4,2]   & QPC-SIM      & QPC-SIM       & 0.95        & 0.90       & 0.85       & 0.80       & 0.74     & 0.74      & 0.71      & 0.71      \\
qOrthoNN & [4,2]   & QPC-SIM      & QPC-QHW       & 0.95        & 0.89       & 0.82        & 0.76       & 0.75      & 0.73      & 0.71      & 0.68      \\
qOrthoNN & [8,2]   & SVB      & SVB       & 0.96        & 0.88       & 0.86        & 0.81       & 0.83      & 0.83      & 0.78      & 0.80      \\
qOrthoNN & [8,2]   & QPC-SIM      & QPC-SIM       & 0.96        & 0.88       & 0.86        & 0.80       & 0.83      & 0.84      & 0.78      & 0.79     \\
qOrthoNN & [8,2]   & QPC-SIM      & QPC-QHW       & 0.96        & 0.89       & 0.84        & 0.83       & 0.81      & 0.77      & 0.75      & 0.74      \\ \hline
\end{tabular}
}
\caption{Results of our experiments on Pneumonia and Retina datasets, reported for the train set and the test set, using AUC and ACC. Methods include quantum-assisted neural networks (qNN) or quantum orthogonal neural network (qOrthoNN) with different layer architectures. The training and inference are done classically via standard forward/backpropagation (CLA), or in a quantum-assisted way both on a quantum simulator (SIM) and on quantum hardware (QHW) for the qNN; and with the Singular-Value Bounded algorithm (SVB), or with a quantum pyramid circuit algorithm both on a quantum simulator (QPC-SIM) or on quantum hardware (QPC-QHW) for the qOrthoNN.
The experiments that do not involve quantum hardware have been repeated ten times: mean values are shown in this table, and the standard deviation is in most cases $\pm 0.01$ and up to $\pm 0.04$. \revised{All results include Error Mitigation (\ref{sec:error}).}}
\label{table}
\end{table*}

As general remarks from these results, one can see that for the Pneumonia-MNIST dataset, both the AUC and ACC are quite close for all different experiments, showing that both quantum simulations and quantum hardware experiments reach performance levels on par with classical neural networks. We also note that the quantum-assisted neural networks achieved somewhat higher level of accuracy than the orthogonal neural networks. For the Retina-MNIST dataset, experiments with 8-dimensional data achieve higher AUC and ACC than the ones with 4-dimensional data. \revised{A potential reason for this difference in accuracy could be that a larger input feature space would be much more informative and help the network to learn more efficiently for a harder task. Comparing the accuracy between the Retina and the Pneumonia task, it can be observed that the Retina dataset presents a more challenging task.} The quantum simulations achieve similar performance to the classical one, while for the case where both the training and the inference were performed on a quantum computer, a drop in the performance is recorded, mainly due to the noise present in the hardware and the higher difficulty of the dataset.\\

We provide now a more detailed description of the different experiments that were performed.

For the quantum-assisted neural networks, we used two architectures, one of size $[4,4,2]$ and the other $[8,4,2]$, meaning the input size is four or eight dimensions respectively, and both has exactly one hidden layer of four nodes. We used the sigmoid as activation function, and performed binary classification. We also used two small orthogonal neural networks of size $[4,2]$ and $[8,2]$, meaning the input size is four or eight dimensions respectively, with no hidden layer. We used the same sigmoid as activation, and performed binary classification. This change in architecture is due to the fact that the circuit complexity for each orthogonal layer is still quite high for current quantum machines, and better quality qubits would be required in order to scale these architectures up. We performed two different classification tasks, one for the Pneumonia-MNIST dataset, and one between class 0 and classes $\{1,2,3,4\}$ of the Retina-MNIST dataset. For each task, training and forward inference were performed in combinations of classical, quantum simulator, and quantum hardware. 

Experiments involving classical methods or quantum simulators have been repeated ten times each to extract mean values and error bars for the AUC and ACC quantities. The mean values appear in Table \ref{table} and the errors are bounded by $\pm 0.01 $ for most cases and reach up to $\pm 0.04$, which reflects the randomness in initialization of the weights and randomness in the quantum estimation procedures. For the hardware experiments, we performed the experiments exactly once per different type of neural network, layer architecture, training method, inference method, and each training and test sets of the Retina-MNIST and Pneumonia-MNIST datasets. Each hardware experiments took between 45 minutes to several hours. The longest one, training the $[4,4,2]$ quantum-assisted neural network, took more than 10 hours, with the majority of time spent not on the actual quantum hardware but on handling a large number of jobs within the quantum cloud service. Training and forward inference with classical methods or the quantum simulator took a few seconds to complete.

In order to more precisely benchmark the performance of the quantum hardware for different types of quantum circuits we used, we provide in the Appendix, Section \ref{sec:additional_results} an analysis where we plot the simulated versus the experimental value of the output of the quantum circuits we used in the quantum-assisted neural networks over the entire test sets. This allows us to see how the quantum hardware behaves on real data. For the 5 qubit experiments the hardware results are very close to the expected ones, while for the 9-qubit experiments we see that with some large probability the hardware execution diverges from simulated results. 

We note also that simulations show that the models trained through quantum methods can be robust. Variation in AUC and ACC between different simulated runs are small and comparable to the classical models. On the other hand, the behavior of the quantum hardware can be quite unstable over time, where repeating an experiment in different days or weeks can provide quite different results. This instability in performance is not due to the randomness in the training or inference method where variation from the simulation results is showed to be very small, namely around $\pm 0.01$, but due to the levels of noise in the hardware that varies between different time periods, between the specific subset of qubits one uses within the same quantum computer, and between different quantum hardware machines.

\paragraph{Simulation results}

Looking at the quantum simulation results, AUC and ACC of the quantum-assisted neural networks match those of the corresponding classical ones. This is to be expected, since the quantum circuits assist the training process but do not change the classical architectures. The main difference stems from quantum procedures which estimate and not compute exactly quantities such as inner products between data points and weights. This way, the quantum-assisted neural-nets are closer to classical FNNs where noise has been injected artificially during training and inference. Such noise injection can actually be beneficial at times, especially in the presence of small training sets, since they make the models more robust and potentially generalize better.

\begin{figure}[!h]
    \centering
    \includegraphics[width=0.5\textwidth]{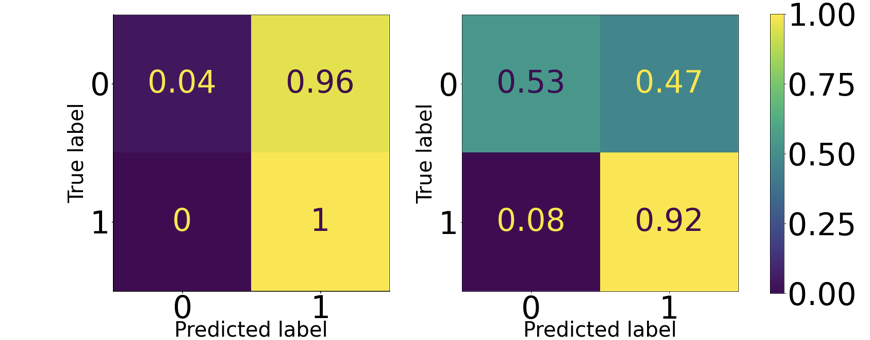}
    \caption{Results of training of an [32,16,2] orthogonal neural network on the Retina-MNIST (classes 0 vs. \{1,2,3,4\}) with training set of size 1080 (486-594) and test size 400 (174-226) with the SVB-based algorithm (left side: ACC(test) = 58.25\%) and the simulated quantum pyramid circuit algorithm (right side: ACC(test) = 75.25\%). 
}
    \label{diffmodel}
\end{figure}

For the quantum orthogonal neural networks, performance varies more compared to the performance of equivalent classical orthogonal neural networks based on Singular Value Decomposition (SVB), and this difference can be explained by the fact that training methods which are completely different. Our new way of training optimizes parameters of the gates of the pyramidal quantum circuit instead of elements of the weight matrices, which allows for a training in time $O(n^2)$ instead of the previously known $O(n^3)$ \cite{jia2019orthogonal}. Moreover, the models produced are often quite different due to the gradient optimization on the landscape of the circuit parameters, and thus can be a powerful source of more accurate models. We provide one such example of a substantially different training in Fig.\ref{diffmodel}, where we see substantial difference in the ACC and confusion matrices, where our new way of training achieves $75\%$ accuracy and the SVB-based one does not train and outputs practically always 1. 

\begin{figure}[!h]
    \centering
      \includegraphics[width=0.5\textwidth]{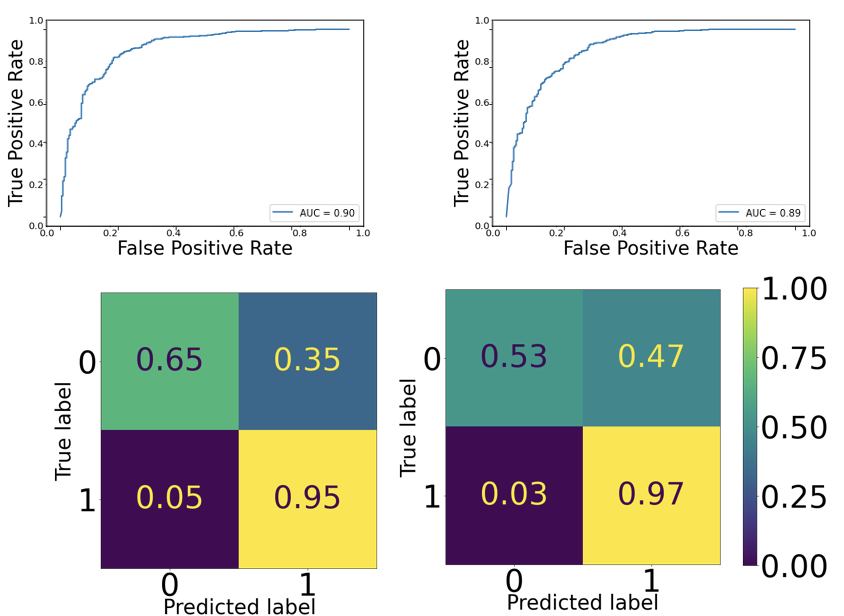}
    \caption{Training of an [784,64,2] quantum-assisted neural network on the Pneumonia-MNIST with training set of size 4708 (1214 – 3494) and test size of 624 (234-390) with classical training (left side: AUC (test set) = 0.90, ACC (test set) = 83.65\%) and the simulated quantum-assisted training algorithm (right side: AUC (test set) = 0.89, ACC (test set) = 80.77\%).
}
    \label{bigsim}
\end{figure}

Enabled by tailor-made quantum simulators designed specifically for the circuits used in this work, we were able to perform larger-scale simulations, which provide strong evidence of the scalability of our methods. In Fig.\ref{bigsim} we provide an example of simulation results for quantum-assisted neural networks on the 784 dimensional images that match the accuracy of the classical NNs. 

%\color{green} AGNES: I wonder if the report of the specificity against sensitivity needs to that graphical so that large when I consider some comments by Tufte   . But they look nice so no need to change them specifically for me. 
%\color{black}

In the Appendix, Section \ref{sec:additional_results}, we provide simulations of the training of the quantum orthogonal neural networks for different layer sizes to show in practice the asymptotic running time of $O(n^2)$ of the quantum pyramid circuit training algorithm. We also provide more details about the number of steps that the quantum-assisted neural networks take in comparison to the classical training.

\paragraph{Hardware results}

Looking at the results of the hardware demonstration, it is clear that current hardware is not ready to perform (medical) image classification in a way that is competitive to classical neural networks, since the hardware does not have sufficient number and quality of qubits. Possibly better algorithms and heuristics are needed to train better and faster models. 
Nevertheless, experiments showed promising results with 5-qubit and to a lesser extent with 9-qubit circuits, providing small-scale confirmation of the proposed methods, while the larger simulations we performed provide more evidence about scalability and future performance. 

One can see the overall results of the simulations and hardware experiments in Table \ref{table}:  classically-trained architectures and quantum forward inference matches the classical and quantum simulator performance for quantum assisted NNs. However, when both training and forward inference are performed on the quantum computer, the $[4,4,2]$ neural network managed to train with a small drop in the accuracy, while the $[8,4,2]$ neural network did not train at all, predicting always the same classification value for the entire dataset. \revised{This accuracy drop for larger and deeper quantum circuits is mainly due to the noise accumulation when using current quantum hardware.}
%\yvonna{but we did not show the results of 8,4,2 QNN with quantum training in Table II.} 

For the orthogonal neural networks, results of the hardware demonstration match the quantum simulation ones, with a small drop only for the Retina-MNIST and the 9-qubit experiments, showing that overall the hardware has sufficient quality to perform the necessary quantum circuits. Note that in this case, we have shown that the training can always be performed optimally classically.

We provide some further results, including AUC, ACC and confusion matrices for a number of the experiments in the Appendix, Section \ref{sec:additional_results}.

\subsection{Discussion}

We introduced two quantum methods for neural networks in this work: the quantum-assisted neural networks, which are classical neural networks that use a quantum computer to assist in their training and inference through quantum procedures for efficiently loading data and performing inner product estimation; and the quantum orthogonal neural networks, a quantum generalization of classical orthogonal neural networks, which is based on unary amplitude encoding and the pyramidal quantum circuit as an ansatz. We provided an optimal training algorithm for both classical and quantum orthogonal neural networks, improving on previously known training methods for such networks.  

We also studied the effectiveness of these quantum methods for medical image classification. 
We used the MedMNIST suite of datasets as standardized and lightweight benchmarks for classification via quantum techniques, and we show via simulations and hardware demonstrations how quantum-assisted neural networks and quantum orthogonal neural networks can be powerful tools for image classification. Our methods can be combined with other types of parametrized quantum circuits that explore larger Hilbert spaces. However, if and how these other techniques can enhance training of models on classical data is still an open question, while our current methods have been designed to retain a clear connection to classical methods, and allow theoretical study of their scalability, performance and training time. Simulations and hardware experiments we performed were extensive, showing the power and limitations of current hardware; we also devise error mitigation techniques that improved the results.

Regarding the potential advantages of quantum neural network techniques in more general:
with respect to the accuracy of the quantum models, we see the potential of quantum neural networks to provide different and at times better models, by having access to different optimization landscapes that could avoid effects like barren plateaus due to quantum properties such as orthogonality. 
%\yvonna{repetitive, consider removing: Note also, that there is a possibility for the quantum-assisted neural networks to actually provide different models via gradient optimization on the landscape of the circuit parameters and not directly on the elements of the weight matrices.} 
We performed gradient computations for the more complex orthogonal neural network case, and the techniques can be readily transferred to this easier case. Whether such different quantum models are advantageous will depend on the specifics of the use case, and further experimentation is certainly needed.

Regarding computation speed, with the coming of faster quantum machines that can perform parallel gates on qubits, we expect quantum training and inference via the log-depth quantum inner product estimation circuits to become quite efficient. However, highly specialized classical hardware such as GPUs and TPUs are extremely efficient in matrix-matrix multiplications, and it is very likely that more powerful quantum techniques will have to be combined with the simpler techniques described here if we would like to also provide speedup with respect to classical training. In addition to provable scalability, the study of quantum neural networks can also lead to novel, faster classical training methods, as is the case for orthogonal neural networks (see Section \ref{sec:quantum_orthoNN}).

\revised{This work has also opened the way for more advanced quantum methods where the same pyramidal quantum circuits can be shown to be closely related to Clifford Algebras and compound matrices where speedups can be more important \cite{KP22}. They can also be used in more complex quantum architectures and provide clear improvements in accuracy  \cite{cherrat2022quantum}.} Furthermore, it is possible to use the proposed orthogonal layers to explore a larger part of the Hilbert space by using higher hamming weight states as inputs. Potential solutions include loading a superposition of more than one data points or a polynomial expansion of the classical data vector.

Last, the application of neural networks is ubiquitous in almost all data domains, not only in image classification. Hence, we expect the quantum methods developed and tested in this work to have much broader impact both in life sciences and beyond. Such methods will of course continue to be refined as the hardware evolves.\\
%At the same time, while orthogonality tends to enhance performance for much deeper architectures \cite{jia2019orthogonal}, here we did not see any tangible improvement with respect to normal fully-connected neural networks for the small-scale simulations we performed. We do expect such advantages to appear for deeper neural networks. Last, the quantum circuits for performing inference have depth linear in the size of the layer, instead of quadratic, which may provide a speedup with a new generation of quantum computers. 

\subsection{Data Availability}
The data supporting the findings is available from the corresponding author upon reasonable request.

\subsection{Author Contribution}
JL, NM, and IK developed the theory of quantum orthogonal neural networks. JL and NM developed code for the quantum orthogonal and assisted neural networks and performed data analysis. NM compiled the quantum circuits, and performed the corresponding experiments. YL and MS formulated the use case and relevant experiments to demonstrate the usefulness of this work. SK has developed code for the different types of data loaders. AP has contributed in the theoretical analysis of the quantum neural networks. IK conceived the project and contributed in all aspects.

\subsection{Acknowledgements}
This work has appeared in a preliminary version as arXiv:2106.07198 on the theory of orthogonal neural networks, and arXiv:2109.01831 on the application to medical image classification. We acknowledge the use of IBM Quantum services for this work. The views expressed are those of the authors, and do not reflect the official policy or position of IBM or the IBM Quantum team. The following members of the Roche pRED Quantum Computing Taskforce also contributed to this work: Marielle van de Pol, Agnes Meyder, Detlef Wolf, Stanislaw Adaszewski.

\bibliographystyle{quantum}
\bibliography{references}

\begin{thebibliography}{10}

\bibitem{HHL}
Aram~W Harrow, Avinatan Hassidim, and Seth Lloyd.
\newblock ``Quantum algorithm for linear systems of equations''.
\newblock \href{https://dx.doi.org/10.1103/PhysRevLett.103.150502}{Physical
  review letters {\bf 103}, 150502}~(2009).

\bibitem{lloyd2013quantum}
Seth Lloyd, Masoud Mohseni, and Patrick Rebentrost.
\newblock ``Quantum algorithms for supervised and unsupervised machine
  learning''~(2013).

\bibitem{lloyd2014quantum}
Seth Lloyd, Masoud Mohseni, and Patrick Rebentrost.
\newblock ``Quantum principal component analysis''.
\newblock \href{https://dx.doi.org/10.1038/nphys3029}{Nature Physics {\bf 10},
  631--633}~(2014).

\bibitem{kerenidis2016quantum}
Iordanis Kerenidis and Anupam Prakash.
\newblock ``Quantum recommendation systems''.
\newblock 8th Innovations in Theoretical Computer Science Conference (ITCS
  2017) {\bf 67}, 49:1--49:21~(2017).
\newblock
  url:~\href{https://doi.org/10.48550/arXiv.1603.08675}{doi.org/10.48550/arXiv.1603.08675}.

\bibitem{kerenidis2019qmeans}
Iordanis Kerenidis, Jonas Landman, Alessandro Luongo, and Anupam Prakash.
\newblock ``q-means: A quantum algorithm for unsupervised machine learning''.
\newblock In Advances in Neural Information Processing Systems 32.
\newblock Pages 4136--4146.
\newblock Curran Associates, Inc.~(2019).
\newblock
  url:~\href{https://dl.acm.org/doi/abs/10.5555/3454287.3454659}{dl.acm.org/doi/abs/10.5555/3454287.3454659}.

\bibitem{lloyd2016quantum}
Seth Lloyd, Silvano Garnerone, and Paolo Zanardi.
\newblock ``Quantum algorithms for topological and geometric analysis of
  data''.
\newblock Nature communications {\bf 7}, 1--7~(2016).
\newblock
  url:~\href{https://doi.org/10.1038/ncomms10138}{doi.org/10.1038/ncomms10138}.

\bibitem{QNN2018}
Edward Farhi and Hartmut Neven.
\newblock ``Classification with quantum neural networks on near term
  processors''~(2018).
\newblock
  url:~\href{https://doi.org/10.48550/arXiv.1802.06002}{doi.org/10.48550/arXiv.1802.06002}.

\bibitem{QCNN2019}
I~Kerenidis, J~Landman, and A~Prakash.
\newblock ``Quantum algorithms for deep convolutional neural networks''.
\newblock \href{https://dx.doi.org/10.48550/arXiv.1911.01117}{EIGHTH
  INTERNATIONAL CONFERENCE ON LEARNING REPRESENTATIONS ICLR}~(2019).

\bibitem{QNN2020}
J~Allcock, CY~Hsieh, I~Kerenidis, and S~Zhang.
\newblock ``Quantum algorithms for feedforward neural networks''.
\newblock \href{https://dx.doi.org/10.1145/3411466}{ACM Transactions on Quantum
  Computing 1 (1), 1-24}~(2020).

\bibitem{CNN2019}
Iris Cong, Soonwon Choi, and Mikhail~D. Lukin.
\newblock ``Quantum convolutional neural networks''.
\newblock \href{https://dx.doi.org/10.1038/s41567-019-0648-8}{Nature Physics
  15}~(2019).

\bibitem{Image2020}
Hector~Ivan Garcıa-Hernandez, Raymundo Torres-Ruiz, and Guo-Hua Sun.
\newblock ``Image classification via quantum machine learning''~(2020).
\newblock
  url:~\href{https://doi.org/10.48550/arXiv.2011.02831}{doi.org/10.48550/arXiv.2011.02831}.

\bibitem{Dressed2020}
Saurabh Kumar, Siddharth Dangwal, and Debanjan Bhowmik.
\newblock ``Supervised learning using a dressed quantum network with "super
  compressed encoding": Algorithm and quantum-hardware-based
  implementation''~(2020).
\newblock
  url:~\href{https://doi.org/10.48550/arXiv.2007.10242}{doi.org/10.48550/arXiv.2007.10242}.

\bibitem{Semisupervised2020}
Kouhei Nakaji and Naoki Yamamoto.
\newblock ``Quantum semi-supervised generative adversarial network for enhanced
  data classification''~(2020).
\newblock
  url:~\href{https://doi.org/10.1038/s41598-021-98933-6}{doi.org/10.1038/s41598-021-98933-6}.

\bibitem{Polyadic2020}
William Cappelletti, Rebecca Erbanni, and Joaquín Keller.
\newblock ``Polyadic quantum classifier''~(2020).
\newblock
  url:~\href{https://doi.org/10.48550/arXiv.2007.14044}{doi.org/10.48550/arXiv.2007.14044}.

\bibitem{Supervised2018}
Vojtech Havlicek, Antonio~D. Córcoles, Kristan Temme, Aram~W. Harrow, Abhinav
  Kandala, Jerry~M. Chow, and Jay~M. Gambetta.
\newblock ``Supervised learning with quantum enhanced feature spaces''~(2018).
\newblock
  url:~\href{https://doi.org/10.1038/s41586-019-0980-2}{doi.org/10.1038/s41586-019-0980-2}.

\bibitem{Hierarchical2018}
Edward Grant, Marcello Benedetti, Shuxiang Cao, Andrew Hallam, Joshua Lockhart,
  Vid Stojevic, Andrew~G. Green, and Simone Severini.
\newblock ``Hierarchical quantum classifiers''~(2018).
\newblock
  url:~\href{https://doi.org/10.1038/s41534-018-0116-9}{doi.org/10.1038/s41534-018-0116-9}.

\bibitem{Kiani20}
Bobak~Toussi Kiani, Agnes Villanyi, and Seth Lloyd.
\newblock ``Quantum medical imaging algorithms''~(2020).
\newblock
  url:~\href{https://doi.org/10.48550/arXiv.2004.02036}{doi.org/10.48550/arXiv.2004.02036}.

\bibitem{cerezo2020variational}
Marco Cerezo, Andrew Arrasmith, Ryan Babbush, Simon~C Benjamin, Suguru Endo,
  Keisuke Fujii, Jarrod~R McClean, Kosuke Mitarai, Xiao Yuan, Lukasz Cincio,
  et~al.
\newblock ``Variational quantum algorithms''~(2020).
\newblock
  url:~\href{https://doi.org/10.1038/s42254-021-00348-9}{doi.org/10.1038/s42254-021-00348-9}.

\bibitem{bharti2021noisy}
Kishor Bharti, Alba Cervera-Lierta, Thi~Ha Kyaw, Tobias Haug, Sumner
  Alperin-Lea, Abhinav Anand, Matthias Degroote, Hermanni Heimonen, Jakob~S
  Kottmann, Tim Menke, et~al.
\newblock ``Noisy intermediate-scale quantum algorithms''.
\newblock Reviews of Modern Physics {\bf 94}, 015004~(2022).
\newblock
  url:~\href{https://doi.org/10.1103/RevModPhys.94.015004}{doi.org/10.1103/RevModPhys.94.015004}.

\bibitem{noirhomme2011far}
Monique Noirhomme-Fraiture and Paula Brito.
\newblock ``Far beyond the classical data models: symbolic data analysis''.
\newblock Statistical Analysis and Data Mining: the ASA Data Science Journal
  {\bf 4}, 157--170~(2011).
\newblock
  url:~\href{https://doi.org/10.1002/sam.10112}{doi.org/10.1002/sam.10112}.

\bibitem{perez2020data}
Adri{\'a}n P{\'e}rez-Salinas, Alba Cervera-Lierta, Elies Gil-Fuster, and
  Jos{\'e}~I Latorre.
\newblock ``Data re-uploading for a universal quantum classifier''.
\newblock Quantum {\bf 4}, 226~(2020).
\newblock
  url:~\href{https://doi.org/10.22331/q-2020-02-06-226}{doi.org/10.22331/q-2020-02-06-226}.

\bibitem{mitarai2018quantum}
Kosuke Mitarai, Makoto Negoro, Masahiro Kitagawa, and Keisuke Fujii.
\newblock ``Quantum circuit learning''.
\newblock \href{https://dx.doi.org/10.1103/PhysRevA.98.032309}{Physical Review
  A {\bf 98}, 032309}~(2018).

\bibitem{schuld2019evaluating}
Maria Schuld, Ville Bergholm, Christian Gogolin, Josh Izaac, and Nathan
  Killoran.
\newblock ``Evaluating analytic gradients on quantum hardware''.
\newblock \href{https://dx.doi.org/10.1103/PhysRevA.99.032331}{Physical Review
  A {\bf 99}, 032331}~(2019).

\bibitem{schuld2021quantum}
Maria Schuld and Francesco Petruccione.
\newblock ``Quantum models as kernel methods''.
\newblock In Machine Learning with Quantum Computers.
\newblock Pages 217--245.
\newblock Springer~(2021).

\bibitem{schuld2021effect}
Maria Schuld, Ryan Sweke, and Johannes~Jakob Meyer.
\newblock ``Effect of data encoding on the expressive power of variational
  quantum-machine-learning models''.
\newblock \href{https://dx.doi.org/10.1103/PhysRevA.103.032430}{Physical Review
  A {\bf 103}, 032430}~(2021).

\bibitem{cong2019quantum}
Iris Cong, Soonwon Choi, and Mikhail~D Lukin.
\newblock ``Quantum convolutional neural networks''.
\newblock Nature Physics {\bf 15}, 1273--1278~(2019).

\bibitem{mcclean2018barren}
Jarrod~R McClean, Sergio Boixo, Vadim~N Smelyanskiy, Ryan Babbush, and Hartmut
  Neven.
\newblock ``Barren plateaus in quantum neural network training landscapes''.
\newblock Nature communications {\bf 9}, 1--6~(2018).
\newblock
  url:~\href{https://doi.org/10.1038/s41467-018-07090-4}{doi.org/10.1038/s41467-018-07090-4}.

\bibitem{marrero2020entanglement}
Carlos~Ortiz Marrero, M{\'a}ria Kieferov{\'a}, and Nathan Wiebe.
\newblock ``Entanglement-induced barren plateaus''.
\newblock PRX Quantum {\bf 2}, 040316~(2021).
\newblock
  url:~\href{https://doi.org/10.1103/PRXQuantum.2.040316}{doi.org/10.1103/PRXQuantum.2.040316}.

\bibitem{QCNNplateaus2020}
Marco Cerezo, Akira Sone, Tyler Volkoff, Lukasz Cincio, and Patrick~J Coles.
\newblock ``Cost function dependent barren plateaus in shallow parametrized
  quantum circuits''.
\newblock Nature communications {\bf 12}, 1--12~(2021).
\newblock
  url:~\href{https://doi.org/10.1038/s41467-021-21728-w}{doi.org/10.1038/s41467-021-21728-w}.

\bibitem{Train2020}
Kunal Sharma, Marco Cerezo, Lukasz Cincio, and Patrick~J Coles.
\newblock ``Trainability of dissipative perceptron-based quantum neural
  networks''.
\newblock Physical Review Letters {\bf 128}, 180505~(2022).
\newblock
  url:~\href{https://doi.org/10.1103/PhysRevLett.128.180505}{doi.org/10.1103/PhysRevLett.128.180505}.

\bibitem{NearestCentroid2021}
S~Johri, S~Debnath, A~Mocherla, A~Singh, A~Prakash, J~Kim, and I~Kerenidis.
\newblock ``Nearest centroid classification on a trapped ion quantum
  computer''~(2021).

\bibitem{jia2019orthogonal}
Kui Jia, Shuai Li, Yuxin Wen, Tongliang Liu, and Dacheng Tao.
\newblock ``Orthogonal deep neural networks''.
\newblock \href{https://dx.doi.org/10.1109/TPAMI.2019.2948352}{IEEE
  transactions on pattern analysis and machine intelligence}~(2019).

\bibitem{wang2020orthogonal}
Jiayun Wang, Yubei Chen, Rudrasis Chakraborty, and Stella~X Yu.
\newblock ``Orthogonal convolutional neural networks''.
\newblock In Proceedings of the IEEE/CVF Conference on Computer Vision and
  Pattern Recognition.
\newblock \href{https://dx.doi.org/10.1109/CVPR42600.2020.01152}{Pages
  11505--11515}.
\newblock ~(2020).

\bibitem{bansal2018can}
Nitin Bansal, Xiaohan Chen, and Zhangyang Wang.
\newblock ``Can we gain more from orthogonality regularizations in training
  deep networks?''.
\newblock \href{https://dx.doi.org/10.5555/3327144.3327339}{Advances in Neural
  Information Processing Systems{\bf 31}}~(2018).

\bibitem{Zhai}
Xiaohua Zhai, Alexander Kolesnikov, Neil Houlsby, and Lucas Beyer.
\newblock ``Scaling vision transformers''~(2021).

\bibitem{KP22}
Iordanis Kerenidis and Anupam Prakash.
\newblock ``Quantum machine learning with subspace states''~(2022).
\newblock
  url:~\href{https://doi.org/10.48550/arXiv.2202.00054}{doi.org/10.48550/arXiv.2202.00054}.

\bibitem{unary2019}
Sergi Ramos-Calderer, Adrián Pérez-Salinas, Diego García-Martín, Carlos
  Bravo-Prieto, Jorge Cortada, Jordi Planagumà, and José~I. Latorre.
\newblock ``Quantum unary approach to option pricing''~(2019).

\bibitem{parallelGates}
Nikodem Grzesiak, Reinhold Blümel, Kenneth Wright, Kristin~M. Beck, Neal~C.
  Pisenti, Ming Li, Vandiver Chaplin, Jason~M. Amini, Shantanu Debnath, Jwo-Sy
  Chen, and Yunseong Nam.
\newblock ``Efficient arbitrary simultaneously entangling gates on a
  trapped-ion quantum computer''.
\newblock \href{https://dx.doi.org/10.1038/s41467-020-16790-9}{Nat Commun,
  11}~(2020).

\bibitem{ZNL21}
Alexander Zlokapa, Hartmut Neven, and Seth Lloyd.
\newblock ``A quantum algorithm for training wide and deep classical neural
  networks''~(2021).
\newblock
  url:~\href{https://doi.org/10.48550/arXiv.2107.09200}{doi.org/10.48550/arXiv.2107.09200}.

\bibitem{lezcano2019cheap}
Mario Lezcano-Casado and David Mart{\i}nez-Rubio.
\newblock ``Cheap orthogonal constraints in neural networks: A simple
  parametrization of the orthogonal and unitary group''.
\newblock In International Conference on Machine Learning.
\newblock Pages 3794--3803.
\newblock PMLR~(2019).
\newblock
  url:~\href{https://doi.org/10.48550/arXiv.1901.08428}{doi.org/10.48550/arXiv.1901.08428}.

\bibitem{leshno1993multilayer}
Moshe Leshno, Vladimir~Ya Lin, Allan Pinkus, and Shimon Schocken.
\newblock ``Multilayer feedforward networks with a nonpolynomial activation
  function can approximate any function''.
\newblock \href{https://dx.doi.org/10.1016/S0893-6080(05)80131-5}{Neural
  networks {\bf 6}, 861--867}~(1993).

\bibitem{hecht1992theory}
Robert Hecht-Nielsen.
\newblock ``Theory of the backpropagation neural network''.
\newblock In Neural networks for perception.
\newblock \href{https://dx.doi.org/10.1109/IJCNN.1989.118638}{Pages 65--93}.
\newblock Elsevier~(1992).

\bibitem{rojas1996backpropagation}
Raul Rojas.
\newblock ``The backpropagation algorithm''.
\newblock In Neural networks.
\newblock \href{https://dx.doi.org/10.1007/978-3-642-61068-4_7}{Pages
  149--182}.
\newblock Springer~(1996).

\bibitem{medmnist}
Jiancheng Yang, Rui Shi, and Bingbing Ni.
\newblock ``Medmnist classification decathlon: A lightweight automl benchmark
  for medical image analysis''~(2020).

\bibitem{pneumonia}
Daniel~S. Kermany, Michael Goldbaum, and et~al.
\newblock ``Identifying medical diagnoses and treatable diseases by image-based
  deep learning''.
\newblock \href{https://dx.doi.org/10.1016/j.cell.2018.02.010}{Cell, vol. 172,
  no. 5, pp. 1122 – 1131.e9,}~(2018).

\bibitem{retina}
Ping Zhang and Bin Sheng.
\newblock ``Deepdr diabetic retinopathy image dataset (deepdrid), "the 2nd
  diabetic retinopathy – grading and image quality estimation challenge"''.
\newblock https://isbi.deepdr.org/data.html~(2020).

\bibitem{noiseNN2017}
Hyeonwoo Noh, Tackgeun You, Jonghwan Mun, and Bohyung Han.
\newblock ``Regularizing deep neural networks by noise: Its interpretation and
  optimization''.
\newblock \href{https://dx.doi.org/10.5555/3295222.3295264}{NeurIPS}~(2017).

\bibitem{ying2019overview}
Xue Ying.
\newblock ``An overview of overfitting and its solutions''.
\newblock In Journal of physics: Conference series.
\newblock \href{https://dx.doi.org/10.1088/1742-6596/1168/2/022022}{Volume
  1168, page 022022}.
\newblock IOP Publishing~(2019).

\bibitem{cherrat2022quantum}
El~Amine Cherrat, Iordanis Kerenidis, Natansh Mathur, Jonas Landman, Martin
  Strahm, and Yun~Yvonna Li.
\newblock ``Quantum vision transformers''~(2022).

\bibitem{readthefineprint}
Scott Aaronson.
\newblock ``Read the fine print''.
\newblock \href{https://dx.doi.org/10.1038/nphys3272}{Nature Physics {\bf 11},
  291--293}~(2015).

\bibitem{nielsen}
Michael~A. Nielsen.
\newblock ``Neural networks and deep learning''.
\newblock Determination Press~(2015).

\end{thebibliography}

\newpage

\appendix

\section{Appendix}

\subsection{Classical Orthogonal Neural Network Training}\label{sec:ortho_nn_backprop_details}
For completeness, we present two algorithms from \cite{jia2019orthogonal} for updating orthogonal matrices. 

The first algorithm is an approximate one, called \emph{Singular Value Bounding} (SVB). It starts by applying the usual gradient descent update on the matrix, therefore making it not orthogonal anymore. Then, the singular values of the new matrix are extracted using Singular Value Decomposition (SVD), their values are manually pushed to be close to 1, and the matrix is recomposed, hence enforcing orthogonality. This method shows less advantage on practical experiments \cite{jia2019orthogonal}. It has a complexity of $O(n^3)$ due to the SVD, and in practice is better than the second algorithm described below. Note that this running time is still asymptotically worse than $O(n^2)$, the running time to perform standard gradient descent.

The second algorithm ensures perfect orthogonality by performing the gradient descent in the manifold of orthogonal matrices, called the Stiefel Manifold. In practice, \cite{jia2019orthogonal} this method showed a substantially advantageous classification results on standard datasets, indicating that perfect orthogonality can be a much more desired property than approximate one. This algorithm requires $O(n^3)$ operations, and is quite prohibitive in practice. We give an informal step-by-step detail of this algorithm:

\begin{enumerate}
    \item Compute the gradient $G$ of the weight matrix $W$.
    \item Project the gradient matrix $G$ in the tangent space,  (The space tangent to the manifold at this point $W$): multiply $G$ by some other matrices based on $W$: $$(I-WW^T)G+\frac{1}{2}W(W^TG-G^TW)$$ This requires several matrix-matrix multiplications. In the case of square $n\times n$ matrices, each has complexity $O(n^3)$. the result of this projection is called the \emph{manifold gradient} $\Omega$.
    \item update $W' = W - \eta\Omega$, where $\eta$ is the chosen learning rate. 
    \item Perform a \emph{retraction} from the tangent space to the manifold. To do so, we multiply $W'$ by $Q$ factor of the \emph{QR decomposition}, obtained using Gram Schmidt orthonormalization, which has complexity $O(2n^3)$. 
\end{enumerate}

\subsection{Classical orthogonal layer from a pyramid circuit}\label{Cpyr}

\begin{figure}[t]
    \centering
    \includegraphics[width=0.5\textwidth]{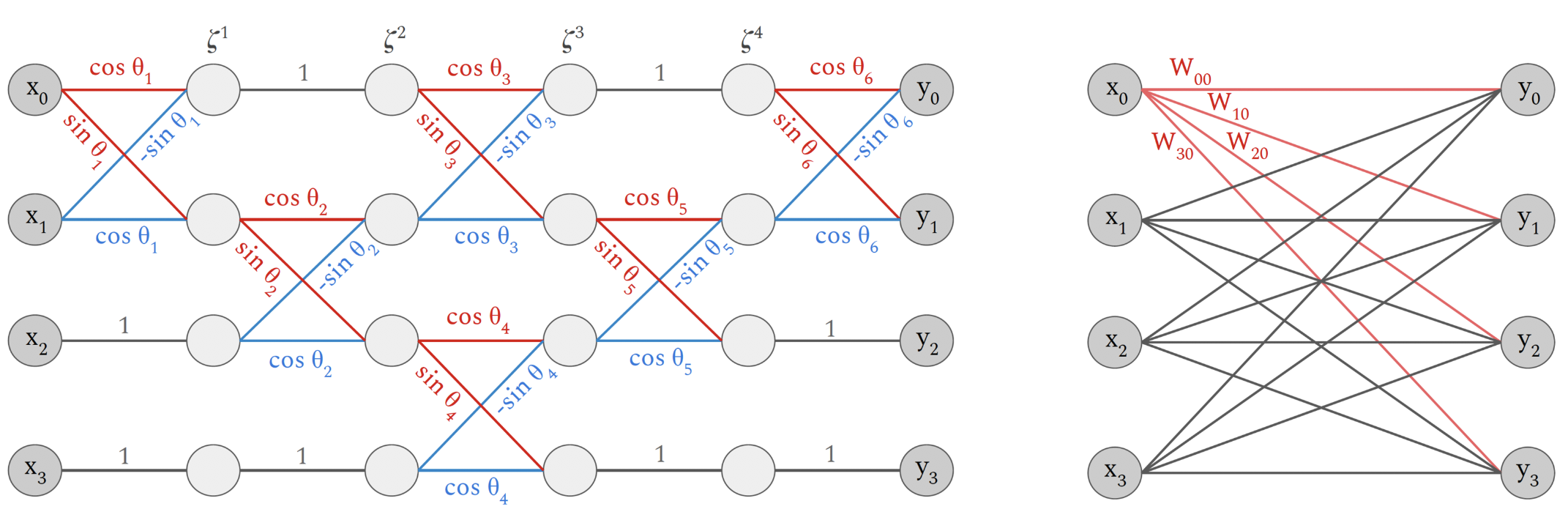}
    \caption{Representation of a single classical orthogonal layer (left) on a $4\times 4$ case ($n$=4). Each connecting line represents a scalar multiplication with the value indicated. \emph{Inner layers} $\zeta^{\lambda}$ are displayed. A \emph{timestep} corresponds to the lines in between two \emph{inner layers} (see Section \ref{sec:trainingquantum} for definitions). The operation performed is equivalent to a mapping $x\mapsto y=Wx$, namely to a normal fully-connected layer (right), with the extra property that the matrix $W$ is orthogonal.}
    \label{fig:quantum_vs_classical_representation}
\end{figure}

As we have said before, the effect of applying a pyramid circuit on a unary quantum state with $n$ qubits can be simulated classically with a small overhead, since we know that the output vector lies in the $n$-dimensional subspace spanned by the unary basis states. This implies that we can define a classical orthogonal layer, based on the simulation of the pyramid circuit, where each $RBS$ gate is replaced by a planar rotation between its two inputs.
This is shown in Fig.\ref{fig:quantum_vs_classical_representation}, where each layer is constituted of $\frac{n(n-1)}{2}$ planar rotations, for a total of $4\times\frac{n(n-1)}{2} = O(n^2)$ basic operations. Therefore, our single orthogonal layer forward pass has the same complexity $O(n^2)$ as the usual matrix multiplication. 

One may still have an advantage in performing the quantum circuit for inference, since the quantum circuit has depth $O(n)$, instead of the $O(n^2)$ classical complexity of the matrix-vector multiplication. More importantly, the main advantage of our method is that we can also now train orthogonal weight matrices classically in time $O(n^2)$, instead of the previously best-known $O(n^3)$. Last, involving input states that are no more unary increases the time of the classical simulation \cite{KP22}. 

\subsection{Mapping orthogonal matrices to pyramids} \label{appendix_proof_pyramid_anlges}
As described in Section \ref{sec:ortho_pyramid_details}, any pyramidal circuit gives rise to a classical orthogonal matrix. To prove that this is a bijective relation, we need to show that any orthogonal matrix can be represented as a pyramidal circuit. As we said before, in fact an orthogonal matrix with determinant $1$ can be mapped into a pyramidal circuit, while any orthogonal matrix with determinant $-1$ can be mapped to a pyramidal circuit plus a single $Z$ gate on the last qubit. Let us be more precise about the last statement. For an orthogonal matrix $O$ with determinant $-1$, we can factorize the matrix as $O^-Z$, where $O^-$ is equal to $O$ apart from the last column where all signs are flipped. The matrix $Z$ is equal to the Identity matrix with only the last diagonal element being $-1$ instead. Now this $Z$ matrix corresponds to the application of the $Z$ gate restricted to the unary basis, and the $O^-$ matrix is now an orthogonal matrix with determinant $1$. It remains to show how an orthogonal matrix with determinant $1$ can be mapped to a pyramid circuit.

As said in Section \ref{sec:QONN_forward}, each element of an orthogonal matrix can be seen as the sum of all possible paths from some input qubit to some output qubit and each term in this sum is a product of cosines and sines of the angles of the gates that one encounters in the path (see below for a simple $3\times 3$ case). Note that finding the exact expression for each element of the matrix is possible but not efficient, but here we are only interested in showing the bijection and in fact we never go back to an orthogonal matrix from a pyramid circuit during our training or inference. 
Knowing the elements of the matrix, one can retrieve the angles themselves by essentially solving a system of equations. More precisely, we traverse the orthogonal matrix column by column from right to left, and going from bottom to top (until before the anti-diagonal element) in each column. 
Since we know exactly the expression in terms of sines and cosines of a subset of angles for each matrix element, we just need to equate it with the corresponding actual value in the orthogonal matrix. Traversing it in this manner always leads to equations with only one unknown angle which can be, therefore, easily retrieved.

We show a simple illustrative example for the $3\times 3$ case (see Fig.\ref{fig:3x3_case}). The equation below shows how we can map the elements of a $3 \times 3$ orthogonal matrix $M = [M_{ij}]$ to the angles of a $3\times 3$ pyramidal circuit, as in Fig.\ref{fig:3x3_case}. $M$ has only 3 free parameters that we will find by traversing the matrix as we described above.

%\noindent
%\begin{bmatrix}
%c(\theta_1)c(\theta_3) - s(\theta_1)c(\theta_2)s(\theta_3) & -s(\theta_1)c(\theta_3) - %c(\theta_1)c(\theta_2)s(\theta_3) & s(\theta_2)s(\theta_3)\\
%s(\theta_1)c(\theta_2)c(\theta_3) + c(\theta_1)s(\theta_3) & c(\theta_1)c(\theta_2)c(\theta_3) - %s(\theta_1)s(\theta_3) & -s(\theta_2)c(\theta_3)\\
%s(\theta_1)s(\theta_2) & c(\theta_1)s(\theta_2) & c(\theta_2)\\
%\end{bmatrix}

\begin{equation}
\begin{bmatrix}
\cdot & \cdot & \cdot\\
\cdot & \cdot & -s(\theta_2)c(\theta_3)\\
\cdot & c(\theta_1)s(\theta_2) & c(\theta_2)\\
\end{bmatrix}
=
\begin{bmatrix}
\cdot & \cdot & \cdot\\
\cdot & \cdot & M_{23}\\
\cdot & M_{32} & M_{33}\\
\end{bmatrix}
\end{equation}

We find the angles using the following set of equations, in this order:

\begin{equation}
    \begin{cases}
    (1) \quad M_{33} = \cos(\theta_2)\\
    (2) \quad M_{23} = -s(\theta_2)c(\theta_3)\\
    (3) \quad M_{32} = c(\theta_1)s(\theta_2)
    \end{cases}
\end{equation}

This can be solved sequentially to find the set of corresponding angles.

\begin{equation}
    \begin{cases}
    \theta_2 = \arccos(M_{33})\\
    \theta_3 = -\arccos(\frac{M_{23}}{\sin(\theta_2)})\\
    \theta_1 = \arccos(\frac{M_{32}}{\sin(\theta_2)})
    \end{cases}
\end{equation}
It is easy to see how this procedure generalizes to the general case.

%\begin{bmatrix}
%\_ & \_ & \_ & \_ & \_ & \_ & \_ & \_\\
%\_ & \_ & \_ & \_ & \_ & \_ & \_ & 7(\theta_{28})\\
%\_ & \_ & \_ & \_ & \_ & \_ & 13(\theta_{21}) & 6(\theta_{27})\\
%\_ & \_ & \_ & \_ & \_ & 18(\theta_{15}) & 12(\theta_{20}) & 5(\theta_{26})\\
%\_ & \_ & \_ & \_ & 22(\theta_{10}) & 17(\theta_{14}) & 11(\theta_{19}) & 4(\theta_{25})\\
%\_ & \_ & \_ & 25(\theta_{6}) & 21(\theta_{9}) & 16(\theta_{13}) & 10(\theta_{18}) & 3(\theta_{24})\\
%\_ & \_ & 27(\theta_{3}) & 24(\theta_{5}) & 20(\theta_{8}) & 15(\theta_{12}) & 9(\theta_{17}) & %2(\theta_{23})\\
%\_ & 28(\theta_{1}) & 26(\theta_{2}) & 23(\theta_{4}) & 19(\theta_{7}) & 14(\theta_{11}) & %8(\theta_{16}) & 1(\theta_{22})\\
%\end{bmatrix}
%
%\begin{figure}[!h]
%    \centering
%\includegraphics[width=0.8\linewidth]{imageortho/circuit_angles_.png}
%    \label{fig:data_loader}
%\caption{The matrix and circuit show how we traverse the elements of the orthogonal matrix and find the %corresponding angles of the pyramid circuit.}
%\end{figure}

\subsection{Error Mitigation}\label{sec:error}

It is important to notice that with our restriction to unary states, strong error mitigation techniques become available. Indeed, as we expect to obtain only quantum superposition of unary states at every layer, we can post process our measurements and discard the ones that present non-unary states (\emph{i.e.} states with more than one qubit in state $\ket{1}$, or the ground state). The most expected error is a bit flip between $\ket{1}$ and $\ket{0}$. The case where two bit flips happen at the same time, which would change a unary state to a different unary state and would thus pass through our error mitigation, is even less probable. This error mitigation procedure can be applied efficiently to the results of a hardware demonstration, and it has been used in the results presented in this paper. \revised{Note that error mitigation techniques involving specific hardware noise models or other techniques, are not included in this work.}

\subsection{Quantum Orthogonal Network: Tomography}\label{sec:tomography}

As shown in Fig.\ref{fig:full_schema}, when using the quantum circuit, the output is a quantum state $\ket{y} = \ket{Wx}$. As often in quantum machine learning \cite{readthefineprint}, it is important to consider the cost of retrieving the classical outputs, using a procedure called tomography. In our case this is even more crucial since, between each layer, the quantum output will be converted into a classical one in order to apply a non-linear function, and then reloaded for the next layer.

Retrieving the amplitudes of a quantum state comes at cost of multiple measurements, which requires running the circuit multiples times, hence adding a multiplicative overhead in the running time. A finite number of samples is also a source of approximation error in the final result. In this work, we will allow for $\ell_{\infty}$ errors \cite{QCNN2019}. The $\ell_{\infty}$ tomography on a quantum state $\ket{y}$ with unary encoding on $n$ qubits requires $O(\log(n)/\delta^2)$ measurements, where $\delta>0$ is the error threshold allowed. For each $j\in [n]$, $|y_j|$ will be obtained with an absolute error $\delta$, and if $|y_j| < \delta$, it will most probably not be measured, hence set to 0. In practice, one would perform as many measurements as is convenient during the experiment, and deduce the equivalent precision $\delta$ from the number of measurements made.   

Note that it is important to obtain the amplitudes of the quantum state, which in our case are positive or negative real numbers, and not just the probabilities of the outcomes, which are the squares of the amplitudes. There are different ways of obtaining the sign of the amplitudes, and we present two different ways below.

Indeed, a simple measurement in the computational basis will only provide us with estimations of the probabilities that are the squares of the quantum amplitudes. In the case of neural networks, it is important to obtain the sign of the layer's components in order to apply certain type of non-linearities. For instance, the ReLu activation function is often used to set all negative components to 0. 

\begin{figure}[h]
    \centering
    \includegraphics[width=0.49\textwidth]{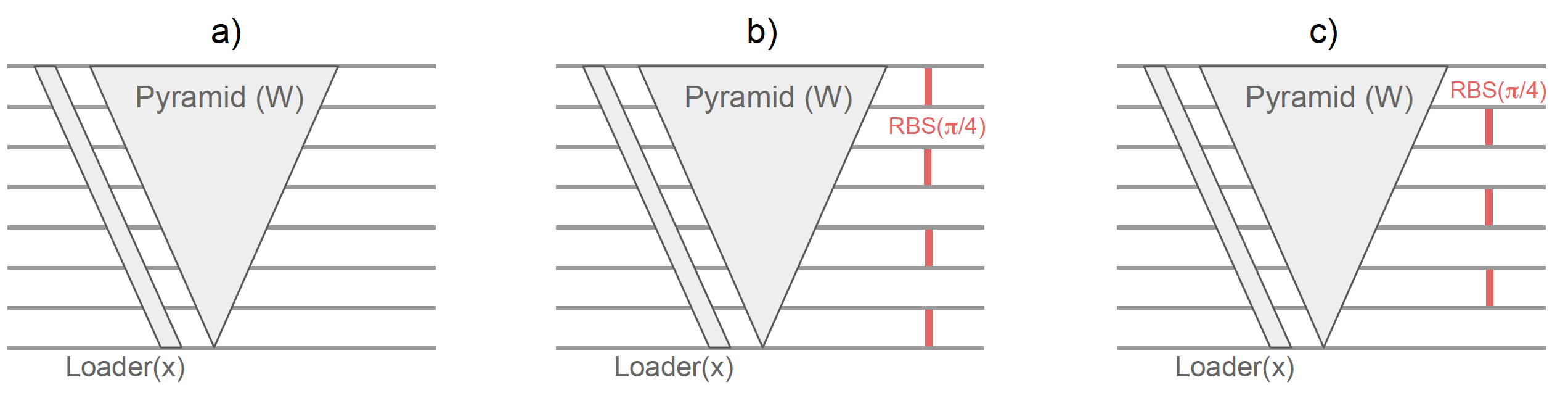}
    \caption{First tomography procedure to retrieve the value and the sign of each component of the resulting vector $\ket{y}=\ket{Wx}$. Circuit a) is the original one, while circuits b) and c) have additional \emph{RBS} gates with angle $\pi/4$ at the end to compare the signs between adjacent components. In all three cases, an $\ell_{\infty}$ tomography is applied.}
    \label{fig:tomography}
\end{figure}

In Fig.\ref{fig:tomography}, we propose a specific enhancement to our circuit to obtain the signs of the vector's components at low cost. The sign retrieval procedure consists of three parts. 
\begin{enumerate}
    \item The circuit is first applied as described above, allowing to retrieve each squared amplitude $y_j^2$ with precision $\delta >0$ using the $\ell_{\infty}$ tomography. The probability of measuring the unary state $\ket{e_1}$ (\emph{i.e.} $\ket{100...}$), is $p(e_1) = y_1^2$. 
    \item We apply the same steps a second time on a modified circuit. It has additional \emph{RBS} gates with angle $\pi/4$ at the end, which will mix the amplitudes pair by pair. The probabilities to measure $\ket{e_1}$ and $\ket{e_2}$ are now given by $p(e_1) = (y_1 + y_2)^2$ and $p(e_2) = (y_1 - y_2)^2$. Therefore, if $p(e_1)>p(e_2)$, we have $sign(y_1)\neq sign(y_2)$, and if $p(e_1)<p(e_2)$, we have $sign(y_1)= sign(y_2)$. The same holds for the pairs ($y_3$, $y_4$), and so on. 
    \item We finally perform the same, where the \emph{RBS} are shifted by one position below. Then we compare the signs of the pairs ($y_2$, $y_3$), ($y_4$, $y_5$) and so on. 
\end{enumerate}

At the end, we are able to recover each value $y_j$ with its sign, assuming that $y_1 > 0$ for instance. This procedure has the benefit of not adding depth to the original circuit, but requires 3 times more runs. The overall cost of the tomography procedure with sign retrieval is given by $\widetilde{O}(n/\delta^2)$.

\begin{figure}[h]
    \centering
    \includegraphics[width=0.5\textwidth]{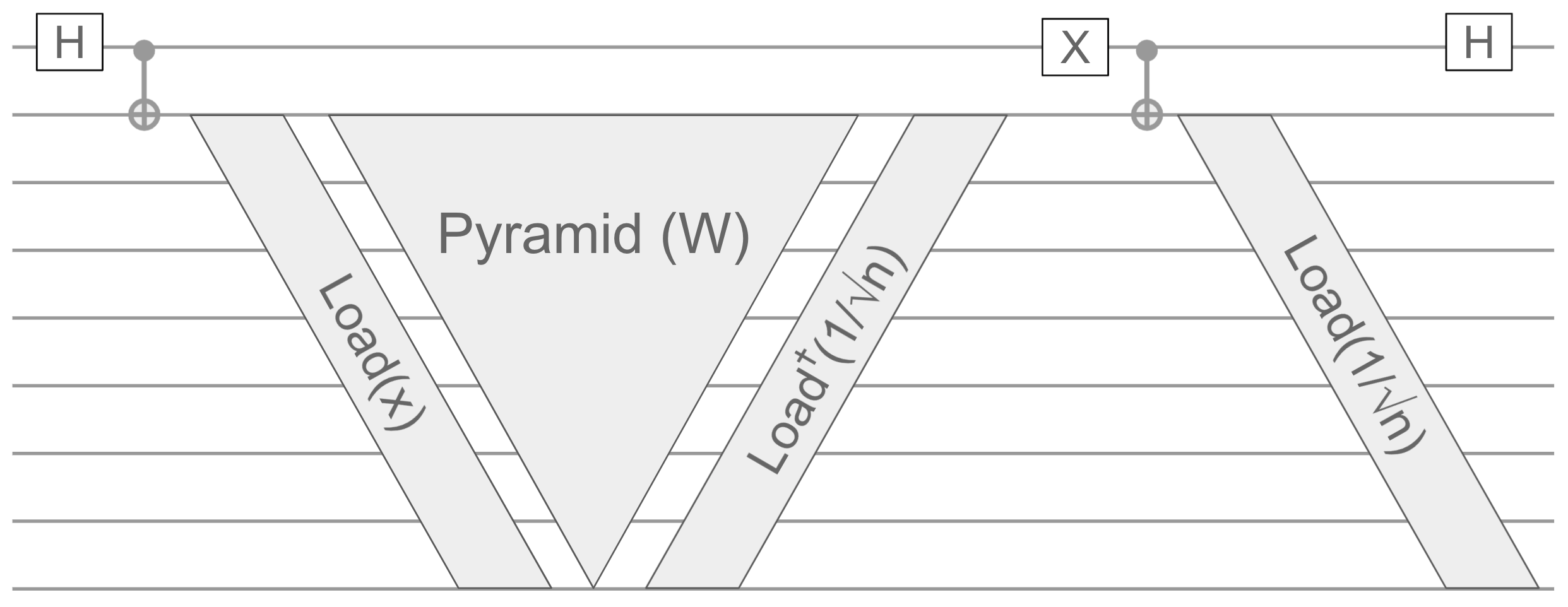}
    \caption{Second tomography procedure to retrieve the value and the sign of each component of the resulting vector $\ket{y}=\ket{Wx}$. For a \emph{rectangular} case with output of size $m$, the two opposite loaders at the end must be on the last $m$ qubits only, and the $CNOT$ gate between them connects the top qubits to the loader's top qubit as well.}
    \label{fig:tomography_2}
\end{figure}

In Fig.\ref{fig:tomography_2} we propose another method to obtain the values of the amplitudes and their signs, which is in fact what we used for the hardware demonstrations. Compared to the above procedure, it relies on one circuit only, but requires an extra qubit and a depth of $3n+O(1)$ instead of $2n+O(1)$.

This circuit performs a Hadamard and CNOT gate in order to initialize the qubits in the state $\frac{1}{\sqrt{2}}\ket{0}\ket{0}+\frac{1}{\sqrt{2}}\ket{1}\ket{e_1}$, where the second register corresponds to the $n$ qubits that will be processed by the pyramidal circuit and the loaders. 

Next, applying the data loader for the normalized input vector $x$ (see Section \ref{sec:data_loading}) and the pyramidal circuit will, according to Eq.(\ref{eq:full_map}), map the state to 
\begin{equation}
    \frac{1}{\sqrt{2}}\ket{0}\ket{0}
    +\frac{1}{\sqrt{2}}\ket{1}\sum_{j=1}^n W_j x \ket{e_j}
\end{equation}

In other words, we performed the pyramid circuits controlled on the first qubit being in state $\ket{1}$. Then, we flip the first qubit with an X gate and perform a controlled loading of the uniform norm-1 vector $(\frac{1}{\sqrt{n}},\cdots,\frac{1}{\sqrt{n}})$.
 For this, we add the adjoint data loader for the state, a CNOT gate and the data loader a second time.
 Recall that if a circuit $U$ is followed by $U^{\dagger}$, it is equivalent to the identity.
Therefore, this will load the uniform state only when the first qubit is in state $\ket{1}$:
\begin{equation}
    \frac{1}{\sqrt{2}}\ket{0}\sum_{j=1}^n W_j x \ket{e_j}
    +
    \frac{1}{\sqrt{2}}\ket{1}\sum_{j=1}^n \frac{1}{\sqrt{n}}\ket{e_j}
\end{equation}

Note that the final transposed loader could be in fact replaced by a \emph{parallel} data loader (see Fig.\ref{loaders}) for shorter depth. 

Finally, a Hadamard gate will mix both parts of the amplitudes on the extra qubit to give us the desired state:
\begin{equation}
    \frac{1}{2}\ket{0}\sum_{j=1}^n \left(W_j x + \frac{1}{\sqrt{n}} \right)\ket{e_j}
    + \frac{1}{2}\ket{1}\sum_{j=1}^n \left(W_j x - \frac{1}{\sqrt{n}} \right)\ket{e_j}
\end{equation}

On this final state, we can see that the difference in the probabilities of measuring the extra qubit in state $0$ or $1$ and rest in the unary state $e_j$ is given by $\Pr[0,e_j] - \Pr[1,e_j] = \frac{1}{4}\left(W_j x + \frac{1}{\sqrt{n}}\right)^2 - \frac{1}{4}\left(W_j x - \frac{1}{\sqrt{n}}\right)^2 = W_j x /\sqrt{n} $. Therefore, for each $j$, we can deduce the sign of $W_j x$ by looking at the most frequent output of the measurement of the first qubit. To deduce as well the value of $W_j x$, we simply use $\Pr[0,e_j]$ or $\Pr[1,e_j]$ depending on the sign found before. For instance, if $W_j x>0$ we have $W_j x = 2\sqrt{\Pr[0,e_j]} - \frac{1}{\sqrt{n}}$.

Combining with the $\ell_{\infty}$ tomography and the non-linearity, the overall cost of this tomography is given by $\widetilde{O}(n/\delta^2)$ as well.

\subsection{Additional details on Quantum Experiments}\label{sec:additional_results}

\subsubsection{Quantum software}

Software development of quantum algorithms was performed using tools from the QC Ware Forge platform, including the quasar language, the data loader and the inner product estimation procedures, as well as the quantum pyramid circuits. \revised{Simulations of quantum circuit were done in the noise-free model.} The final circuits were translated into qiskit circuits that were then sent to the IBM hardware. 

The datasets were downloaded from the MedMNIST repository \cite{medmnist} and pre-processed using the sklearn implementation of PCA. 

For benchmarking as accurately as possibly against classical fully-connected neural networks, we used the code from \cite{nielsen} for training classical neural networks. For the quantum-assisted neural networks, we adapted this code to use a quantum procedure for the dot product computations. For the orthogonal neural networks, we implemented our new training algorithm for quantum orthogonal neural networks (See Section \ref{sec:quantum_orthoNN}), and we also developed the code for classical orthogonal neural networks based on singular value decomposition, following \cite{jia2019orthogonal}. 

\subsubsection{Optimizations for the hardware demonstration}

\begin{figure}[h]
\centering
\begin{subfigure}{.5\textwidth}
  \centering
  \includegraphics[width=\linewidth]{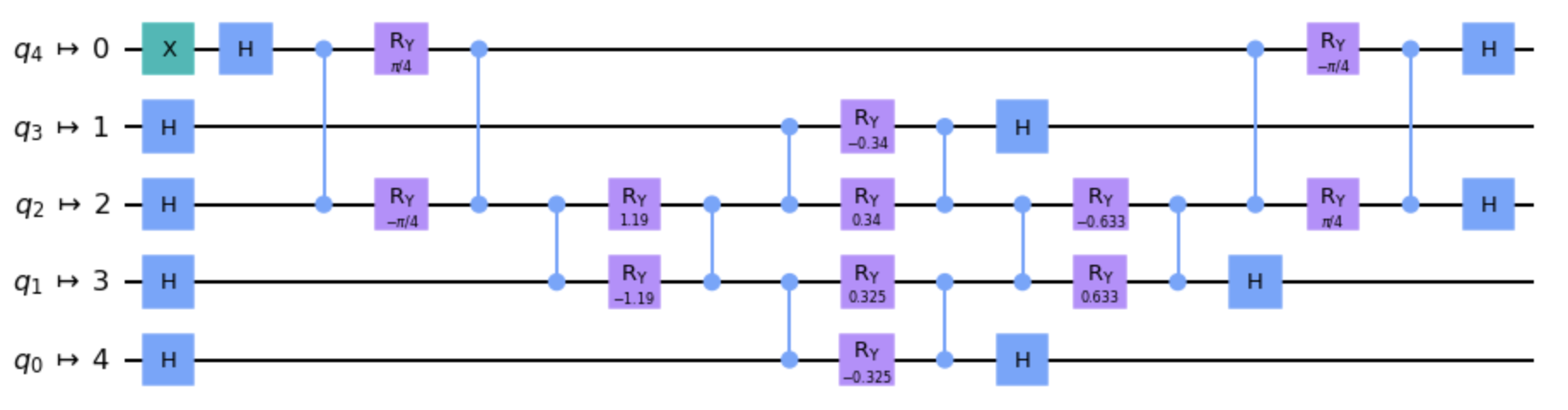}
  \caption{}
  \label{fig:QONNcircuit_rectangular_ibm}
\end{subfigure}%
\newline
\begin{subfigure}{.5\textwidth}
  \centering
  \includegraphics[width=\linewidth]{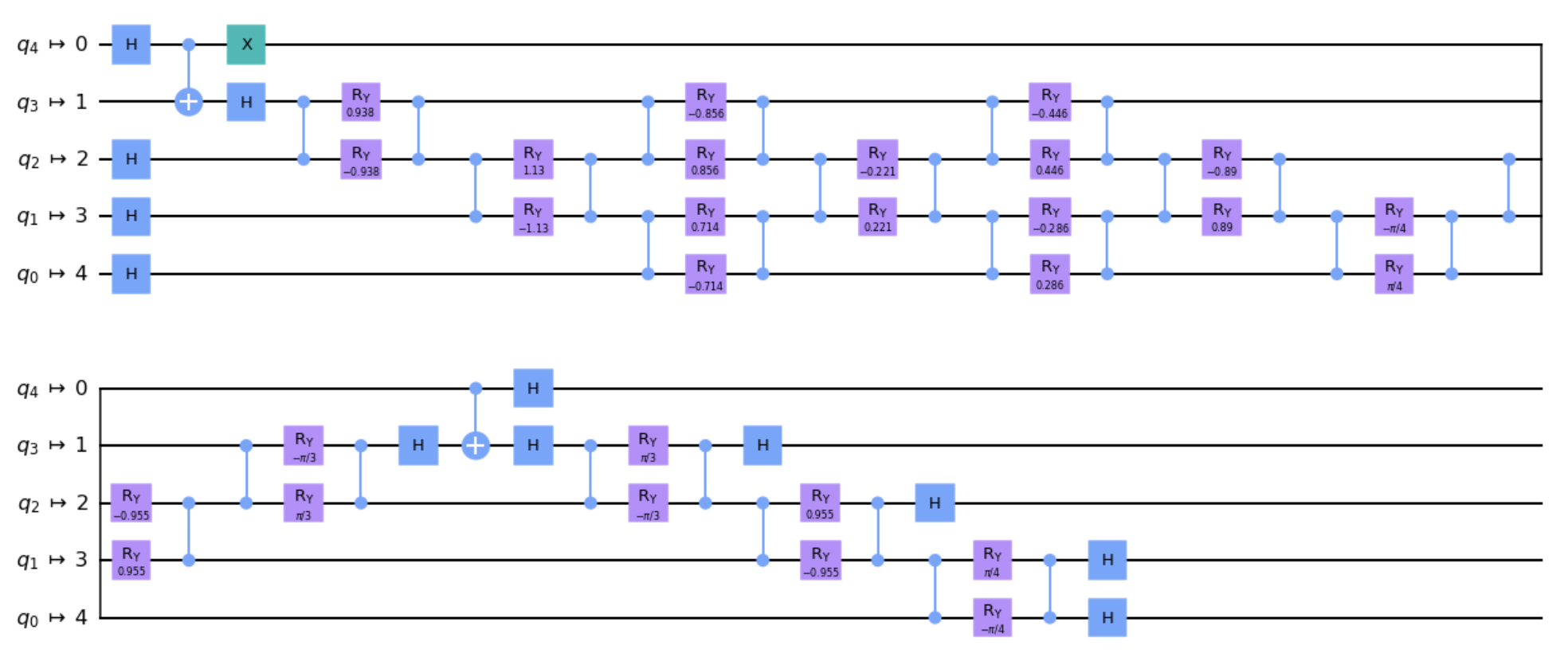}
  \caption{}
\end{subfigure}
\caption{ The optimized quantum circuits on qiskit for (a) an [4,2] quantum-assisted NN layer and (b) an [4,2] quantum orthogonal NN layer}
\label{final_circuitIBM}
\end{figure}

Before describing the experimental results, we outline briefly optimizations of our circuit design.

First, using unary encoding for loading the data is very useful in order to mitigate errors that arise from the hardware. In fact, all results that correspond to outcomes which are not unary strings can be discarded and this dramatically increases the accuracy of the computation, as we see in the experimental results below.

Further optimizations were performed with respect to the layout of the hardware machines and the translation of the RBS gates into native hardware gates. 
We provide in Fig. \ref{final_circuitIBM} the final qiskit circuits that were used for one $[4,2]$ layer of the quantum-assisted neural network and of the quantum orthogonal neural network, where an efficient compilation of the original circuits have led to a reduction of the number of gates, in particular by removing a number of Hadamard gates that were appearing as pairs. 
The purple boxes correspond to the $R_y$ single qubit gates with the corresponding parameter noted within the box.  
Descriptions of this $[8,4]$ layer of the quantum-assisted neural network and the $[8,2]$ layer of the quantum orthogonal neural network can be found in the Appendix, Section \ref{sec:additional_results}.

\subsubsection{Additional quantum circuits}

Circuits for one $[8,4]$ layer of the quantum-assisted neural network and one $[8,2]$ layer of the quantum orthogonal neural network are shown below. 

\begin{figure}[!h]
    \centering
    \includegraphics[width=0.5\textwidth]{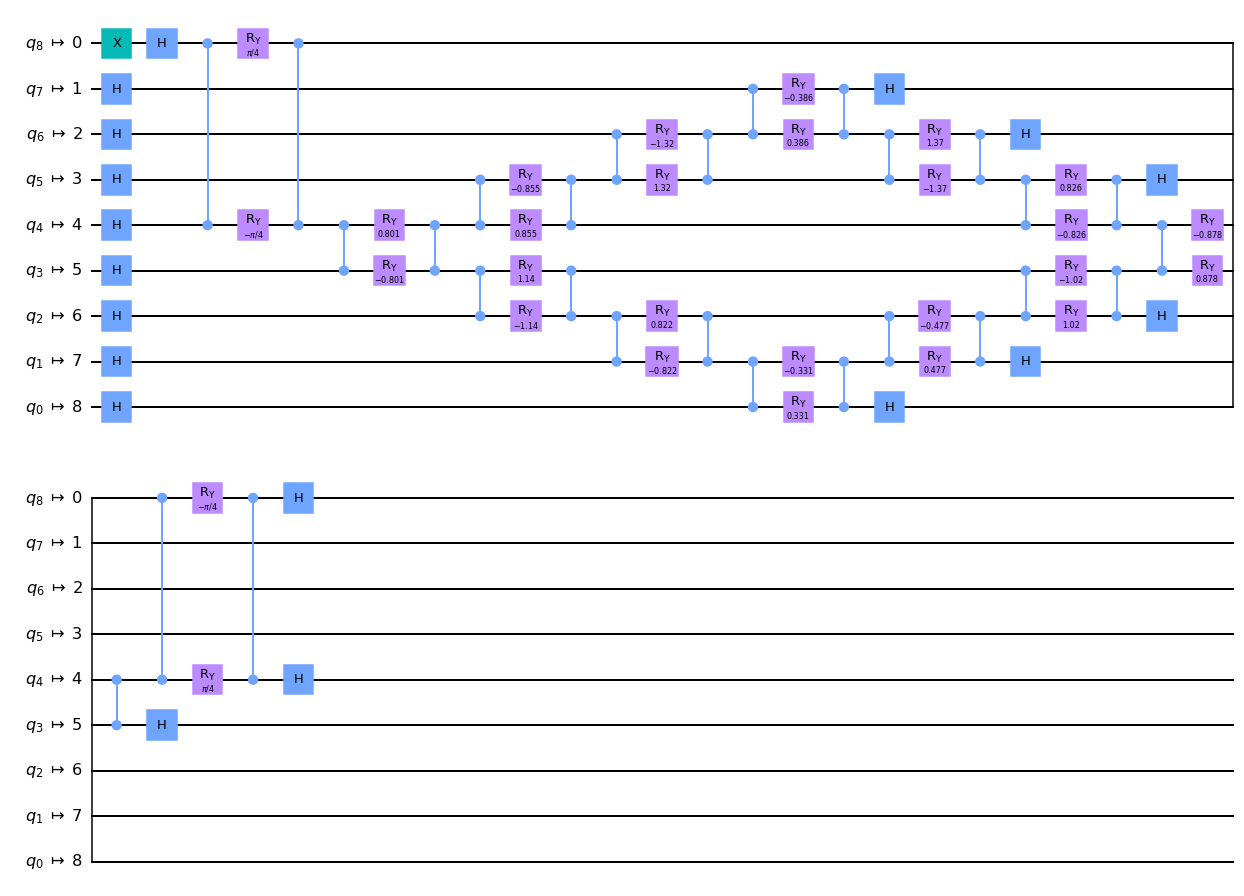}
    \caption{The optimized qiskit circuit for an [8,4] quantum-assisted NN layer}
    \label{qiskit8-4}
\end{figure}

\begin{figure}[!h]
    \centering
    \includegraphics[width=0.5\textwidth]{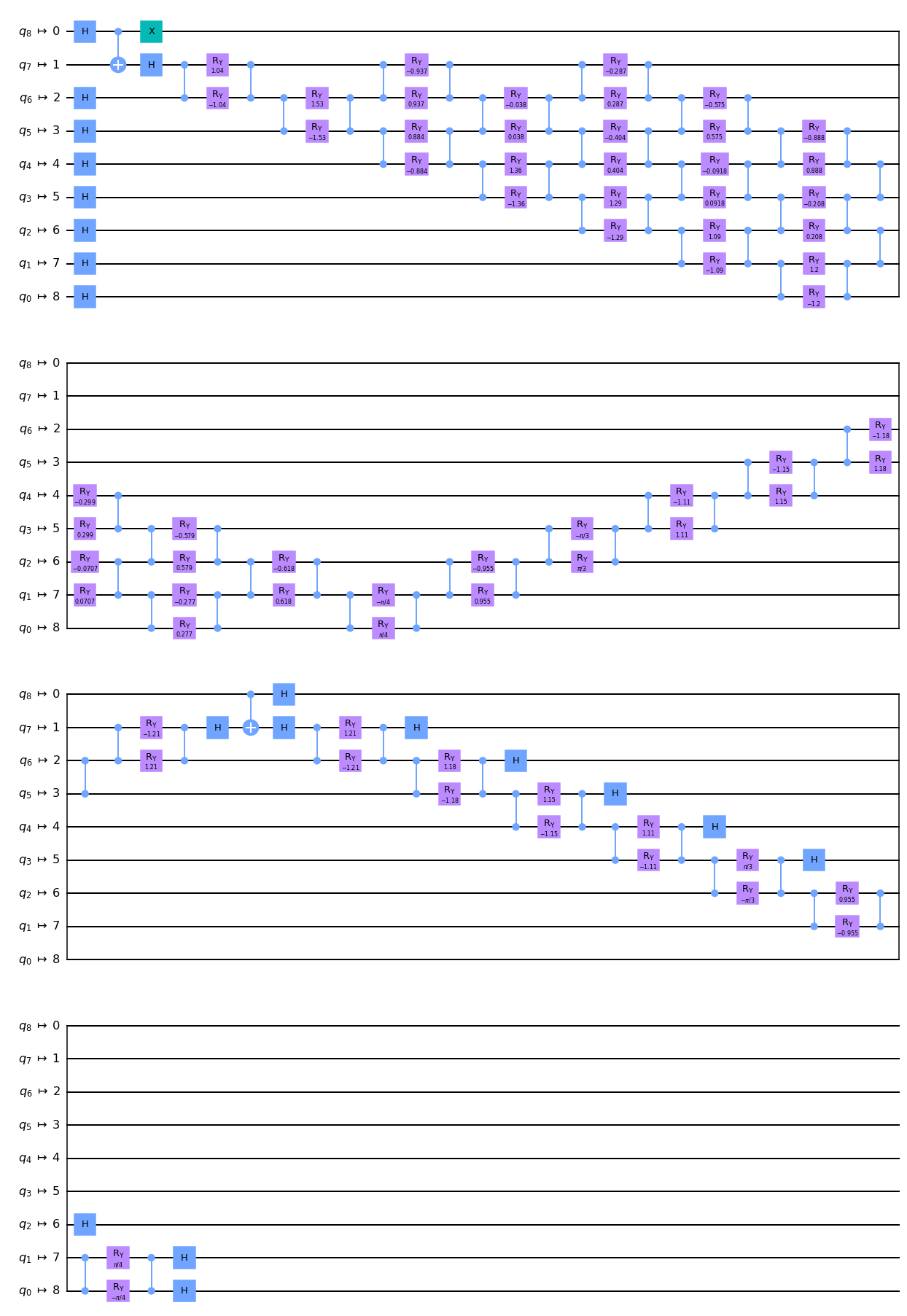}
    \caption{The optimized qiskit circuit for an [8,2] quantum orthogonal NN layer}
    \label{qiskit8-2}
\end{figure}

\subsubsection{Additional simulation results}

\begin{figure}[!h]
    \centering
    \includegraphics[width=0.4\textwidth]{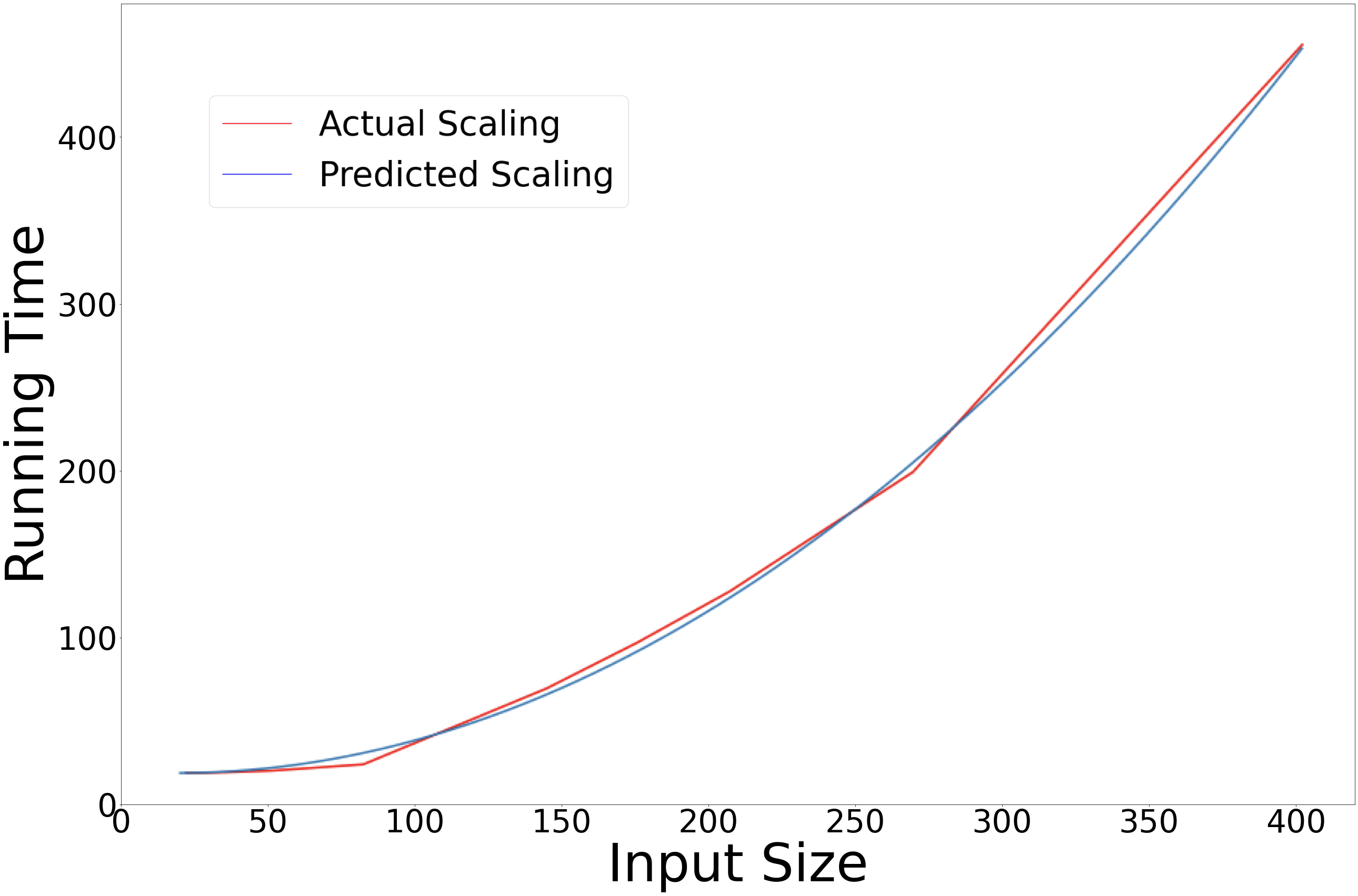}
    \caption{Experimental scaling of the running time for growing dimension $n$ of training an $[n,n,2]$ orthogonal neural network and the predicted scaling 0.006($0.5 n^2 + 1.5n - 3$).}
    \label{time}
\end{figure}

As shown in Section \ref{sec:trainingquantum}, run time of training of the orthogonal neural networks based on the quantum pyramid circuit scales linearly with respect to the number of parameters. This is corroborated by our results shown in Fig.\ref{time}. In particular, we trained $[n,n,2]$ quantum orthogonal neural networks, for different integer values of $n \in [2,392]$ and saw that the running time grows with respect to the number of parameters as $0.5 n^2 + 1.5n - 3$. This is asymptotically better than the previously known training algorithms for orthogonal neural networks that run in time $O(n^3)$. 

\begin{figure}[!h]
    \centering
    \includegraphics[width=0.4\textwidth]{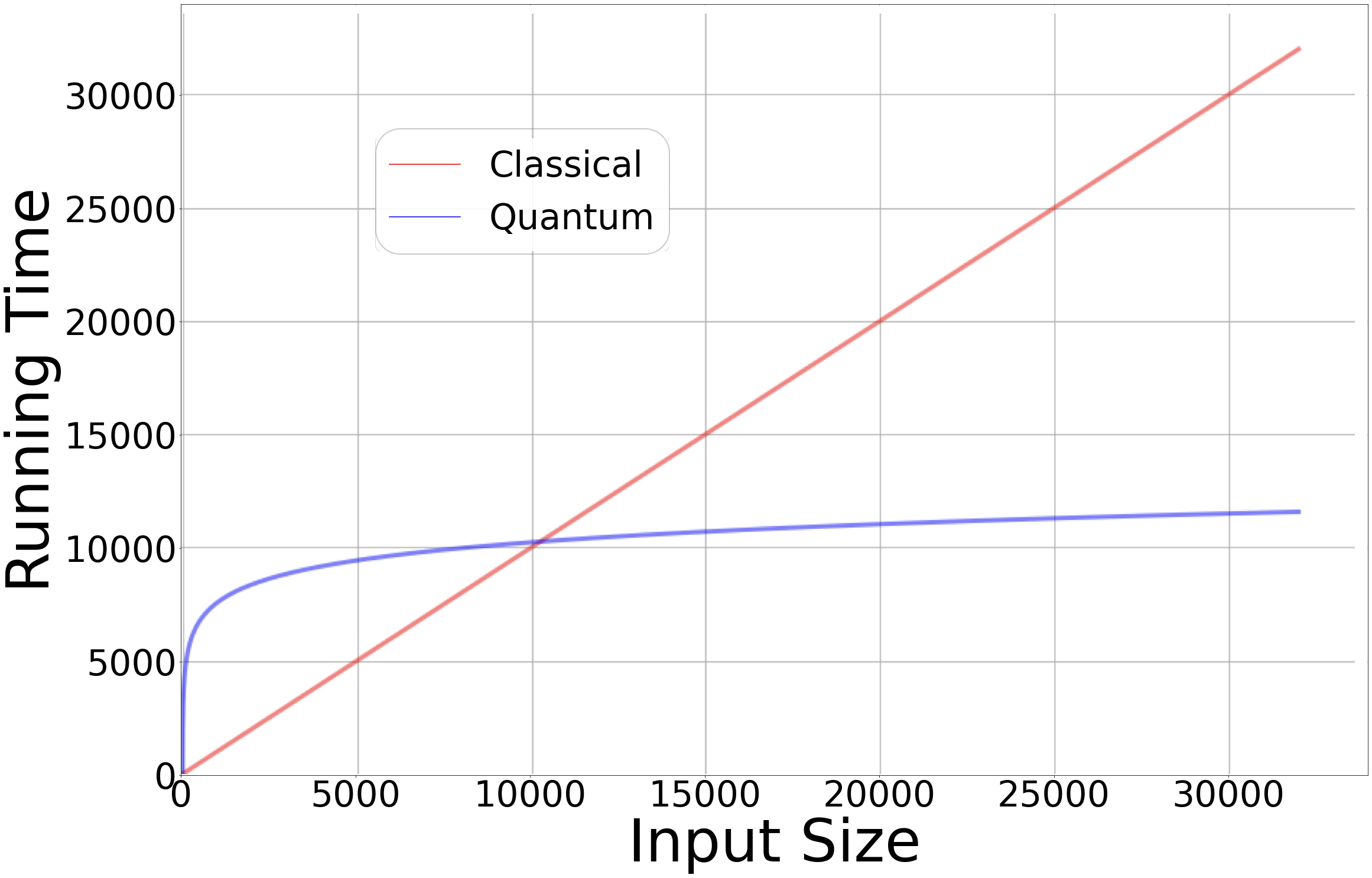}
    \caption{Scaling of quantum steps for the inner 
product estimation between two n-dim vectors  
(blue: $400(2\log n-1)$)  versus for the classical inner 
product computation (red: $n$). The crossover point is for $100 \times 100$ pixel images.}
    \label{time_qnn}
\end{figure}

Note that the run time of quantum-assisted neural networks can be analyzed theoretically as in \cite{NearestCentroid2021}, where the main difference compared to classical fully-connected neural networks is on the computation of the inner product between vectors. While on a single CPU an inner-product computation between two $n$-dimensional vectors takes $n$ steps, on a single quantum processing unit with the ability to perform parallel quantum gates, inner product estimation uses a quantum circuit of depth only $2\log(n)-1$. These shallow quantum circuits need to be repeated a number of times (shots) in order to get an accurate estimation of the inner product. By applying Chernoff bounds on estimation of a binomial distribution, we see that an $\epsilon$-accurate result in the quantum circuit needs to be repeated $O(1/\epsilon^2)$ times. From trial and error, we found that repeating the quantum circuits 400 times (independent of the dimension $n$) suffices to get the desired accuracy. In Fig.\ref{time_qnn} we compare the scaling of the number of steps on a quantum chip versus a single classical and see that for images of size $100 \times 100$ the number of quantum steps required starts to become smaller. A smaller number of theoretical steps does not imply necessarily a faster running time, since here we assume the calculations to be performed by a single quantum processor which can apply gates on different qubits in parallel (see e.g. \cite{parallelGates}) and we only compare to a single CPU. As we have said, using GPUs or TPUs can substantially speedup the time to perform a large number of inner products, and hence we do not think that speed should be the primary goal of quantum neural networks. One should also take into account the time to apply one quantum or one classical step, which can vary from machine to machine.

\subsubsection{Additional hardware results}

\begin{figure}[!h]
    \centering
    \includegraphics[width=0.5\textwidth]{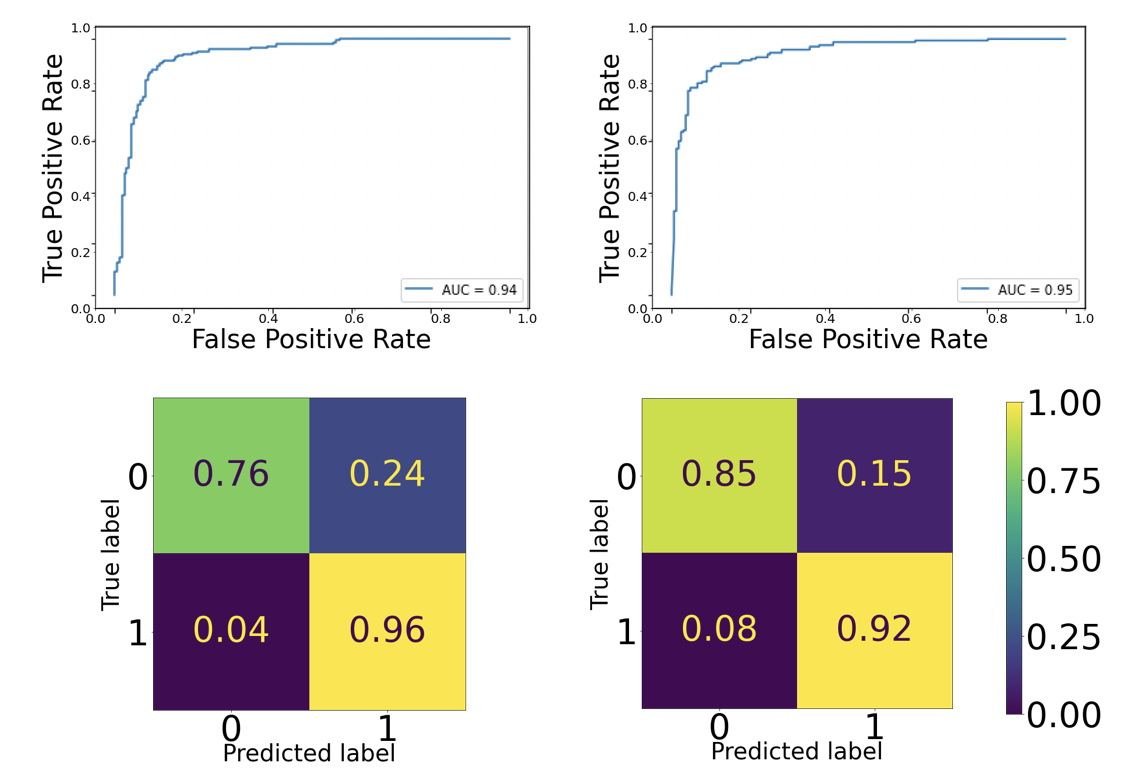}
    \caption{Training of an [8,4,2] quantum-assisted neural network on the Pneumonia-MNIST with training set of size 2428 (1214 – 1214)  and test size 336 (168-168) with training on a quantum simulator and inference on a quantum hardware (left: AUC (test set) = 0.94, ACC = (test set) 86.01\%), and on the quantum simulator(right: AUC = (test set) 0.95, ACC (test set) = 88.39\%).
}
    \label{hwqNNPneumonia}
\end{figure}

\begin{figure}[!h]
    \centering
    \includegraphics[width=0.5\textwidth]{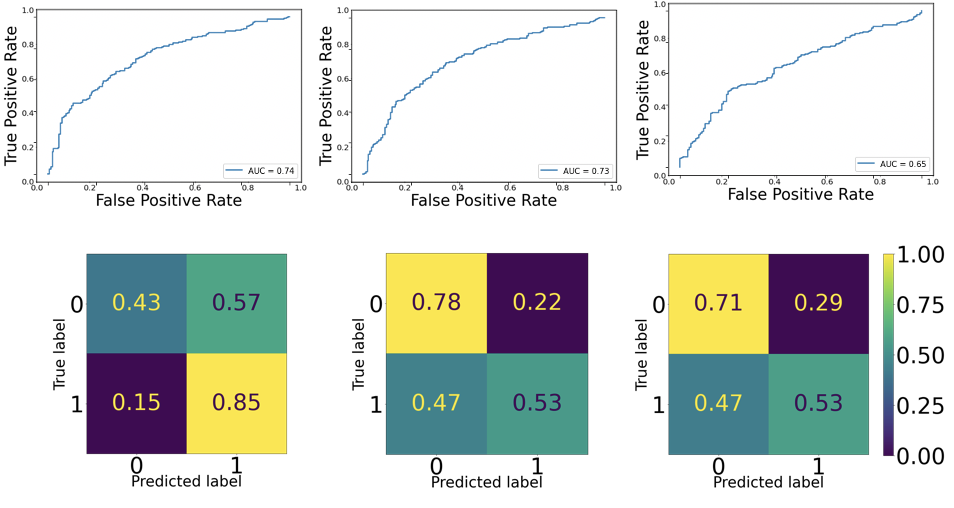}
    \caption{Training of an [4,4,2] quantum-assisted neural network on the Retina-MNIST with training set of size 100 (50-50) and test size 400 (174-226) with both training and testing on a quantum simulator (left: AUC (test set) = 0.74, ACC (test set) = 66.5\%), the training on a quantum simulator and the inference on the quantum hardware (middle: AUC (test set) = 0.73, ACC (test set) = 63.75\%), and last with both the training and testing on the quantum hardware (right: AUC (test set) = 0.65, ACC (test set) = 60\%).}
    \label{hwqNNRetina}
\end{figure}

\begin{figure}[!h]
    \centering
    \includegraphics[width=0.5\textwidth]{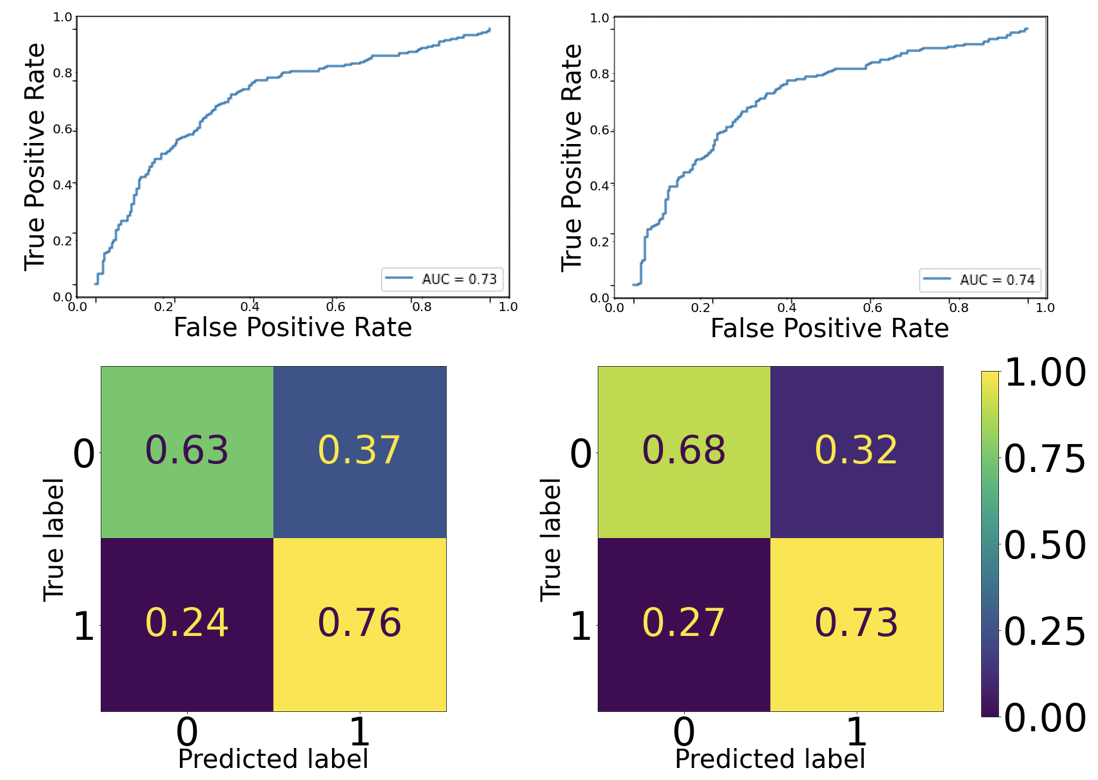}
    \caption{Training of an [4,2] quantum orthogonal neural network on the Retina-MNIST with training set of size 1080 (486-594) and test size 400 (174-226) with testing on a quantum simulator (left: AUC (test set) = 0.73, ACC (test set) = 70\%), and testing on the quantum hardware (right: AUC (test set) = 0.74, ACC (test set) = 70.75\%).
}
    \label{hwqOrthoNNRetina}
\end{figure}

In Fig.\ref{hwqNNPneumonia}, \ref{hwqNNRetina}, and \ref{hwqOrthoNNRetina} AUC curves, ACC and confusion matrices for some of the experiments we performed are shown. Examples include both Pneumonia and Retina datasets, quantum-assisted and orthogonal neural networks, and for training and forward inference on simulators or quantum hardware.

We also present here an analysis for the quantum circuits we used in order to perform the quantum-assisted neural networks, namely the quantum inner product estimation circuits, which are useful for other applications beyond training neural networks. 
In particular, we present inference results aggregated from 624 data points found in the test set of Pneumonia-MNIST and from 400 data points found in the test set of Retina-MNIST. Results are shown for both the [4,4,2] and [8,4,2] quantum-assisted neural networks, and we check the estimated value of the first node in the final layer of the neural network (which corresponds to the first class of the binary classification task) on the simulator versus on the real hardware. Weights of the trained model were obtained from the quantum simulator. This allows us to run the same quantum circuits a large number of times on different inputs and get an estimate of how well the quantum hardware can perform this particular application task.

\begin{figure}[!h]
    \centering
    \includegraphics[width=0.4\textwidth]{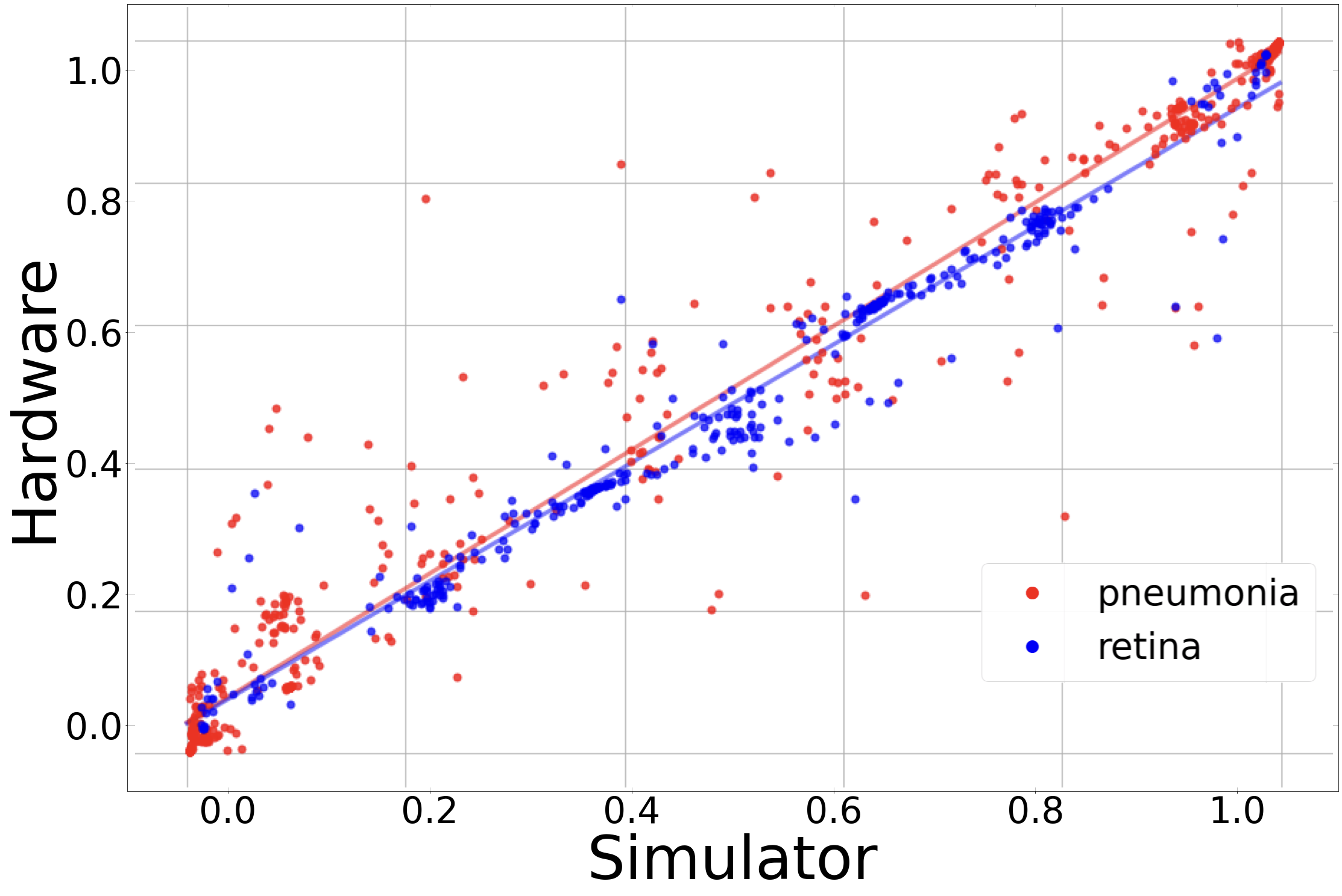}
    \caption{Experimental versus simulated value of the first node of the final layer of the [4,4,2] qNN. Pneumonia: slope=$0.94 \pm 0.09$, intercept=$0.04\pm 0.06$; Retina: slope=$0.90\pm 0.20$, intercept=$0.04 \pm 0.14$
}
    \label{4-hw}
\end{figure}

\begin{figure}[!h]
    \centering
    \includegraphics[width=0.4\textwidth]{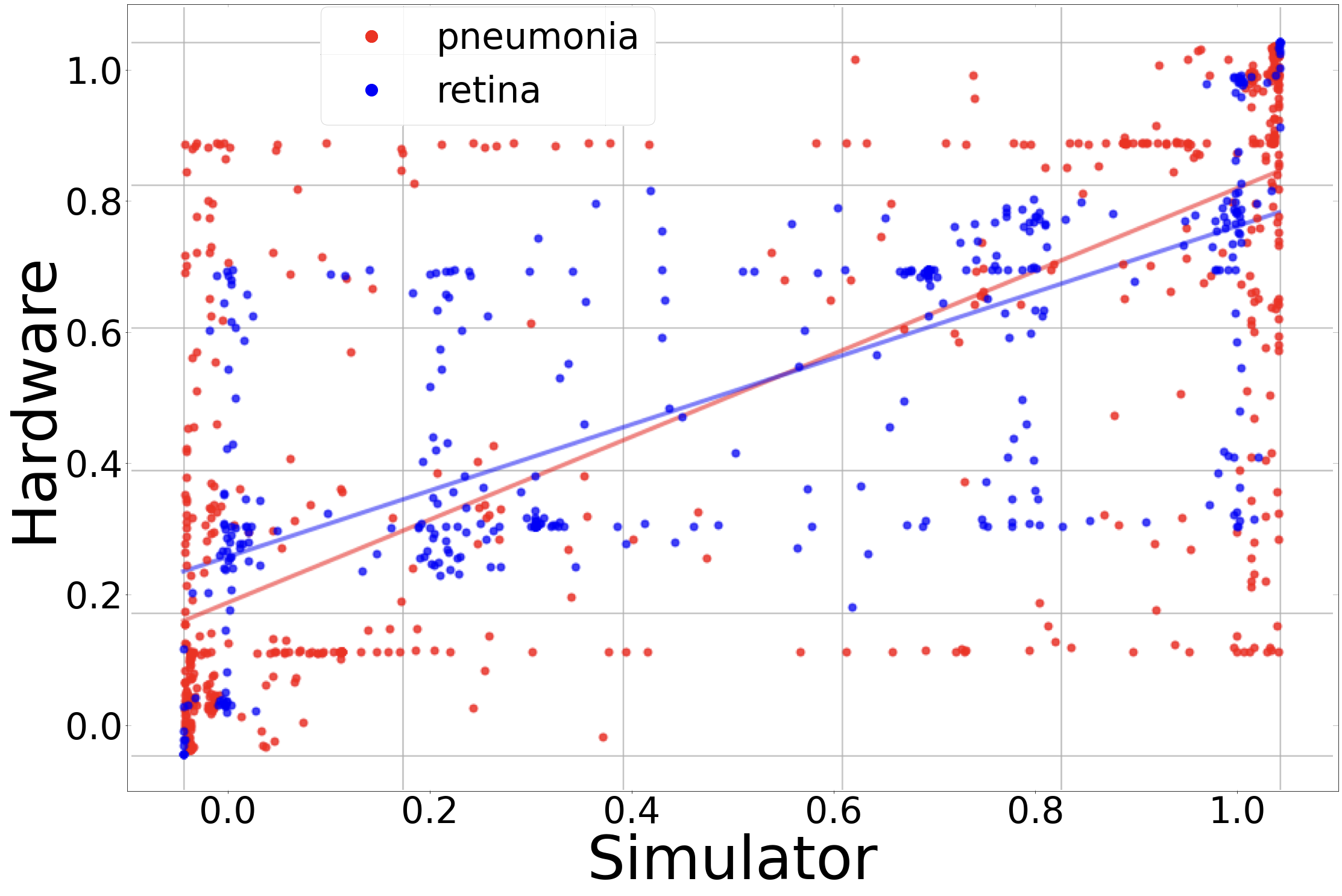}
    \caption{Experimental versus simulated value of the first node of the final layer of the [8,4,2] qNN. Pneumonia: slope=$0.63 \pm 0.09$, intercept=$0.19\pm 0.06$; Retina: slope=$0.50\pm 0.15$, intercept=$0.26 \pm 0.10$}
    \label{8-hw}
\end{figure}

In Fig.\ref{4-hw}, we see that for the $[4,4,2]$ architecture, results of the hardware executions are quite close to the simulation for both data sets. In Fig.\ref{8-hw}, we see that for the $[8,4,2]$ architecture, while most of the hardware results agree with the simulations, there is a large fraction of the points where an error occurs in the quantum circuit, causing the results between simulation and hardware execution to diverge. This divergence does not necessarily translate to a drop in accuracy, as reflected in Table \ref{table}, since some points that were misclassified by the simulator can now be classified correctly on the quantum hardware due to possible errors in the quantum circuits. It is quite clear that the larger circuits we used are pushing the boundaries of what the current quantum machines can perform reliably, and one would need better and more robust hardware to increase confidence on how quantum machine learning techniques can help in image classification tasks.  

\end{document}